\documentclass[aps,prl,twocolumn,superscriptaddress]{revtex4-2}

\usepackage{soul} 
\usepackage{graphicx}
\usepackage{xcolor}
\usepackage{times}
\usepackage[caption=false]{subfig}
\usepackage{amsmath}
\usepackage{slashed}%
\usepackage{bbold}%
\usepackage{bm}
\usepackage{mathtools}
\usepackage{physics}
\usepackage{amsthm}
\usepackage{newunicodechar}
\usepackage{geometry}
\usepackage{simpler-wick}
\usepackage[compat=1.1.0]{tikz-feynman}
\usepackage{feynmp-auto}
\usepackage{amssymb}
\usepackage{placeins}
\newcommand{\eqgraph}[3]{%
  \begin{gathered}
  \raisebox{0pt}[\dimexpr\height+#1][\dimexpr\depth+#2]{\ignorespaces#3\unskip}%
  \end{gathered}
}

\def\iv{\textbf{\textit{v}}}

\def\ik{\textbf{\textit{k}}}
\def\iA{\textbf{\textit{A}}}

\def\iq{\textbf{\textit{q}}}

\def\E{\E_\ik}

\def\MM{{\mathcal{M}}}

\def\iv{{i\nu_n}}
\def\io{{i\omega_n}}
\def\ioo{(i\omega_n)^2}

\def\Ag{{\text{A}_{1\textrm{g}}}}
\def\Bg{\text{B}_{1\textrm{g}}}

\def\TK{{\Tilde{K}}}
\def\TE{{\Tilde{E}}}
\def\TI{{\Tilde{\mathcal{I}}}}

\tikzset{
  double dashed/.style={
    double, 
    decoration={snake, segment length=6pt, amplitude=0pt, post length=1pt, pre length=1pt},
    postaction={decorate},
    thick,
    dashed,
    }}

\theoremstyle{thmstyleone}%

\theoremstyle{thmstyletwo}%

\theoremstyle{thmstylethree}%

\begin{document}

\title{Direct observation of the Higgs particle in a superconductor by non-equilibrium Raman scattering}

\author{Tomke E. Glier}
\email[]{tglier@physnet.uni-hamburg.de}
\affiliation{Institute of Nanostructure and Solid State Physics, Universität Hamburg, Hamburg, 22761, Germany.}
\affiliation{Present Address: Department of Physics, University of Colorado Boulder, Boulder, Colorado, 80309, USA}

\author{Sida Tian}
\affiliation{Max Planck Institute for Solid State Research, Stuttgart, 70569, Germany.}

\author{Mika Rerrer}
\affiliation{Institute of Nanostructure and Solid State Physics, Universität Hamburg, Hamburg, 22761, Germany.}

\author{Lea Westphal}
\affiliation{Institute of Nanostructure and Solid State Physics, Universität Hamburg, Hamburg, 22761, Germany.}
\affiliation{Present Address: Forschungsneutronenquelle Heinz Maier-Leibnitz (FRM II), Technische Universität München, Garching b. München, 85748, Germany}

\author{Garret Lüllau}
\affiliation{Institute of Nanostructure and Solid State Physics, Universität Hamburg, Hamburg, 22761, Germany.}

\author{Liwen Feng}
\affiliation{Institute of Solid State and Materials Physics, TUD Dresden University of Technology, Dresden, 01062, Germany.}

\author{Jakob Dolgner}
\affiliation{Max Planck Institute for Solid State Research, Stuttgart, 70569, Germany.}

\author{Rafael Haenel}
\affiliation{Max Planck Institute for Solid State Research, Stuttgart, 70569, Germany.}

\author{Marta Zonno}
\affiliation{Max Planck Institute for Solid State Research, Stuttgart, 70569, Germany.}
\affiliation{Quantum Matter Institute, University of British Columbia, Vancouver, BC V6T 1Z4, Canada.}
\affiliation{Department of Physics \& Astronomy, University of British Columbia, Vancouver, BC V6T 1Z1, Canada.}

\author{Hiroshi Eisaki}
\affiliation{Nanoelectronics Research Institute, National Institute of Advanced Industrial Science and Technology, Tsukuba, Ibaraki 305-8568, Japan.}

\author{Martin Greven}
\affiliation{School of Physics and Astronomy, University of Minnesota, Minneapolis, MN, USA.}

\author{Andrea Damascelli}
\affiliation{Quantum Matter Institute, University of British Columbia, Vancouver, BC V6T 1Z4, Canada.}
\affiliation{Department of Physics \& Astronomy, University of British Columbia, Vancouver, BC V6T 1Z1, Canada.}

\author{Stefan Kaiser}
\email[]{stefan.kaiser@tu-dresden.de}
\affiliation{Institute of Solid State and Materials Physics, TUD Dresden University of Technology, Dresden, 01062, Germany.}

\author{Dirk Manske}
\email[]{d.manske@fkf.mpg.de}
\affiliation{Max Planck Institute for Solid State Research, Stuttgart, 70569, Germany.}

\author{Michael Rübhausen}
\email[]{mruebhau@physnet.uni-hamburg.de}
\affiliation{Institute of Nanostructure and Solid State Physics, Universität Hamburg, Hamburg, 22761, Germany.}

\date{November 30, 2024}

\begin{abstract}
Even before its role in electroweak symmetry breaking, the Anderson-Higgs mechanism was introduced to explain the Meissner effect in superconductors. Spontaneous symmetry-breaking yields massless phase modes representing the low-energy excitations of the Mexican-Hat potential. Only in superconductors the phase mode is shifted towards higher energies owing to the gauge field of the charged condensate. This results in a low-energy excitation spectrum governed by the Higgs mode. Consequently, the Meissner effect signifies a macroscopic quantum condensate in which a photon acquires mass, representing a one-to-one analogy to high-energy physics. We report on the direct observation of the Higgs particle in the high-temperature superconductor Bi-2212 by developing an innovative technique to study its symmetries and energies after a "soft quench" of the Mexican-Hat potential. Population inversion of the metastable Higgs particle induced by an initial laser pulse allows identifying the polarization-dependent Higgs modes as an additional anti-Stokes Raman-scattering signal. Within Ginzburg-Landau theory, the Higgs-mode energy is connected to the Cooper-pair coherence length. Within a BCS weak-coupling model we develop a quantitative and coherent description of single-particle and two-particle channels. This opens the avenue for Higgs Spectroscopy in quantum condensates and provides a unique pathway to control and explore Higgs physics.
\end{abstract}

\maketitle

The Higgs mode is a Raman-active excitation, and was observed in NbSe$_2$ in 1980 by R. Sooryakumar and M.V. Klein.\cite{Sooryakumar1980} The microscopic nature of this discovery was pointed out later by Y. Nambu to P. Higgs, and was seen by both as a first observation of the Higgs mode in experimental physics.\cite{Higgs2007} However, the Raman cross section of the Higgs mode in superconductors is generally very small, which results in Higgs modes that remain invisible in most Raman experiments. In NbSe$_2$, a unique interplay between a soft phonon in the charge-density wave state and the Higgs mode in the superconducting (SC) state leads to a distinct and sharp mode slightly below $2\Delta$.\cite{Feng2023, Cea2014, Littlewood1982} This interplay was studied as a function of temperature and pressure.\cite{Measson2014, Grasset2018} Very recently, the coupling between charge-density wave and Higgs mode was also studied by time-resolved spectroscopy of amplitude- and phase-sensitive high harmonics.\cite{Feng2023, Chu2023} Up to now, the observation of the Higgs mode in Raman scattering has been limited to the particular case of NbSe$_2$. In non-charge-density-wave systems the Mexican-Hat potential must be specifically quenched, for instance by a light pulse.\cite{Matsunaga2014}

Spontaneous Raman spectroscopy on high-temperature superconductors has been used with regard to excitation-energy resonances and dynamics of the SC gap feature. Static Raman spectra of Bi$_{2}$Sr$_{2}$CaCu$_2$O$_{8+\delta}$ (Bi-2212) in B$_{1\textrm{g}}$ symmetry show distinct resonances between 2~eV and 3.5~eV indicating a multicomponent origin of the excitation spectrum close to $2\Delta$.\cite{Budelmann2005} Furthermore, transient time-resolved Raman scattering in the SC state enables the study of pair-breaking (PB) excitations and the dynamics of the SC order parameter.\cite{Saichu2009} By utilizing the Bose factor, time-resolved Stokes-anti-Stokes Raman scattering has been applied as a stroboscopic tool to determine transient temperatures and melting processes in highly excited states of phonons.\cite{Pellatz2021, Han2019}

Advances in THz-laser technology have enabled the investigation of Higgs modes in several classes of superconductors via non-equilibrium THz spectroscopy.\cite{Matsunaga2014, Katsumi2018, Chu2020, Vaswani2021, Wang2022, Reinhoffer2022, Katsumi2023, Shimano2020, Kim2023a} These experiments involve either an impulsive excitation of the Higgs mode based on a \textit{quench} of the SC state or a \textit{drive} of the Higgs mode, resulting in coherent oscillations or high-harmonic generation, respectively. Experiments on s-wave superconductors show that Higgs modes are indeed stable excitations.\cite{Shimano2020} However, in d-wave superconductors, it was expected and measured that Higgs modes are metastable due to interactions with remanent nodal quasiparticles.\cite{Barlas2013, Katsumi2023, Schwarz2020} The theoretical models are based on time-dependent Ginzburg-Landau approaches and time-dependent BCS theories in the framework of a pseudospin model.\cite{Vaswani2021, Tsuji2015, Hannibal2015, Schwarz2020} The important influence of disorder and impurity effects on high-harmonic generation was investigated from a theoretical point of view.\cite{Benfatto2023, Seibold2021, Udina2022} Furthermore, the SC state exhibits a manifold of low-energy excitations, such as pair breaking, Josephson plasmons, Bardasis-Schrieffer, or Leggett modes.\cite{Sun2020, Gabriele2021, Sellati2023, Barlas2013}  Some of them are difficult to distinguish from the Higgs mode in an experimental data set. Thus, it is of great importance that the Higgs modes can be classified based on the symmetry of the SC condensate, the symmetry of the quench, and the symmetry of the Higgs excitation.\cite{Schwarz2020a} This unique diversity of physical properties enables the exploration and control of Higgs physics in an unprecedented way.

    \begin{figure*}[htp!]
	\includegraphics[width=0.75\linewidth
	]{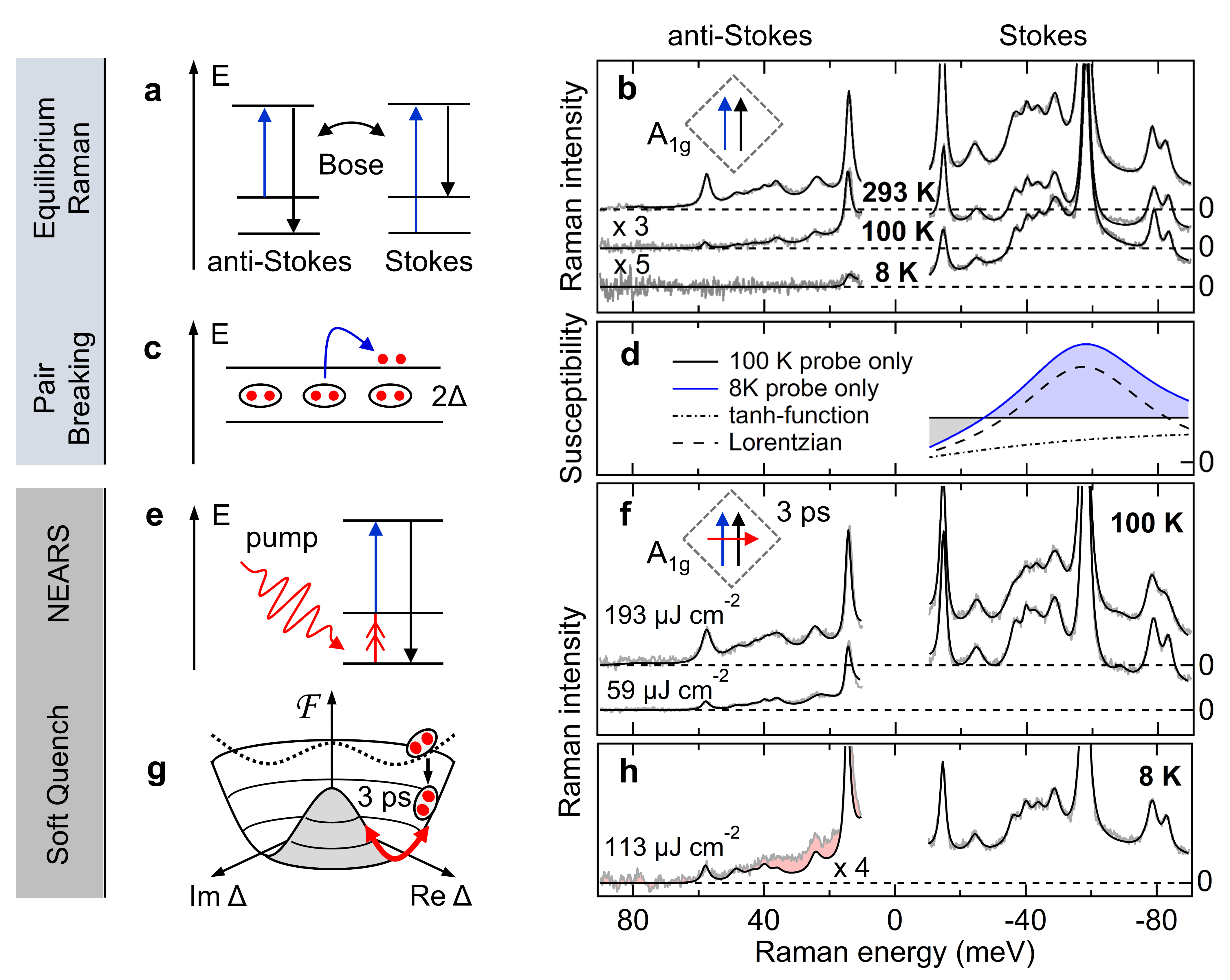}
	\caption{\label{fig:fig1}Equilibrium (a-d) vs. non-equilibrium (e-h) resonant Raman scattering. (a) Energy diagrams of Stokes and anti-Stokes Raman scattering. The scattering cross-sections of equilibrium Stokes (energy-loss) and anti-Stokes (energy-gain) scattering are linked by the Bose function. The Raman probe excitation is depicted in blue, the scattered light is shown in black. (b) Stokes and anti-Stokes equilibrium Raman data of Bi-2212 (T$_{\textrm{c}}$ = 92~K\cite{Eisaki2004}, see Fig. S3) in an A$_{1\textrm{g}}$ scattering configuration at indicated base temperatures. The measured intensities are shown in gray, and the black solid lines represent fits to the data (see SI S.6). The dashed lines represent zero intensity for the three displayed data sets. The anti-Stokes Raman spectrum at 8~K is vanishing because the anti-Stokes Bose-function at 8~K is effectively zero. (c) The Raman response is dominated by PB of Cooper pairs. (d) Electronic Raman susceptibility in probe-only measurements as extracted from the parameterization (see SI S.6, Fig. S10). At 100~K the electronic background is constant (black solid line). At 8~K, the PB feature dominates the response (blue solid line). The PB feature consists by a tanh-function (dash-dotted line) and a Lorentzian (dashed line). (e)~After a pump (red wiggly line), the populated Higgs mode (red double arrow) can be detected by an annihilation process (anti-Stokes). (f)  Pump-probe Stokes and anti-Stokes Raman spectra at a fluence of 59~µJ~cm$^{-2}$ and 193~µJ~cm$^{-2}$ at a delay of 3 ps. Data were taken in A$_{1\textrm{g}}$ symmetry at 100~K base temperature. (g) The modification of the free-energy landscape (dotted line) induced by the pump (\textit{soft quench}) enables the annihilation of the Higgs oscillations (solid red line) at a delay time of 3 ps. (h) Below T$_\textrm{c}$, at 8~K base temperature, and an exemplary fluence of 113~µJ~cm$^{-2}$ the anti-Stokes intensity cannot be described via the Stokes susceptibility and a new NEARS feature occurs (red), which we attribute to the A$_{1\textrm{g}}$ Higgs mode. Please note that at low temperatures, higher fluences were not applied to remain in the superconducting state.}
    \end{figure*}

\subsection{Non-Equilibrium Anti-Stokes Raman Scattering}

Here we resolve the ambiguity of the identification of Higgs modes in superconductors by introducing a new spectroscopic technique: Non-Equilibrium Anti-Stokes Raman Scattering (NEARS) that makes use of the metastable character of the Higgs particle. NEARS utilizes a so-called \textit{soft quench} of the Mexican-Hat potential, as we will detail below, to populate Higgs modes of different symmetries and probes them by anti-Stokes Raman scattering. The conventional spontaneous Raman scattering signal is proportional to a four-photon Green's function where Stokes and anti-Stokes signals are generated simultaneously. Compared to conventional Raman scattering, non-equilibrium Raman scattering overpopulates the lowest-energy excitation of the superconductor, which is the Higgs particle, as a consequence of the relaxation process of the free-energy landscape. This population inversion (see SI Fig. S9) can be measured by a comparison between the simultaneous Stokes and anti-Stokes signals, i.e. energy-loss and energy-gain data of the spontaneous Raman process.

Conventional Raman scattering excites quasiparticles leading to energy-loss spectroscopic features on the Stokes side (see Fig. 1a-d). In superconductors, this technique is sensitive to low-energy excitations, such as PB excitations\cite{Munnikes2011, Devereaux1995}, see Fig. 1c and d, density-correlation functions of Josephson plasmons\cite{Sellati2023}, Leggett modes\cite{Blumberg2007}, and Bardasis-Schrieffer modes\cite{Maiti2016,Bohm2018}. Our aim is to measure the relaxation of a superconductor and the concomitant population of Higgs modes in the quasi-static limit and in nearly thermal equilibrium.

In this work, NEARS is used to study the high-temperature superconductor Bi-2212 (optimally doped, T$_{\textrm{c}}$ = 92 K, see SI Fig. S3).\cite{Eisaki2004} To verify the condition of quasi-equilibrium in the non-pumped data, we use the fact that Stokes and anti-Stokes scattering intensities in equilibrium are linked by the Bose function. We measure anti-Stokes and Stokes data and convert the Stokes spectrum to the corresponding anti-Stokes response by applying the Bose function. Matching Stokes and anti-Stokes spectra over a large low-energy range ensures that we have determined the temperature of the sample under laser illumination, excluded any background artifacts from the pulsed laser source, and avoided unwanted self-excitation effects induced by the pulsed Raman probe (see Fig. S5 and S8).\cite{Bock1995} Indeed, in Fig. 1b, a perfect agreement between the converted and measured anti-Stokes data is observed when assuming a heating of 9~K due to the Raman probe. 
We apply a scattering geometry that probes A$_{1\textrm{g}}$ symmetry with incidence and scattered light fields parallel to each other and rotated 45$^\circ$ to the CuO$_2$ plane of Bi-2212, as indicated in the pictogram of Fig. 1b. An incident photon energy of 3~eV (400~nm), a laser power of 4.8~mW, and a pulse duration of 1.2~ps were used. In the SC state, at low temperatures, and on the Stokes side, we observe the well-known phonons and excitations around twice the SC gap energy ($2\Delta \approx 60$~meV) due to the predominant PB process of Bi-2212 as expected for a sample with T$_{\textrm{c}} \approx$ 92~K.\cite{Budelmann2005, Klein2006} As shown in Fig. 1d, the Raman susceptibility of the PB feature is described by a tanh-function and a Lorentzian according to \cite{Budelmann2005}. The PB process is sketched in Fig. 1c. At 8~K, the anti-Stokes side is essentially dark as no quasiparticles are thermally excited at T $\approx$ 0~K and, hence, no annihilation of excitations can occur (see 8~K data in Fig. 1b). 

The NEARS experiment is illustrated in Fig. 1e. The pump quenches the Mexican-Hat potential, but allows it to relax so that the Higgs mode populates as shown in Fig. 1g. We call this scenario a \textit{soft quench}, which controls the inversion population of the metastable Higgs mode via the fluence. In the normal state (Fig. 1f), with a pump (1.2~ps, 1.55~eV (800~nm)) orthogonal to the Raman probe polarizations and with a time delay of 3~ps between pump and probe, we can still apply the quasi-equilibrium approach. We can convert the measured Stokes spectra to the measured anti-Stokes data by using an additional pump heating of 3~K/mW (see Fig. S8). This is in contrast to many other experiments that use peak powers three to five orders of magnitude higher and thus explore the physics of hot electrons and hot phonons in the \textit{hard quench} regime.\cite{Perfetti2007, Graf2011, Smallwood2012, Toda2014} At 3~ps delay, fast electronic and phononic responses have already decayed.\cite{Perfetti2007, Toda2014, Smallwood2012} Thus, following Kasha's Rule, the soft quench populates the lowest metastable excitation of the superconductor, which is the Higgs mode.\cite{Kasha1968}

    \begin{figure*}[htp!]
	\includegraphics[width=0.8\linewidth
		]{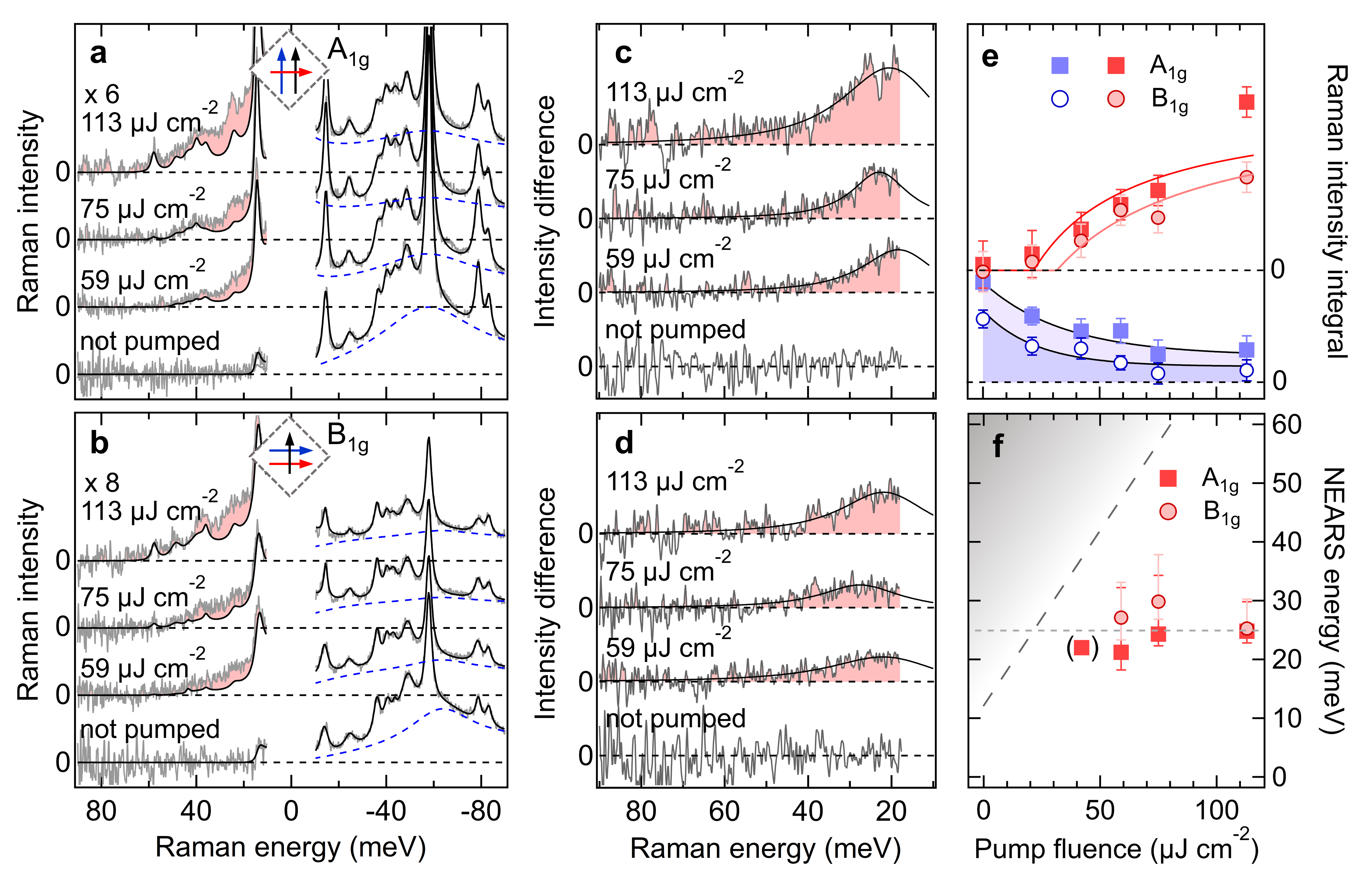}
	\caption{\label{fig:fig2}Higgs modes as a function of fluence. (a) A$_{1\textrm{g}}$ Pump-probe Stokes and anti-Stokes Raman spectra at 8~K base temperature and a time delay of 3~ps for selected fluences between 0~µJ~cm$^{-2}$ (not pumped) and 113~µJ~cm$^{-2}$. The complete data set is displayed in Fig. S6. Anti-Stokes Raman intensities are multiplied with a factor of 6 for better visibility. Data is shown in gray and black solid lines represent fits to the Raman intensities (see SI S.6). The dashed horizontal lines mark the zero intensity for the data sets at different fluences. The parameterized PB response on the Stokes side is highlighted in blue. With increasing pump fluence, gap filling occurs and the PB amplitude is decreased. The anti-Stokes spectra are presented together with the fits to the Stokes side converted to the anti-Stokes side by the Bose function (black solid line). Quasi-equilibrium temperatures used for the fits are listed in Table S2. At fluences larger than 50~µJ~cm$^{-2}$ the in-gap electronic response is dominated by a new NEARS feature (A$_{1\textrm{g}}$ Higgs mode) resulting in a difference signal compared to the Stokes signal (red). (b) B$_{1\textrm{g}}$ Pump-probe Stokes and anti-Stokes Raman spectra at 8 K analogous to (a). (c) A$_{1\textrm{g}}$ difference signal between anti-Stokes data and Stokes-fit (corresponding to red area in a). Solid black lines represent fits following eq. \ref{eq:GL_Higgs_res_main} (see text and SI S.4 for details). (d) B$_{1\textrm{g}}$ difference intensities analogous to (c). (e) Integrated Raman intensities of the Higgs modes (red) and of the PB response (blue) as a function of fluence. The Higgs mode intensity integrals are the integrated anti-Stokes difference spectra shown in c and d. Solid black lines are exponential guide to the eyes. The solid red lines represents the equation of inversion population (see SI S.3) with a critical fluence of $21.7 \pm 5.3$~µJ~cm$^{-2}$ for A$_{1\textrm{g}}$ and $31.6 \pm 2.3$~µJ~cm$^{-2}$ for B$_{1\textrm{g}}$ symmetry. (f) Excitation energy of the Higgs modes as a function of fluence. The equilibrium anti-Stokes intensity is limited at higher energies due to thermal Bose factors of the Raman intensity (see also Fig. S4). This is depicted as gray-shaded area.}
    \end{figure*}

Indeed, in the SC state we identify an extra signal on the anti-Stokes side (see Fig. 1h) which can be assigned to the population of the Higgs modes. This feature is solely present on the anti-Stokes side, while on the pumped Stokes side we can identify a persistent suppressed PB peak clearly indicating that the sample is still in its SC state 3~ps after the pump (see also S.5 and Fig. S12). We argue in the SI (S.7) that alternative excitations such as Josephson plasmons and Bardasis-Schrieffer modes cannot be responsible for our observations. Instead, the experimental results are in agreement with Higgs excitations. Within the phenomenological Ginzburg-Landau theory we can connect the energy of the Higgs excitation to the Cooper-pair coherence length. Within a BCS weak-coupling model we further develop a quantitative and coherent description of the pair-breaking excitations in the single-particle channel together with the Higgs excitations in the two-particle channel.

\subsection{Fluence dependence of NEARS data} 
Figure~2 presents the fluence dependence of the Bi-2212 NEARS spectra at 8~K and at 3~ps delay in A$_{1\textrm{g}}$ and B$_{1\textrm{g}}$ symmetry, as shown in the pictograms of Fig. 2a and b. The dashed squares mark the orientation of the CuO$_2$ planes. The pump is applied along the diagonals of the CuO$_2$ planes and carries a non-zero in-plane momentum due to its incidence angle of 21.8$^\circ$ (see Fig. S2).\cite{Schulz2005} In this configuration, one expects Higgs modes in both B$_{1\textrm{g}}$ and A$_{1\textrm{g}}$ probe symmetries.\cite{Schwarz2020a} The direct comparison between Stokes and anti-Stokes responses allows us to discriminate excitations around $2\Delta$ from NEARS features below $2\Delta$. 
Fig. 2a and b show Stokes and anti-Stokes Raman intensities for exemplary fluences between 0 µJ cm$^{-2}$ (not pumped) and 113 µJ cm$^{-2}$. Raman intensities corresponding to the parameterized PB feature on the Stokes side are depicted in blue. One can clearly see that the excitations around $2\Delta$ in both B$_{1\textrm{g}}$ and A$_{1\textrm{g}}$ Raman probe symmetry get suppressed with increasing fluence, but remain non-zero even at the highest fluence, which demonstrates that Bi-2212 remains in the SC state (see also SI S.5 and Fig. S12). The anti-Stokes spectra are presented together with the phononic and electronic Raman intensity fitted to the Stokes side and converted to the anti-Stokes side by the Bose function. The utilized quasi-equilibrium temperatures are listed in Table S2 (see also SI S.5) and are confirmed by evaluating the superconductivity-induced features of the Stokes spectrum as a function of fluence (see Fig. S12). The additional signal, which we attribute to the Higgs mode is highlighted in red. This feature on the anti-Stokes side increases in intensity with increasing fluence. We associate this with the enhanced inversion population of the metastable Higgs mode as a consequence of the increased strength of the soft quench. In a three-level picture of population inversion (shown in Fig. S9a), an excited state is populated with a short lifetime $\tau_{\textrm{relax}}$\cite{Perfetti2007} by quenching the Mexican-Hat. Subsequently a metastable lower-energy state (i.e. the Higgs mode) with a longer lifetime $\tau_{\textrm{Higgs}}>\tau_{\textrm{relax}}$ is populated. Population inversion occurs if $N_{\textrm{Higgs}} > N_{\textrm{initial}}$. After exceeding a critical fluence that is required for population inversion, the anti-Stokes intensity of the metastable Higgs excitation scales with the ratio $(N_{\textrm{Higgs}}-N_{\textrm{initial}})/N_{\textrm{total}}$.

Fig. 2c and d show the differences between the anti-Stokes data (gray in a and b) and converted Stokes fits (black solid lines in a and b), respectively. In Fig. 2e, we show the integrated Raman intensity of the Higgs modes and the PB excitation as a function of fluence. In both probe symmetries, we find an increase of the Higgs mode intensity and a concomitant decrease of the PB. The integrated intensities of the anti-Stokes Higgs difference signals as a function of fluence agree very well with the equation of population inversion (red lines) with a critical fluence of $21.7 \pm 5.3$~µJ~cm$^{-2}$ for A$_{1\textrm{g}}$ and $31.6 \pm 2.3$~µJ~cm$^{-2}$ for B$_{1\textrm{g}}$ symmetry (see SI S.3, eq. S24, and Fig. S9b). 

\subsection*{Higgs response in Ginzburg-Landau theory}

A phenomenological model of the Higgs response can be obtained within Ginzburg-Landau theory utilizing a Klein-Gordon like Lagrangian. The equations of motion are derived from a Mexican-Hat potential $F(\Psi) = \alpha |\Psi|^2 + \frac{\beta}{2} |\Psi|^4$ ($\alpha < 0$) with amplitude and phase fluctuations.\cite{puviani_current-assisted_2020} Since the phase fluctuations are gauged out by the Anderson-Higgs mechanism, we can calculate the Green's function of the Higgs mode through the equations of motion in the \textit{optical} ${\bf q} \rightarrow 0$ limit assuming a $\delta$ function-like quench due to a change in $\beta$ (see SI S.4). The result is a Lorentzian response $I(\omega)$

\begin{equation}
    \label{eq:GL_Higgs_res_main}
	I(\omega) =  I_0 \frac{\gamma \omega }{(\omega^2-2\abs{\alpha})^2+(\gamma \omega)^2} \ ,
\end{equation}

where $\gamma$ is the phenomenological width and $2 \alpha$ corresponds to the energy of the Higgs mode. The SC coherence length $\xi= \sqrt{\hbar^2/({|\alpha|4m^*})}$ 
is given by the Higgs-mode energy $\alpha$.

Fig. 2f shows the fluence dependence of the symmetry-dependent excitation energies derived by fitting the NEARS difference data with eq. \ref{eq:GL_Higgs_res_main} as shown in Fig. 2c and d. In A$_{1\textrm{g}}$ and B$_{1\textrm{g}}$ symmetry the Higgs mode arises at around $2\alpha = 0.168\cdot2\Delta_0 = 10.24 $~meV, corresponding to an energy of $\omega_H = \sqrt{2\alpha} = 0.41\cdot 2\Delta_0 = 25$~meV. Using established values $m^*/m_e$ for optimally-doped cuprates of the order of $m^* = 10 \ m_e$,\cite{Legros2019} we obtain in-plane coherence lengths of smaller than 5~nm in agreement with other experimental observations.\cite{Hwang2021} The Higgs-mode energy is only weakly dependent on fluence as expected for a population quench.

    \begin{figure}[h]
		\includegraphics[width=0.9\linewidth
		]{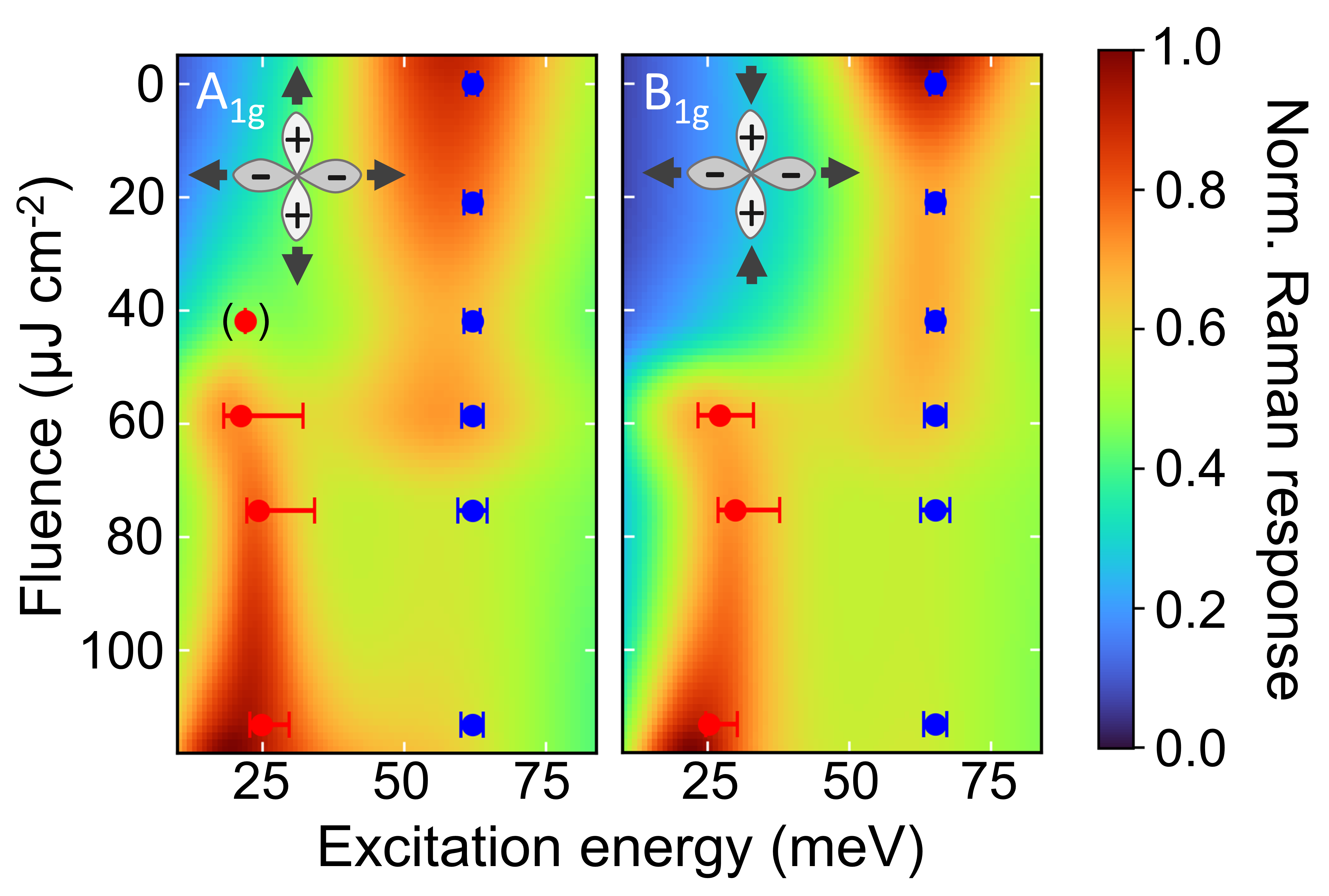}
		\caption{\label{fig:fig3}Unified excitation spectrum of the superconductor Bi-2212. The NEARS maps of Bi-2212 show the Raman response of the Higgs particle at around 25 meV together with the PB excitation Raman susceptibility around 60 meV in A$_{1\textrm{g}}$ (left) and B$_{1\textrm{g}}$ (right) geometry at a time delay of 3 ps. To create the NEARS maps, the fits from Fig. 2 were interpolated as a function of fluence. Raman energies of the utilized Lorentzians are shown as red and blue data points, respectively. The symmetry of the Higgs particle for the respective pump/probe configuration is indicated at the top right.}
    \end{figure}
    
In Figure 3, we jointly show the PB excitations of the single-particle channel together with the Higgs modes of the two-particle channel (see also Figs. 1 and 2 and Fig. S7) yielding a unified excitation landscape of the SC state in Bi-2212. These NEARS maps represent a superposition of the fluence-dependent superconductivity-induced excitations from both the Stokes and anti-Stokes spectra in A$_{1\textrm{g}}$ and B$_{1\textrm{g}}$ symmetry, respectively, identifying the A$_{1\textrm{g}}$ and B$_{1\textrm{g}}$ modes of the Higgs particle. The non-quenched energy landscape is dominated by excitations around $2\Delta$.\cite{Sellati2023, Munnikes2011} With increasing fluence, the $2\Delta$ excitations gradually weaken, but remain at constant energy, indicating that Bi-2212 remains in its SC state. The excitation of the Higgs particle is an in-gap excitation around 0.4~$2\Delta$ and increases in strength with fluence, showing a broadening at the highest fluence.  

\subsection*{Higgs BCS weak-coupling reponse theory}

To interpret the NEARS data in a microscopic framework, we utilize a mean-field weak-coupling BCS theory, as discussed in detail in SI S.2. The superconductor exhibits two main contributions to its electronic Raman response in the energy range of the gap: the quasiparticle/PB response due to single-particle excitations and the collective modes' response.\cite{Cea2016} The Higgs mode is the lowest-energy excitation in the two-particle channel and therefore plays a prominent role. The PB Raman vertices for B$_{1\textrm{g}}$ and A$_{1\textrm{g}}$ can be expressed as $\gamma_{\textrm{B}_{1\textrm{g}}} = \gamma_{b} \cos(2\phi)$ and $\gamma_{\textrm{A}_{1\textrm{g}}} = (1+b_0\xi)(\gamma_0 + \gamma_1 \cos(4\phi) + \gamma_2 \cos(8\phi))$, respectively.\cite{Devereaux1995} $\gamma_0, \gamma_1, \gamma_2$ are the coefficients in the generalized Fermi-surface harmonic expansion of $\gamma_{\textrm{A}_{1\textrm{g}}}$, and $\gamma_b$ is the coefficient in the expansion of $\gamma_{\textrm{B}_{1\textrm{g}}}$. We choose polar coordinates whose angle $\phi$ is defined as described in Fig. S14. The energy dependence of $\gamma_{\textrm{A}_{1\textrm{g}}}$ is encoded in $b_0$. The deviation from a constant density of state in the thin energy hull around the Fermi surface, inside which we assume the net-attractive electron-electron interaction, is represented by $b_1$: $N(\xi) = N_F(1+b_1\xi)$. Please note that $b_0$ and $b_1$ break particle-hole symmetry in the Raman vertices and the density of states, accordingly. Within BCS theory, the breaking of particle-hole symmetry is required in order to obtain a finite cross-section of the Higgs mode.

In Figure 4a and b we compare the results from our calculations with the equilibrium experimental data. In a first step, the lowest-order contribution to the B$_{1\textrm{g}}$ PB susceptibility is evaluated as shown in the Feynman diagram of Fig. 4a (see SI eq. S13).\cite{Devereaux1995} We obtain an excellent agreement between the parameterized, phonon-subtracted quasiparticle susceptibilities from unpumped measurements and the microscopically calculated quasiparticle response for a coefficient $\gamma_b = 0.1116 \pm 0.0003$, an electronic lifetime $\eta = 0.225 \pm 0.003$, and a superconducting order parameter $\Delta_0 = 30.50~\pm~0.05$~meV. Since $\eta$ and $\Delta_0$ are expected to be shared amongst all response functions, we keep $\eta$ and $\Delta_0$ constant for all further calculations. 

NEARS utilizes a probe energy of 3.1~eV. It is known that the PB peak changes as a function of incident photon energy.\cite{Budelmann2005} We model the Coulomb-screened A$_{1\textrm{g}}$ response in terms of Fermi-surface harmonics with the expansion parameters $\gamma_1$ and $\gamma_2$ (see SI S.2). Diagrammatically, the A$_{1\textrm{g}}$ PB response modified by Coulomb screening is shown in Fig. 4b (see eq. S17). Fitting $\gamma_1$ and $\gamma_2$ to the experimental phonon-subtracted A$_{1\textrm{g}}$ susceptibility yields $\gamma_1=0.117 \pm 0.001$, $\gamma_2 =0.054 \pm 0.001$ with $\eta = 0.225 =$ const, and $\Delta_0 = 30.50$~meV = const. 

The lowest-order contribution to the equilibrium Higgs response is given by the Feynman diagram shown in Fig. 4c (SI S.2).\cite{Cea2014, puviani_current-assisted_2020} With the established parameters $\gamma_1$, $\gamma_2$, $\eta$, and $\Delta_0$ we evaluate the A$_{1\textrm{g}}$ response of the Higgs excitation following eq. S21 and show the resulting susceptibility together with the NEARS Higgs feature at 75~µJ~cm$^{-2}$ in Fig. 4c. The position of the Higgs excitation within the weak-coupling BCS model relative to $2\Delta$ depends on the symmetry of the order parameter as well as on $r = b_1/(b_0+b_1)$.\cite{Cea2016} Increasing the parameter $r$ shifts the peak of the Higgs response towards lower energies. The solid red line in Fig. 4c shows the calculated Higgs response for $r = 0.99$ and $\gamma_0=0.1$. However, Bi-2212 is a strong-coupling superconductor that goes beyond the BCS weak-coupling model. Moreover, Fig. 4c represents a non-equilibrium state that cannot be adequately captured by an equilibrium calculation. In an equilibrium calculation performed in the clean limit, we find that the Higgs susceptibility is approximately three orders of magnitude weaker than the PB peak. The NEARS experiment, on the other hand, probes the Higgs modes following population inversion in a non-equilibrium state.

Important conclusions can be drawn from our calculations. Fig. 4a and b show an excellent agreement between the microscopic theory and our phonon-subtracted electronic equilibrium susceptibilities. This is remarkable, since there is no A$_{1\textrm{g}}$ problem\cite{Devereaux1995, Krantz_1995} in our data taken with an incident photon energy of 3.1~eV. In addition, the Higgs particle is an in-gap excitation that can appear at energies below $2 \Delta$ depending on the breaking of particle-hole symmetry. 

    \begin{figure*}[htp!]
	\includegraphics[width=0.9\linewidth
		]{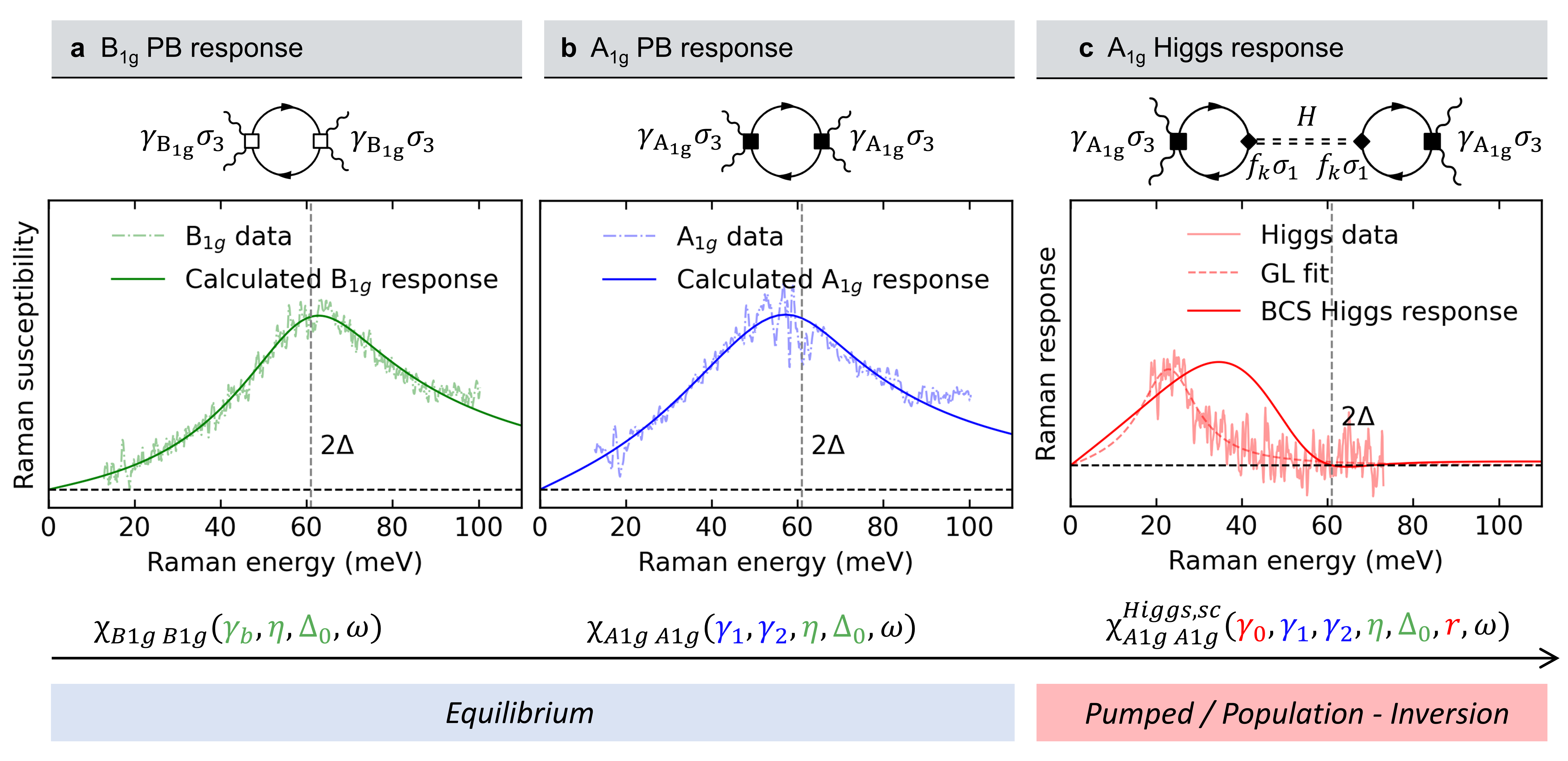}
	\caption{\label{fig:fig4}Comparison between NEARS data and BCS model calculations. (a) The experimental B$_{1\textrm{g}}$ electronic Raman susceptibility on the Stokes side (dash-dotted line) is extracted by parameterization and phonon subtraction as described in SI S.6. The B$_{1\textrm{g}}$ PB Raman response is depicted diagrammatically in the upper part of panel (see also eq. S13). We find the parameters $\gamma_b= 0.1116 \pm 0.0003$, $\eta= 0.225 \pm 0.003$, and $\Delta_0 = 30.50 \pm 0.05$ meV from the B$_{1\textrm{g}}$ fit (green). (b) The A$_{1\textrm{g}}$ PB Raman response is depicted diagrammatically with filled squares representing the Coulomb-screened vertices (see also eq. S17). We keep $\eta$ and $\Delta_0$ from the B$_{1\textrm{g}}$ fit, and fit $\gamma_1$ and $\gamma_2$ to the A$_{1\textrm{g}}$ experimental data (blue, $\gamma_1=0.117 \pm 0.001$, $\gamma_2 = 0.054 \pm 0.001$ with $\eta = 0.225$, $\Delta_0 = 30.50$ meV.). Please note that no information of $\gamma_0$ can be obtained due to the complete screening of the isotropic A$_{1\textrm{g}}$ component.\cite{Devereaux1995} We assume that $b_0$ and $b_1$ are small ($\Delta_0 b_0, \Delta_0 b_1 \ll 1$), therefore A$_{1\textrm{g}}$ has negligible dependence on $b_0, b_1$. (c) Finally, the A$_{1\textrm{g}}$ Higgs response is evaluated with the free parameters $\gamma_0$ and $r = b_1/(b_0+b_1)$ (see Feynman diagram and eq. S21). All other parameters ($\gamma_1$, $\gamma_2$, $\eta$, $\Delta_0$) are fixed. The red solid line corresponds to $\gamma_0$ = 0.1 and $r = 0.99$. The experimental Higgs Raman response corresponds to the 75~µJ~cm$^{-2}$ A$_{1\textrm{g}}$ data presented in Fig. 2c. The dashed line represents the fit according to eq. \ref{eq:GL_Higgs_res_main} derived within Ginzburg-Landau theory. Please note that due to the population of the Higgs mode by the soft quench, the Higgs response is a non-equilibrium response and enhanced in our experiment compared to the model. We therefore scale the intensities accordingly.}
    \end{figure*}


\subsection*{Conclusions} In conclusion, by introducing NEARS we find new in-gap excitations in a high-T$_\textrm{C}$ superconductor which can be assigned to the Higgs particle. This technique goes beyond conventional Raman spectroscopy for the identification of the symmetry of SC order parameters. NEARS experiments determine energy and life time enabling a first quantitative approach to the dynamics of the Higgs particle in a superconductor. NEARS makes Higgs spectroscopy applicable to many materials classes characterized by the interplay of superconductivity and competing or coexisting orders.\cite{Bardasis1961, Norman2011, Service2020} Higgs spectroscopy depends on both the symmetry of the quench and the symmetry of the Raman polarization probing the condensate. The energy and symmetry properties of the SC Higgs field are material-dependent and enable to study and control Higgs physics. Since the Meissner effect requires the presence of the Higgs field, the observation of a Higgs particle can serve as a novel criterion for superconductivity.

\section{Acknowledgments}
\begin{acknowledgments}
The authors thank Lara Benfatto (Università di Roma “La Sapienza”), Roberto Merlin (University of Michigan), Peter Abbamonte and Lance Cooper (University of Illinois Urbana-Champaign), Kenneth Burch (Boston College), Peter Littlewood (The University of Chicago), Jim Freericks (Georgetown University), and Herbert Fotso (University at Buffalo) for inspiring and productive discussions and input.

We acknowledge funding from Max Planck-UBC-UTokyo Center for Quantum Materials; Canada First Research Excellence Fund, Quantum Materials and Future Technologies Program; Natural Sciences and Engineering Research Council of Canada (NSERC); Canada Foundation for Innovation (CFI); Department of National Defence (DND); British Columbia Knowledge Development Fund (BCKDF); Canada Research Chairs Program; CIFAR Quantum Materials Program; U.S. Department of Energy through the University of Minnesota Center for Quantum Materials, Grant No. DE-SC0016371; Bundesministerium für Bildung und Forschung via 05K19GU5 and 05K22GU2; Deutsche Forschungsgemeinschaft (DFG) through SFB 1143 (project id 247310070); the Würzburg-Dresden Cluster of Excellence on Complexity and Topology in Quantum Matter – ct.qmat (EXC 2147, project id 390858490); and Funding by the European Union (ERC, T-Higgs, GA 101044657). (Views and opinions expressed are however those of the author(s) only and do not necessarily reflect those of the European Union or the European Research Council Executive Agency. Neither the European Union nor the granting authority can be held responsible for them.)

\end{acknowledgments}

\bibliography{Literature}

\begin{thebibliography}{10}
\expandafter\ifx\csname url\endcsname\relax
  \def\url#1{\texttt{#1}}\fi
\expandafter\ifx\csname urlprefix\endcsname\relax\def\urlprefix{URL }\fi
\providecommand{\bibinfo}[2]{#2}
\providecommand{\eprint}[2][]{\url{#2}}

\bibitem{Sooryakumar1980}
\bibinfo{author}{Sooryakumar, R.} \& \bibinfo{author}{Klein, M.~V.}
\newblock \bibinfo{title}{{Raman scattering by superconducting-gap excitations
  and their coupling to charge-density waves}}.
\newblock \emph{\bibinfo{journal}{Phys. Rev. Lett.}}
  \textbf{\bibinfo{volume}{45}}, \bibinfo{pages}{660 -- 662}
  (\bibinfo{year}{1980}).
\newblock \urlprefix\url{https://doi.org/10.1103/PhysRevLett.45.660}.

\bibitem{Higgs2007}
\bibinfo{author}{Higgs, P.}
\newblock \bibinfo{title}{{Prehistory of the Higgs boson}}.
\newblock \emph{\bibinfo{journal}{Comptes Rendus Physique}}
  \textbf{\bibinfo{volume}{8}}, \bibinfo{pages}{970 -- 972}
  (\bibinfo{year}{2007}).
\newblock \urlprefix\url{https://doi.org/10.1016/j.crhy.2006.12.006}.

\bibitem{Feng2023}
\bibinfo{author}{Feng, L.} \emph{et~al.}
\newblock \bibinfo{title}{{Dynamical interplay between superconductivity and
  charge density waves: A nonlinear terahertz study of coherently driven
  2H-NbSe$_2$}}.
\newblock \emph{\bibinfo{journal}{Phys. Rev. B}}
  \textbf{\bibinfo{volume}{108}}, \bibinfo{pages}{L100504}
  (\bibinfo{year}{2023}).
\newblock \urlprefix\url{https://doi.org/10.1103/PhysRevB.108.L100504}.

\bibitem{Cea2014}
\bibinfo{author}{Cea, T.} \& \bibinfo{author}{Benfatto, L.}
\newblock \bibinfo{title}{{Nature and Raman signatures of the Higgs amplitude
  mode in the coexisting superconducting and charge-density-wave state}}.
\newblock \emph{\bibinfo{journal}{Phys. Rev. B}} \textbf{\bibinfo{volume}{90}},
  \bibinfo{pages}{224515} (\bibinfo{year}{2014}).
\newblock \urlprefix\url{https://doi.org/10.1103/PhysRevB.90.224515}.

\bibitem{Littlewood1982}
\bibinfo{author}{Littlewood, P.} \& \bibinfo{author}{Varma, C.}
\newblock \bibinfo{title}{Amplitude collective modes in superconductors and
  their coupling to charge-density waves}.
\newblock \emph{\bibinfo{journal}{Phys. Rev. B}} \textbf{\bibinfo{volume}{26}},
  \bibinfo{pages}{4883 -- 4893} (\bibinfo{year}{1982}).
\newblock \urlprefix\url{https://doi.org/10.1103/PhysRevB.26.4883}.

\bibitem{Measson2014}
\bibinfo{author}{M{\'{e}}asson, M.-A.} \emph{et~al.}
\newblock \bibinfo{title}{{Amplitude Higgs mode in the 2H-NbSe$_2$
  superconductor}}.
\newblock \emph{\bibinfo{journal}{Phys. Rev. B}} \textbf{\bibinfo{volume}{89}},
  \bibinfo{pages}{060503} (\bibinfo{year}{2014}).
\newblock \urlprefix\url{https://doi.org/10.1103/PhysRevB.89.060503}.

\bibitem{Grasset2018}
\bibinfo{author}{Grasset, R.} \emph{et~al.}
\newblock \bibinfo{title}{{Higgs-mode radiance and charge-density-wave order in
  2H-NbSe$_2$}}.
\newblock \emph{\bibinfo{journal}{Phys. Rev. B}} \textbf{\bibinfo{volume}{97}},
  \bibinfo{pages}{094502} (\bibinfo{year}{2018}).
\newblock \urlprefix\url{https://doi.org/10.1103/PhysRevB.97.094502}.

\bibitem{Chu2023}
\bibinfo{author}{Chu, H.} \emph{et~al.}
\newblock \bibinfo{title}{{Fano interference between collective modes in
  cuprate high-T$_\textrm{C}$ superconductors}}.
\newblock \emph{\bibinfo{journal}{Nat Commun}} \textbf{\bibinfo{volume}{14}},
  \bibinfo{pages}{1343} (\bibinfo{year}{2023}).
\newblock \urlprefix\url{https://doi.org/10.1038/s41467-023-36787-4}.

\bibitem{Matsunaga2014}
\bibinfo{author}{Matsunaga, R.} \emph{et~al.}
\newblock \bibinfo{title}{{Light-induced collective pseudospin precession
  resonating with Higgs mode in a superconductor}}.
\newblock \emph{\bibinfo{journal}{Science}} \textbf{\bibinfo{volume}{345}},
  \bibinfo{pages}{1145 -- 1149} (\bibinfo{year}{2014}).
\newblock \urlprefix\url{https://doi.org/10.1126/science.1254697}.

\bibitem{Budelmann2005}
\bibinfo{author}{Budelmann, D.} \emph{et~al.}
\newblock \bibinfo{title}{{Gaplike excitations in the superconducting state of
  Bi$_2$Sr$_2$CaCu$_2$O$_8$ studied by resonant Raman scattering}}.
\newblock \emph{\bibinfo{journal}{Phys. Rev. Lett.}}
  \textbf{\bibinfo{volume}{95}}, \bibinfo{pages}{057003}
  (\bibinfo{year}{2005}).
\newblock \urlprefix\url{https://doi.org/10.1103/PhysRevLett.95.057003}.

\bibitem{Saichu2009}
\bibinfo{author}{Saichu, R.~P.} \emph{et~al.}
\newblock \bibinfo{title}{{Two-component dynamics of the order parameter of
  high temperature Bi$_2$Sr$_2$CaCu$_2$O$_{8+\delta}$ superconductors revealed
  by time-resolved Raman scattering}}.
\newblock \emph{\bibinfo{journal}{Phys. Rev. Lett.}}
  \textbf{\bibinfo{volume}{102}}, \bibinfo{pages}{177004}
  (\bibinfo{year}{2009}).
\newblock \urlprefix\url{https://doi.org/10.1103/PhysRevLett.102.177004}.

\bibitem{Pellatz2021}
\bibinfo{author}{Pellatz, N.} \emph{et~al.}
\newblock \bibinfo{title}{{Relaxation timescales and electron-phonon coupling
  in optically pumped YBa$_2$Cu$_3$O$_{6+x}$ reveraled by time-resolved Raman
  scattering}}.
\newblock \emph{\bibinfo{journal}{Phys. Rev. B}}
  \textbf{\bibinfo{volume}{104}}, \bibinfo{pages}{L180505}
  (\bibinfo{year}{2021}).
\newblock \urlprefix\url{https://doi.org/10.1103/PhysRevB.104.L180505}.

\bibitem{Han2019}
\bibinfo{author}{Han, S.} \emph{et~al.}
\newblock \bibinfo{title}{{Incoherent phonon population and exciton-exciton
  annihilation dynamics in monolayer WS$_2$ revealed by time-resolved Resonance
  Raman scattering}}.
\newblock \emph{\bibinfo{journal}{Optics Express}}
  \textbf{\bibinfo{volume}{27}}, \bibinfo{pages}{29949} (\bibinfo{year}{2019}).
\newblock \urlprefix\url{https://doi.org/10.1364/OE.27.029949}.

\bibitem{Katsumi2018}
\bibinfo{author}{Katsumi, K.} \emph{et~al.}
\newblock \bibinfo{title}{{Higgs mode in the d-wave superconductor
  Bi$_2$Sr$_2$CaCu$_2$O$_{8+x}$ driven by an intense Terahertz pulse}}.
\newblock \emph{\bibinfo{journal}{Phys. Rev. Lett.}}
  \textbf{\bibinfo{volume}{120}}, \bibinfo{pages}{117001}
  (\bibinfo{year}{2018}).
\newblock \urlprefix\url{https://doi.org/10.1103/PhysRevLett.120.117001}.

\bibitem{Chu2020}
\bibinfo{author}{Chu, H.} \emph{et~al.}
\newblock \bibinfo{title}{{Phase-resolved Higgs response in superconducting
  cuprates}}.
\newblock \emph{\bibinfo{journal}{Nat Commun}} \textbf{\bibinfo{volume}{11}},
  \bibinfo{pages}{1793} (\bibinfo{year}{2020}).
\newblock \urlprefix\url{https://doi.org/10.1038/s41467-020-15613-1}.

\bibitem{Vaswani2021}
\bibinfo{author}{Vaswani, C.} \emph{et~al.}
\newblock \bibinfo{title}{{Light quantum control of persisting Higgs modes in
  iron-based superconductors}}.
\newblock \emph{\bibinfo{journal}{Nat Commun}} \textbf{\bibinfo{volume}{12}},
  \bibinfo{pages}{258} (\bibinfo{year}{2021}).
\newblock \urlprefix\url{https://doi.org/10.1038/s41467-020-20350-6}.

\bibitem{Wang2022}
\bibinfo{author}{Wang, Z.-X.} \emph{et~al.}
\newblock \bibinfo{title}{{Transient Higgs oscillations and high-order
  nonlinear light-Higgs coupling in a terahertz wave driven NbN
  superconductor}}.
\newblock \emph{\bibinfo{journal}{Phys. Rev. B}}
  \textbf{\bibinfo{volume}{105}}, \bibinfo{pages}{L100508}
  (\bibinfo{year}{2022}).
\newblock \urlprefix\url{https://doi.org/10.1103/PhysRevB.105.L100508}.

\bibitem{Reinhoffer2022}
\bibinfo{author}{Reinhoffer, C.} \emph{et~al.}
\newblock \bibinfo{title}{{High-order nonlinear terahertz probing of the
  two-band superconductor MgB$_2$: Third- and fifth-order harmonic
  generation}}.
\newblock \emph{\bibinfo{journal}{Phys. Rev. B}}
  \textbf{\bibinfo{volume}{106}}, \bibinfo{pages}{214514}
  (\bibinfo{year}{2022}).
\newblock \urlprefix\url{https://doi.org/10.1103/PhysRevB.106.214514}.

\bibitem{Katsumi2023}
\bibinfo{author}{Katsumi, K.} \emph{et~al.}
\newblock \bibinfo{title}{Revealing novel aspects of light-matter coupling by
  terahertz two-dimensional coherent spectroscopy: The case of the amplitude
  mode in superconductors}.
\newblock \emph{\bibinfo{journal}{Phys. Rev. Lett.}}
  \textbf{\bibinfo{volume}{132}}, \bibinfo{pages}{256903}
  (\bibinfo{year}{2024}).
\newblock \urlprefix\url{https://doi.org/10.1103/PhysRevLett.132.256903}.

\bibitem{Shimano2020}
\bibinfo{author}{Shimano, R.} \& \bibinfo{author}{Tsuji, N.}
\newblock \bibinfo{title}{Higgs mode in superconductors}.
\newblock \emph{\bibinfo{journal}{Annu. Rev. Condens. Matter Phys.}}
  \textbf{\bibinfo{volume}{11}}, \bibinfo{pages}{103 -- 124}
  (\bibinfo{year}{2020}).
\newblock
  \urlprefix\url{https://doi.org/10.1146/annurev-conmatphys-031119-050813}.

\bibitem{Kim2023a}
\bibinfo{author}{Kim, M.-J.} \emph{et~al.}
\newblock \bibinfo{title}{Tracing the dynamics of superconducting order via
  transient terahertz third-harmonic generation}.
\newblock \emph{\bibinfo{journal}{Science Advances}}
  \textbf{\bibinfo{volume}{10}}, \bibinfo{pages}{eadi7598}
  (\bibinfo{year}{2024}).
\newblock \urlprefix\url{https://doi.org/10.1126/sciadv.adi7598}.

\bibitem{Barlas2013}
\bibinfo{author}{Barlas, Y.} \& \bibinfo{author}{Varma, C.~M.}
\newblock \bibinfo{title}{{Amplitude or Higgs modes in d-wave
  superconductors}}.
\newblock \emph{\bibinfo{journal}{Phys. Rev. B}} \textbf{\bibinfo{volume}{87}},
  \bibinfo{pages}{054503} (\bibinfo{year}{2013}).
\newblock \urlprefix\url{https://doi.org/10.1103/PhysRevB.87.054503}.

\bibitem{Schwarz2020}
\bibinfo{author}{Schwarz, L.} \& \bibinfo{author}{Manske, D.}
\newblock \bibinfo{title}{{Theory of driven Higgs oscillations and
  third-harmonic generation in unconventional superconductors}}.
\newblock \emph{\bibinfo{journal}{Phys. Rev. B}}
  \textbf{\bibinfo{volume}{101}}, \bibinfo{pages}{184519}
  (\bibinfo{year}{2020}).
\newblock \urlprefix\url{https://doi.org/10.1103/PhysRevB.101.184519}.

\bibitem{Tsuji2015}
\bibinfo{author}{Tsuji, N.} \& \bibinfo{author}{Aoki, H.}
\newblock \bibinfo{title}{{Theory of Anderson pseudospin resonance with Higgs
  mode in superconductors}}.
\newblock \emph{\bibinfo{journal}{Phys. Rev. B}} \textbf{\bibinfo{volume}{92}},
  \bibinfo{pages}{064508} (\bibinfo{year}{2015}).
\newblock \urlprefix\url{https://doi.org/10.1103/PhysRevB.92.064508}.

\bibitem{Hannibal2015}
\bibinfo{author}{Hannibal, S.} \emph{et~al.}
\newblock \bibinfo{title}{{Quench dynamics of an ultracold Fermi gas in the BCS
  regime: Spectral properties and confinement-induced breakdown of the Higgs
  mode}}.
\newblock \emph{\bibinfo{journal}{Phys. Rev. A}} \textbf{\bibinfo{volume}{91}},
  \bibinfo{pages}{043630} (\bibinfo{year}{2015}).
\newblock \urlprefix\url{https://doi.org/10.1103/PhysRevA.91.043630}.

\bibitem{Benfatto2023}
\bibinfo{author}{Benfatto, L.}, \bibinfo{author}{Castellani, C.} \&
  \bibinfo{author}{Seibold, G.}
\newblock \bibinfo{title}{Linear and nonlinear current response in disordered
  d-wave superconductors}.
\newblock \emph{\bibinfo{journal}{Phys. Rev. B}}
  \textbf{\bibinfo{volume}{108}}, \bibinfo{pages}{134508}
  (\bibinfo{year}{2023}).
\newblock \urlprefix\url{https://doi.org/10.1103/PhysRevB.108.134508}.

\bibitem{Seibold2021}
\bibinfo{author}{Seibold, G.}, \bibinfo{author}{Udina, M.},
  \bibinfo{author}{Castellani, C.} \& \bibinfo{author}{Benfatto, L.}
\newblock \bibinfo{title}{Third harmonic generation from collective modes in
  disordered superconductors}.
\newblock \emph{\bibinfo{journal}{Phys. Rev. B}}
  \textbf{\bibinfo{volume}{103}}, \bibinfo{pages}{014512}
  (\bibinfo{year}{2021}).
\newblock \urlprefix\url{https://doi.org/10.1103/PhysRevB.103.014512}.

\bibitem{Udina2022}
\bibinfo{author}{Udina, M.} \emph{et~al.}
\newblock \bibinfo{title}{{THz non-linear optical response in cuprates:
  predominance of the BCS response over the Higgs mode}}.
\newblock \emph{\bibinfo{journal}{Faraday Discussions}}
  \textbf{\bibinfo{volume}{237}}, \bibinfo{pages}{168 -- 185}
  (\bibinfo{year}{2022}).
\newblock \urlprefix\url{http://doi.org/10.1039/D2FD00016D}.

\bibitem{Sun2020}
\bibinfo{author}{Sun, Z.}, \bibinfo{author}{Fogler, M.~M.},
  \bibinfo{author}{Basov, D.~N.} \& \bibinfo{author}{Millis, A.~J.}
\newblock \bibinfo{title}{{Collective modes and terahertz near-field response
  of superconductors}}.
\newblock \emph{\bibinfo{journal}{Phys. Rev. Res.}}
  \textbf{\bibinfo{volume}{2}}, \bibinfo{pages}{023413} (\bibinfo{year}{2020}).
\newblock \urlprefix\url{https://doi.org/10.1103/PhysRevResearch.2.023413}.

\bibitem{Gabriele2021}
\bibinfo{author}{Gabriele, F.}, \bibinfo{author}{Udina, M.} \&
  \bibinfo{author}{Benfatto, L.}
\newblock \bibinfo{title}{{Non-linear Terahertz driving of plasma waves in
  layered cuprates}}.
\newblock \emph{\bibinfo{journal}{Nat Commun}} \textbf{\bibinfo{volume}{12}},
  \bibinfo{pages}{752} (\bibinfo{year}{2021}).
\newblock \urlprefix\url{https://doi.org/10.1038/s41467-021-21041-6}.

\bibitem{Sellati2023}
\bibinfo{author}{Sellati, N.}, \bibinfo{author}{Gabriele, F.},
  \bibinfo{author}{Castellani, C.} \& \bibinfo{author}{Benfatto, L.}
\newblock \bibinfo{title}{{Generalized Josephson plasmons in bilayer
  superconductors}}.
\newblock \emph{\bibinfo{journal}{Phys. Rev. B}}
  \textbf{\bibinfo{volume}{108}}, \bibinfo{pages}{014503}
  (\bibinfo{year}{2023}).
\newblock \urlprefix\url{https://doi.org/10.1103/PhysRevB.108.014503}.

\bibitem{Schwarz2020a}
\bibinfo{author}{Schwarz, L.} \emph{et~al.}
\newblock \bibinfo{title}{{Classification and characterization of
  nonequilibrium Higgs modes in unconventional superconductors}}.
\newblock \emph{\bibinfo{journal}{Nat Commun}} \textbf{\bibinfo{volume}{11}},
  \bibinfo{pages}{287} (\bibinfo{year}{2020}).
\newblock \urlprefix\url{https://doi.org/10.1038/s41467-019-13763-5}.

\bibitem{Eisaki2004}
\bibinfo{author}{Eisaki, H.} \emph{et~al.}
\newblock \bibinfo{title}{{Effect of chemical inhomogeneity in bismuth-based
  copper oxide superconductors}}.
\newblock \emph{\bibinfo{journal}{Phys. Rev. B}} \textbf{\bibinfo{volume}{69}},
  \bibinfo{pages}{064512} (\bibinfo{year}{2004}).
\newblock \urlprefix\url{https://doi.org/10.1103/PhysRevB.69.064512}.

\bibitem{Munnikes2011}
\bibinfo{author}{Munnikes, N.} \emph{et~al.}
\newblock \bibinfo{title}{{Pair breaking versus symmetry breaking: Origin of
  the Raman modes in superconducting cuprates}}.
\newblock \emph{\bibinfo{journal}{Phys. Rev. B}} \textbf{\bibinfo{volume}{84}},
  \bibinfo{pages}{144523} (\bibinfo{year}{2011}).
\newblock \urlprefix\url{https://doi.org/10.1103/PhysRevB.84.144523}.

\bibitem{Devereaux1995}
\bibinfo{author}{Devereaux, T.~P.} \& \bibinfo{author}{Einzel, D.}
\newblock \bibinfo{title}{{Electronic Raman scattering in superconductors as a
  probe of anisotropic electron pairing}}.
\newblock \emph{\bibinfo{journal}{Phys. Rev. B}} \textbf{\bibinfo{volume}{51}},
  \bibinfo{pages}{16336 -- 16357} (\bibinfo{year}{1995}).
\newblock \urlprefix\url{https://doi.org/10.1103/PhysRevB.51.16336}.

\bibitem{Blumberg2007}
\bibinfo{author}{Blumberg, G.} \emph{et~al.}
\newblock \bibinfo{title}{{Observation of Leggett's collective mode in a
  multiband MgB$_2$ superconductor}}.
\newblock \emph{\bibinfo{journal}{Phys. Rev. Lett.}}
  \textbf{\bibinfo{volume}{99}}, \bibinfo{pages}{227002}
  (\bibinfo{year}{2007}).
\newblock \urlprefix\url{https://doi.org/10.1103/PhysRevLett.99.227002}.

\bibitem{Maiti2016}
\bibinfo{author}{Maiti, S.}, \bibinfo{author}{Maier, T.~A.},
  \bibinfo{author}{B{\"{o}}hm, T.}, \bibinfo{author}{Hackl, R.} \&
  \bibinfo{author}{Hirschfeld, P.~J.}
\newblock \bibinfo{title}{{Probing the pairing interaction and multiple
  Bardasis-Schrieffer modes using Raman spectroscopy}}.
\newblock \emph{\bibinfo{journal}{Phys. Rev. Lett.}}
  \textbf{\bibinfo{volume}{117}}, \bibinfo{pages}{257001}
  (\bibinfo{year}{2016}).
\newblock \urlprefix\url{https://doi.org/10.1103/PhysRevLett.117.257001}.

\bibitem{Bohm2018}
\bibinfo{author}{Böhm, T.} \emph{et~al.}
\newblock \bibinfo{title}{{Microscopic origin of Cooper pairing in the
  iron-based superconductor Ba$_{1 \textrm{-} x}$K$_x$Fe$_2$As$_2$}}.
\newblock \emph{\bibinfo{journal}{npj Quant Mater}}
  \textbf{\bibinfo{volume}{3}}, \bibinfo{pages}{48} (\bibinfo{year}{2018}).
\newblock \urlprefix\url{https://doi.org/10.1038/s41535-018-0118-z}.

\bibitem{Bock1995}
\bibinfo{author}{Bock, A.}
\newblock \bibinfo{title}{{Laser heating of YBa$_2$Cu$_3$O$_7$ films in Raman
  experiments}}.
\newblock \emph{\bibinfo{journal}{Phys. Rev. B}} \textbf{\bibinfo{volume}{51}},
  \bibinfo{pages}{15506 -- 15518} (\bibinfo{year}{1995}).
\newblock \urlprefix\url{https://doi.org/10.1103/PhysRevB.51.15506}.

\bibitem{Klein2006}
\bibinfo{author}{Klein, M.} \emph{et~al.}
\newblock \bibinfo{title}{{Resonance Raman study of 2$\Delta$-gap like features
  in superconducting Bi-2212 and YBCO}}.
\newblock \emph{\bibinfo{journal}{Journal of Physics and Chemistry of Solids}}
  \textbf{\bibinfo{volume}{67}}, \bibinfo{pages}{298 -- 301}
  (\bibinfo{year}{2006}).
\newblock \urlprefix\url{https://doi.org/10.1016/j.jpcs.2005.10.161}.

\bibitem{Perfetti2007}
\bibinfo{author}{Perfetti, L.} \emph{et~al.}
\newblock \bibinfo{title}{{Ultrafast electron relaxation in superconducting
  ${\mathrm{Bi}}_{2}{\mathrm{Sr}}_{2}{\mathrm{CaCu}}_{2}{\mathrm{O}}_{8+\ensuremath{\delta}}$
  by time-resolved photoelectron spectroscopy}}.
\newblock \emph{\bibinfo{journal}{Phys. Rev. Lett.}}
  \textbf{\bibinfo{volume}{99}}, \bibinfo{pages}{197001}
  (\bibinfo{year}{2007}).
\newblock \urlprefix\url{https://doi.org/10.1103/PhysRevLett.99.197001}.

\bibitem{Graf2011}
\bibinfo{author}{Graf, J.} \emph{et~al.}
\newblock \bibinfo{title}{Nodal quasiparticle meltdown in ultrahigh-resolution
  pump–probe angle-resolved photoemission}.
\newblock \emph{\bibinfo{journal}{Nature Phys}} \textbf{\bibinfo{volume}{7}},
  \bibinfo{pages}{805–809} (\bibinfo{year}{2011}).
\newblock \urlprefix\url{https://doi.org/10.1038/nphys2027}.

\bibitem{Smallwood2012}
\bibinfo{author}{Smallwood, C.~L.} \emph{et~al.}
\newblock \bibinfo{title}{{Tracking Cooper pairs in a cuprate superconductor by
  ultrafast angle-resolved photoemission}}.
\newblock \emph{\bibinfo{journal}{Science}} \textbf{\bibinfo{volume}{336}},
  \bibinfo{pages}{1137--1139} (\bibinfo{year}{2012}).
\newblock \urlprefix\url{https://doi.org/10.1126/science.1217423}.

\bibitem{Toda2014}
\bibinfo{author}{Toda, Y.} \emph{et~al.}
\newblock \bibinfo{title}{{Rotational symmetry breaking in
  Bi$_{2}$Sr$_{2}$CaCu$_{2}$O$_{8+\ensuremath{\delta}}$ probed by polarized
  femtosecond spectroscopy}}.
\newblock \emph{\bibinfo{journal}{Phys. Rev. B}} \textbf{\bibinfo{volume}{90}},
  \bibinfo{pages}{094513} (\bibinfo{year}{2014}).
\newblock \urlprefix\url{https://doi.org/10.1103/PhysRevB.90.094513}.

\bibitem{Kasha1968}
\bibinfo{author}{Henry, B.} \& \bibinfo{author}{Kasha, M.}
\newblock \bibinfo{title}{Radiationless molecular electronic transitions}.
\newblock \emph{\bibinfo{journal}{Annu. Rev. Phys. Chem.}}
  \textbf{\bibinfo{volume}{19}}, \bibinfo{pages}{161--192}
  (\bibinfo{year}{1968}).
\newblock \urlprefix\url{https://doi.org/10.1146/annurev.pc.19.100168.001113}.

\bibitem{Schulz2005}
\bibinfo{author}{Schulz, B.} \emph{et~al.}
\newblock \bibinfo{title}{{Fully reflective deep ultraviolet to near infrared
  spectrometer and entrance optics for resonance Raman spectroscopy}}.
\newblock \emph{\bibinfo{journal}{Rev. Sci. Instrum.}}
  \textbf{\bibinfo{volume}{76}}, \bibinfo{pages}{073107}
  (\bibinfo{year}{2005}).
\newblock \urlprefix\url{https://doi.org/10.1063/1.1946985}.

\bibitem{puviani_current-assisted_2020}
\bibinfo{author}{Puviani, M.}, \bibinfo{author}{Schwarz, L.},
  \bibinfo{author}{Zhang, X.-X.}, \bibinfo{author}{Kaiser, S.} \&
  \bibinfo{author}{Manske, D.}
\newblock \bibinfo{title}{{Current-assisted Raman activation of the Higgs mode
  in superconductors}}.
\newblock \emph{\bibinfo{journal}{Phys. Rev. B}}
  \textbf{\bibinfo{volume}{101}}, \bibinfo{pages}{220507}
  (\bibinfo{year}{2020}).
\newblock \urlprefix\url{https://doi.org/10.1103/PhysRevB.101.220507}.

\bibitem{Legros2019}
\bibinfo{author}{Legros, A.} \emph{et~al.}
\newblock \bibinfo{title}{{Universal T-linear resistivity and Planckian
  dissipation in overdoped cuprates}}.
\newblock \emph{\bibinfo{journal}{Nature Phys.}} \textbf{\bibinfo{volume}{15}},
  \bibinfo{pages}{142–147} (\bibinfo{year}{2019}).
\newblock \urlprefix\url{https://doi.org/10.1038/s41567-018-0334-2}.

\bibitem{Hwang2021}
\bibinfo{author}{Hwang, J.}
\newblock \bibinfo{title}{Superconducting coherence length of hole-doped
  cuprates obtained from electron–boson spectral density function}.
\newblock \emph{\bibinfo{journal}{Sci Rep}} \textbf{\bibinfo{volume}{11}},
  \bibinfo{pages}{11668} (\bibinfo{year}{2021}).
\newblock \urlprefix\url{https://doi.org/10.1038/s41598-021-91163-w}.

\bibitem{Cea2016}
\bibinfo{author}{Cea, T.}, \bibinfo{author}{Castellani, C.} \&
  \bibinfo{author}{Benfatto, L.}
\newblock \bibinfo{title}{{Nonlinear optical effects and third-harmonic
  generation in superconductors: Cooper pairs versus Higgs mode contribution}}.
\newblock \emph{\bibinfo{journal}{Phys. Rev. B}} \textbf{\bibinfo{volume}{93}},
  \bibinfo{pages}{180507} (\bibinfo{year}{2016}).
\newblock \urlprefix\url{https://link.aps.org/doi/10.1103/PhysRevB.93.180507}.

\bibitem{Krantz_1995}
\bibinfo{author}{Krantz, M.} \& \bibinfo{author}{Cardona, M.}
\newblock \bibinfo{title}{{Raman scattering by electronic excitations in
  semiconductors and in high T$_C$ superconductors}}.
\newblock \emph{\bibinfo{journal}{J Low Temp Phys}}
  \textbf{\bibinfo{volume}{99}}, \bibinfo{pages}{205 -- 221}
  (\bibinfo{year}{1995}).
\newblock \urlprefix\url{https://doi.org/10.1007/BF00752289}.

\bibitem{Bardasis1961}
\bibinfo{author}{Bardasis, A.} \& \bibinfo{author}{Schrieffer, J.~R.}
\newblock \bibinfo{title}{{Excitons and plasmons in superconductors}}.
\newblock \emph{\bibinfo{journal}{Phys. Rev.}} \textbf{\bibinfo{volume}{121}},
  \bibinfo{pages}{1050 -- 1062} (\bibinfo{year}{1961}).
\newblock \urlprefix\url{https://doi.org/10.1103/PhysRev.121.1050}.

\bibitem{Norman2011}
\bibinfo{author}{Norman, M.~R.}
\newblock \bibinfo{title}{{The challenge of unconventional superconductivity}}.
\newblock \emph{\bibinfo{journal}{Science}} \textbf{\bibinfo{volume}{332}},
  \bibinfo{pages}{196 -- 200} (\bibinfo{year}{2011}).
\newblock \urlprefix\url{https://doi.org/10.1126/science.1200181}.

\bibitem{Service2020}
\bibinfo{author}{Service, R.~F.}
\newblock \bibinfo{title}{{At last, room temperature superconductivity
  achieved}}.
\newblock \emph{\bibinfo{journal}{Science}} \textbf{\bibinfo{volume}{370}},
  \bibinfo{pages}{273 -- 274} (\bibinfo{year}{2020}).
\newblock \urlprefix\url{https://doi.org/10.1126/science.370.6514.273}.

\bibitem{Pekker2015}
\bibinfo{author}{Pekker, D.} \& \bibinfo{author}{Varma, C.~M.}
\newblock \bibinfo{title}{{Amplitude/Higgs modes in condensed matter physics}}.
\newblock \emph{\bibinfo{journal}{Annual Review of Condensed Matter Physics}}
  \textbf{\bibinfo{volume}{6}}, \bibinfo{pages}{269 -- 297}
  (\bibinfo{year}{2015}).
\newblock
  \urlprefix\url{https://doi.org/10.1146/annurev-conmatphys-031214-014350}.

\bibitem{Nambu1960}
\bibinfo{author}{Nambu, Y.}
\newblock \bibinfo{title}{Quasi-particles and gauge invariance in the theory of
  superconductivity}.
\newblock \emph{\bibinfo{journal}{Phys. Rev.}} \textbf{\bibinfo{volume}{117}},
  \bibinfo{pages}{648 -- 663} (\bibinfo{year}{1960}).
\newblock \urlprefix\url{https://doi.org/10.1103/PhysRev.117.648}.

\bibitem{Peskin}
\bibinfo{author}{Peskin, M.} \& \bibinfo{author}{Schroeder, D.}
\newblock \emph{\bibinfo{title}{An Introduction To Quantum Field Theory}}
  (\bibinfo{publisher}{Taylor and Fracis Group}, \bibinfo{year}{1995}).

\bibitem{Zirnbauer2021}
\bibinfo{author}{Zirnbauer, M.~R.}
\newblock \bibinfo{title}{Particle–hole symmetries in condensed matter}.
\newblock \emph{\bibinfo{journal}{Journal of Mathematical Physics}}
  \textbf{\bibinfo{volume}{62}} (\bibinfo{year}{2021}).
\newblock \urlprefix\url{https://doi.org/10.1063/5.0035358}.

\bibitem{Chiu2016}
\bibinfo{author}{Chiu, C.-K.}, \bibinfo{author}{Teo, J. C.~Y.},
  \bibinfo{author}{Schnyder, A.~P.} \& \bibinfo{author}{Ryu, S.}
\newblock \bibinfo{title}{Classification of topological quantum matter with
  symmetries}.
\newblock \emph{\bibinfo{journal}{Rev. Mod. Phys.}}
  \textbf{\bibinfo{volume}{88}}, \bibinfo{pages}{035005}
  (\bibinfo{year}{2016}).
\newblock \urlprefix\url{https://doi.org/10.1103/RevModPhys.88.035005}.

\bibitem{markiewicz_one-band_2005}
\bibinfo{author}{Markiewicz, R.~S.}, \bibinfo{author}{Sahrakorpi, S.},
  \bibinfo{author}{Lindroos, M.}, \bibinfo{author}{Lin, H.} \&
  \bibinfo{author}{Bansil, A.}
\newblock \bibinfo{title}{{One-band tight-binding model parametrization of the
  high-T$_C$ cuprates including the effect of k$_z$ dispersion}}.
\newblock \emph{\bibinfo{journal}{Phys. Rev. B}} \textbf{\bibinfo{volume}{72}},
  \bibinfo{pages}{054519} (\bibinfo{year}{2005}).
\newblock \urlprefix\url{https://doi.org/10.1103/PhysRevB.72.054519}.

\bibitem{devereaux_inelastic_2007}
\bibinfo{author}{Devereaux, T.~P.} \& \bibinfo{author}{Hackl, R.}
\newblock \bibinfo{title}{Inelastic light scattering from correlated
  electrons}.
\newblock \emph{\bibinfo{journal}{Rev. Mod. Phys.}}
  \textbf{\bibinfo{volume}{79}}, \bibinfo{pages}{175 -- 233}
  (\bibinfo{year}{2007}).
\newblock \urlprefix\url{https://doi.org/10.1103/RevModPhys.79.175}.

\bibitem{monien_theory_1990}
\bibinfo{author}{Monien, H.} \& \bibinfo{author}{Zawadowski, A.}
\newblock \bibinfo{title}{{Theory of Raman scattering with final-state
  interaction in high-T$_C$ BCS superconductors: Collective modes}}.
\newblock \emph{\bibinfo{journal}{Phys. Rev. B}} \textbf{\bibinfo{volume}{41}},
  \bibinfo{pages}{8798--8810} (\bibinfo{year}{1990}).
\newblock \urlprefix\url{https://doi.org/10.1103/PhysRevB.41.8798}.

\bibitem{shimoda2013}
\bibinfo{author}{Shimoda, K.}
\newblock \emph{\bibinfo{title}{Introduction to Laser Physics}}.
\newblock Springer Series in Optical Sciences (\bibinfo{publisher}{Springer
  Berlin Heidelberg}, \bibinfo{year}{2013}).
\newblock \urlprefix\url{https://books.google.de/books?id=yqPxCAAAQBAJ}.

\bibitem{Greiter2005}
\bibinfo{author}{Greiter, M.}
\newblock \bibinfo{title}{Is electromagnetic gauge invariance spontaneously
  violated in superconductors?}
\newblock \emph{\bibinfo{journal}{Annals of Physics}}
  \textbf{\bibinfo{volume}{319}}, \bibinfo{pages}{217--249}
  (\bibinfo{year}{2005}).
\newblock \urlprefix\url{https://doi.org/10.1016/j.aop.2005.03.008}.

\bibitem{Anderson1958}
\bibinfo{author}{Anderson, P.}
\newblock \bibinfo{title}{{Coherent excited states in the theory of
  superconductivity: gauge invariance and the Meissner effect}}.
\newblock \emph{\bibinfo{journal}{Phys. Rev.}} \textbf{\bibinfo{volume}{110}},
  \bibinfo{pages}{827 -- 835} (\bibinfo{year}{1958}).
\newblock \urlprefix\url{https://doi.org/10.1103/PhysRev.110.827}.

\bibitem{Anderson1963}
\bibinfo{author}{Anderson, P.~W.}
\newblock \bibinfo{title}{Plasmons, gauge invariance, and mass}.
\newblock \emph{\bibinfo{journal}{Phys. Rev.}} \textbf{\bibinfo{volume}{130}},
  \bibinfo{pages}{439--442} (\bibinfo{year}{1963}).
\newblock \urlprefix\url{https://doi.org/10.1103/PhysRev.130.439}.

\bibitem{Freericks2021}
\bibinfo{author}{Freericks, J.} \& \bibinfo{author}{Kemper, A.~F.}
\newblock \bibinfo{title}{{What do the two times in two-time correlation
  functions mean for interpreting tr-ARPES?}}
\newblock \emph{\bibinfo{journal}{Journal of Electron Spectroscopy and Related
  Phenomena}} \textbf{\bibinfo{volume}{251}}, \bibinfo{pages}{147104}
  (\bibinfo{year}{2021}).
\newblock \urlprefix\url{https://doi.org/10.1016/j.elspec.2021.147104}.

\bibitem{Deveraux2018}
\bibinfo{author}{Wang, Y.}, \bibinfo{author}{Devereaux, T.~P.} \&
  \bibinfo{author}{Chen, C.-C.}
\newblock \bibinfo{title}{{Theory of time-resolved Raman scattering in
  correlated systems: Ultrafast engineering of spin dynamics and detection of
  thermalization}}.
\newblock \emph{\bibinfo{journal}{Phys. Rev. B}} \textbf{\bibinfo{volume}{98}},
  \bibinfo{pages}{245106} (\bibinfo{year}{2018}).
\newblock \urlprefix\url{https://link.aps.org/doi/10.1103/PhysRevB.98.245106}.

\bibitem{Cyrot1973}
\bibinfo{author}{Cyrot, M.}
\newblock \bibinfo{title}{Ginzburg-landau theory for superconductors}.
\newblock \emph{\bibinfo{journal}{Reports on Progress in Physics}}
  \textbf{\bibinfo{volume}{36}}, \bibinfo{pages}{103} (\bibinfo{year}{1973}).
\newblock \urlprefix\url{https://dx.doi.org/10.1088/0034-4885/36/2/001}.

\bibitem{Scalapino2009}
\bibinfo{author}{Scalapino, D.~J.} \& \bibinfo{author}{Devereaux, T.~P.}
\newblock \bibinfo{title}{{Collective d-wave exciton modes in the calculated
  Raman spectrum of Fe-based superconductors}}.
\newblock \emph{\bibinfo{journal}{Phys. Rev. B}} \textbf{\bibinfo{volume}{80}},
  \bibinfo{pages}{140512} (\bibinfo{year}{2009}).
\newblock \urlprefix\url{https://doi.org/10.1103/PhysRevB.80.140512}.

\bibitem{Chubukov2009}
\bibinfo{author}{Chubukov, A.~V.}, \bibinfo{author}{Eremin, I.} \&
  \bibinfo{author}{Korshunov, M.~M.}
\newblock \bibinfo{title}{{Theory of Raman response of a superconductor with
  extended s-wave symmetry: Application to the iron pnictides}}.
\newblock \emph{\bibinfo{journal}{Phys. Rev. B}} \textbf{\bibinfo{volume}{79}},
  \bibinfo{pages}{220501} (\bibinfo{year}{2009}).
\newblock \urlprefix\url{https://doi.org/10.1103/PhysRevB.79.220501}.

\end{thebibliography}

\onecolumngrid{
\section{Materials and Methods}
\renewcommand{\thefigure}{M\arabic{figure}}
\setcounter{figure}{0}

\renewcommand{\thetable}{M\arabic{table}}
\setcounter{table}{0}

\renewcommand{\theequation}{M\arabic{equation}}
\setcounter{equation}{0}

\subsection*{Raman instrument}
Spontaneous Raman measurements were performed on the UT-3 Raman spectrometer (see Fig. S1).\cite{Schulz2005} This triple-grated spectrometer is fully achromatic due to the use of reflective optics. Excellent stray light rejection is achieved via an entrance objective with a large numerical aperture of 0.5 in a Cassegrain-type design, small focal points due to aberration-free off-axis paraboloids in combination with bilateral slits of the pre-monochromator, and an additional relay-stage equipped with two monolateral slits for setting an asymmetric bandpass for Stokes- and anti-Stokes measurements. A beam block in the entrance objective blocks reflected and emitted light in an angle of 21.8° to the vertical (see Fig. S2). In this way, a low-frequency cutoff of less than 5 cm$^{-1}$ can be achieved.\cite{Schulz2005} However, in the current experiment, the low-frequency cutoff is limited to the natural Fourier-broadening of the laser line of the pulsed laser source to approx. 80 cm$^{-1}$. We utilized a pulsed Ti:Sapphire laser system, Tsunami model HP fs 15 WP (Spectra Physics Lasers Inc., California) at a fundamental wavelength of 802 nm with a second harmonic generation (SHG) unit generating the probe wavelength of 401 nm. The pulse duration was 1.2 $\pm$ 0.1 ps monitored with an autocorrelator (APE GmbH, Berlin, Germany). The 401 nm and 802 nm beams were guided over two separate beam paths as shown in Figure S1. For time delay scans, a motorized delay line in the pump beam path was used. A $\lambda /2$ waveplate in the probe beam path allows symmetry-dependent studies. We have employed A$_{2\textrm{g}}$+B$_{1\textrm{g}}$ polarization by using crossed polarization and A$_{1\textrm{g}}$+B$_{2\textrm{g}}$ symmetry by parallel polarization between incident and scattered light with respect to the a and b axes in the CuO$_2$ planes. By rotating the sample by 45$^\circ$ the A$_{2\textrm{g}}$+B$_{2\textrm{g}}$ signal was measured for reference (see Fig. S10) showing that contributions from A$_{2\textrm{g}}$ and B$_{2\textrm{g}}$ are small. Neutral density filter units/wheels were used for fluence dependence. 
Figure S2 shows the details of the entrance objective consisting of four on-axis parabolic mirrors, which focuses the light into the first monochromator. Pump and probe beam reach the sample at an angle of 21.8$^\circ$. By this, we apply a non-zero in-plane momentum to the sample, which causes symmetry breaking and activation of the Higgs mode in B$_{1\textrm{g}}$ symmetry.

\subsection*{Transient Stokes and anti-Stokes Raman measurements}
To characterize the beam spot size and shape and align the spatial overlap of the probe and the pump spot, a DFK 23GM021 industrial camera (The Imaging Source, Bremen, Germany) with a pixel size of 3.75 µm was positioned at the focal point of the entrance objective of the UT-3. We used a probe spot (401 nm) of FWHM = 19.5 µm x 11.1 µm (hor x ver) and a pump spot of FWHM = 22.5 µm x 16.9 µm (hor x ver). The applied probe power was 4.75 $\pm$ 0.15 mW resulting in a fluence of 35.08 $\pm$ 1.15 µJ cm$^{-2}$ at a repetition rate of 80 MHz of the Tsunami system. For the pump, fluences of 20 µJ cm$^{-2}$ to 113 µJ cm$^{-2}$ were used (see Table S2 for details). In order to establish temporal overlap of the two pulses, an ultrafast diode UPD-50-UP (Alphalas, Göttingen, Germany) was placed in the focal point of the entrance objective, where probe and pump beam were focused. A Picoscope 6402B by Pico Technology (Cambridgeshire, United Kingdom) was used to monitor the pulses and find rough temporal overlap. We then conducted reference measurements on highly oriented pyrolytic graphite (HOPG) by Alfa Aesar, Thermo Fisher Scientific (Massachusetts, USA), to calibrate the delay line.
Probe-only, pump-probe, pump-only, and background measurements were taken one after the other with an integration time of 30 min each. In general, we conducted 3 repetitions per measurement to improve the signal-to-noise ratio and ensure stability over the measurement time. Stokes and anti-Stokes measurements were carried out with two different settings of the spectrometer bandpass and monolateral slits.

\subsection*{Sample}
The investigated sample is an Y$_{0.08}$-substituted Bi-2212 crystal with a T$_\textrm{c}$ of 92 K (see Figure S3) with the nominal composition Bi$_{2.00}$Sr$_{2.00}$Ca$_{0.92}$Y$_{0.08}$Cu$_2$O$_{8+\delta}$. It is slightly underdoped as compared to T$_{\textrm{c,max}}$ = 96 K. The sample was annealed at 500 $^\circ$C in Argon. For more details on sample growth see previous work.\cite{Eisaki2004} A continuous flow LHe Konti-Cryostat Spectro (CryoVac, Troisdorf, Germany) was utilized to cool the sample down to 8 K base temperature. The used cryostat is an exchange-gas cryostat with active cooling from the side of the laser impinging the sample. At a typical optical penetration depth of 20 nm - 100 nm, this leads to an effectively base-temperature independent laser heating.

\subsection*{Data treatment}
Raman spectra have been corrected for the static background detector signal and the spectral response of the spectrometer. The data was then normalized to the respective probe laser power and integration time. For NEARS data, no Bose-function correction was applied to the data itself, since the division Bose factor for anti-Stokes spectra converges to zero (see Figure S4). We, therefore, plot the Raman intensity instead of the Raman susceptibility (see Figure 1 and 2 in the main text). However, by utilizing the linkage of Stokes and anti-Stokes data via the Bose-function (see equation S1), one can calculate the anti-Stokes spectrum from the measured Stokes spectrum, and analyze the difference between the calculated anti-Stokes and the measured anti-Stokes data. As a key result of this work, we find no difference signal for all non-pumped data and pump-probe data above T$_\textrm{c}$ (see Fig. 1 in the main text). However, in the superconducting state and in the pump-probe measurement, we obtain a difference signal in the anti-Stokes data, which can be attributed to an overpopulation of the excited Higgs state. NEARS maps are derived from the data by plotting the superposition of the Raman susceptibility of the in-gap NEARS feature obtained from anti-Stokes data together with the Raman susceptibility of the PB feature (Stokes side) in an interpolation 2D color plot as a function of excitation energy (see also parameterization method of the PB feature in SI S.4). For this, we utilize the python class scipy.interpolate.interp2d. In order to make this new representation of non-equilibrium Raman data more accessible to the reader, Figure S7 shows the Raman susceptibility of the NEARS feature on an energy-gain (anti-Stokes) axis, together with the superconductivity induced Raman susceptibility on the Stokes side (PB peak) on an energy-loss axis. Raman spectra (Fig. 1 and 2) were fitted using the Levenberg–Marquardt fitting routine of Igor Pro (Version 6.3). To fit the electonic Raman response within our BCS model (Fig. 4), we use a Trust Region Reflective (trf) algorithm (scipy.optimize.least\_squares).

\subsection*{Phenomenological Model and BCS Theory}
A phenomenological Ginzburg-Landau model can describe the charged bosonic condensate by using a Klein-Gordon like Lagrangian with a Ginzburg-Landau Mexican-hat potential $F(\Psi) = \alpha |\Psi|^2 + \frac{\beta}{2} |\Psi|^4$ ($\alpha < 0$) and for small fluctuations of the Higgs amplitude ($H$) around the ground state $|\Psi_0| = \sqrt{\frac{-\alpha}{\beta}}$.\cite{puviani_current-assisted_2020}

The Higgs mode, characterizing the low-energy excitation spectrum of the condensate, is Raman active and couples quadratically to the vector potential. We can calculate the equation of motion for the Higgs mode by using the Euler-Lagrange equations in the $q \rightarrow 0$ limit (see SI S.4). In order to account for the non-equilibrium experimental conditions, we can quench the order parameter within the Ginzburg-Landau theory by quenching $\alpha$, $\beta$, or both of them, since $\Psi_0$ itself depends on the ratio of $\sqrt{\frac{|\alpha|}{\beta}}$. If we quench $\alpha$ we will change the frequency of the Higgs mode to lower energy. As shown in Fig. 2f, we do not observe this behavior in the experiment. If we quench $\beta$, we will reduce the superfluid density and not change the frequency of the Higgs mode. This case fits our experimental observations. We can calculate the Green's function in the ${\bf q} \rightarrow 0$ case by assuming a $\delta$ function-like quench due to a change in $\beta$. See SI S.4 for more details.

We further compare the experimental data to a microscopic BCS weak-coupling theory. A weak-coupling Hamiltonian is utilized

\begin{equation}
    \mathcal{H}(t)=\sum_{\bm{k}, \sigma} \xi_{\bm{k}-\textbf{A}(t)} c^\dagger_{\bm{k}, \sigma} c_{\bm{k}, \sigma} - \sum_{\bm{k}, \bm{k}'} V_{\bm{k}, \bm{k}'} c^\dagger_{\bm{k}, \uparrow} c^\dagger_{-\bm{k}, \downarrow} c_{-\bm{k}', \downarrow} c_{\bm{k}', \uparrow}  \, .
\end{equation}

The electron dispersion $\xi_{\bm{k}} = \epsilon_{\bm{k}} - \epsilon_F$ is measured relative to the Fermi level and $c^\dagger_{\bm{k}, \sigma}$ and $c_{\bm{k}, \sigma}$ represent the electron creation or annihilation operators. A separable pairing interaction $V_{\bm{k},\bm{k}'}=V f_{\bm{k}} f_{\bm{k}'}$ with strength $V$ and symmetry function $f_k$ is used. The coupling to light is obtained by the expansion of the minimal coupling up to second order in $\bm{A}$

\begin{equation}
    \xi_{\bm{k}-\bm{A}(t)} = \xi_{\bm{k}} - \sum_i \partial_i \xi_{\bm{k}} A_i(t) + \frac{1}{2} \sum_{i,j} \partial^2_{ij} \xi_{\bm{k}} A_i(t) A_j(t) + \mathcal{O}(A(t)^3) \, .
\end{equation}

The pair-breaking Raman vertices for B$_{1\textrm{g}}$ and A$_{1\textrm{g}}$ can be expressed as $\gamma_{\textrm{B}_{1\textrm{g}}} = \gamma_{b} \cos(2\phi)$ and $\gamma_{\textrm{A}_{1\textrm{g}}} = (1+b_0\xi)(\gamma_0 + \gamma_1 \cos(4\phi) + \gamma_2 \cos(8\phi))$, respectively, as outlined in the main text.\cite{Devereaux1995}

The lowest order contribution to the B$_{1\textrm{g}}$ pair-breaking susceptibility is not Coulomb-screened \cite{Devereaux1995} and can be diagrammatically represented as shown in Fig. 4a. The information about the light-matter interaction and the symmetry channel, in particular, is entirely contained in the Raman vertex function, $\gamma_{\mathrm{B}_{1\mathrm{g}}}\sigma_3$. With the dimensionless frequency $x = {\omega}/{2\Delta_0} + i\eta$ we can write algebraically

\begin{equation}
    \chi_{\Bg \Bg}(\iq=0,x)/N_F = \expval{\gamma_{\Bg}^2} =\gamma_b^2 I_2(x) \, ,
\end{equation}

with the integrals $I_n(x)$ defined as

\begin{equation}
\begin{aligned}
        \expval{\cos^{2(n-1)}(2\phi)} &= N_F \int_{-\xi_D}^{\xi_D} d\xi \int_0^{2\pi}d\phi  \frac{2\Delta_\ik^2\cos^{2(n-1)}(2\phi)}{E_\ik (4E_\ik^2 - \ioo)}  
       \\
       &= N_F \int_0^{2\pi} d\phi \frac{f_\ik^{2n}/x^2}{\sqrt{f_\ik^2/x^2-1}} \tan^{-1} \left( \frac{x}{\sqrt{f_\ik^2-x^2}}\right)\\
       &=  2 N_F \int_0^1 dt \frac{t^{2n}/x^2}{\sqrt{1-t^2}\sqrt{1-t^2/x^2}} \mathcal{D}(x,t)  \coloneq I_n(x) \\
       \text{with}\quad \mathcal{D}(x,t) &= \left( \text{sgn}(\Re(x))i\pi + \ln \left[ \frac{1-\sqrt{1-t^2/x^2}}{1+\sqrt{1-t^2/x^2}}\right]\right)\,.
\end{aligned}
\end{equation}

The A$_{1\textrm{g}}$ response is modified by Coulomb screening of the charge fluctuations. The algebraic expression for the screened susceptibility,
\begin{equation}
{\chi^{sc}_{\Ag \Ag} = \chi_{\Ag \Ag} - \frac{\chi_{\Ag \sigma_3} \chi_{\sigma_3 \Ag}}{\chi_{\sigma_3 \sigma_3}}}\,,
\end{equation}
can be expressed after Fermi-surface harmonics expansion as 

\begin{equation}
    \begin{split}
        \chi^\text{sc}_{\Ag\Ag}
        &=64\gamma_2^2 I_5 + (32\gamma_1 \gamma_2 - 128\gamma_2^2) I_4  
         + (4\gamma_1^2 - 48\gamma_1\gamma_2 + 80\gamma_2^2) I_3 \\
         &\quad + (-4\gamma_1^2 + 20\gamma_1\gamma_2 - 16\gamma_2^2) I_2 
         +(\gamma_1^2  - 2\gamma_1\gamma_2 + \gamma_2^2) I_1\\
         &\quad - \frac{\left(8\gamma_2 I_3 + (2\gamma_1-8\gamma_2) I_2 - \gamma_1 I_1\right)^2}{ I_1}\,.
    \end{split}
\end{equation}
The peak position of the pair-breaking excitation in A$_{1\textrm{g}}$ symmetry is strongly dependent on the admixture of the higher-order Fermi-surface harmonics $\gamma_1, \gamma_2$.
 
The lowest-order contribution to the Higgs response is given by the Feynman diagram shown in Fig. 4c.\cite{Cea2014, puviani_current-assisted_2020} The homogeneous Higgs propagator, assuming an isotropic Fermi-surface and $T \rightarrow 0$ can be expressed as

\begin{equation}
    H^{-1}(\io,\iq=0) = \sum_\ik \frac{f_\ik^2(4\Delta_\ik^2-\io^2)}{2E_\ik(4E^2_\ik-\io^2)}\tanh(\beta E_\ik/2) \, .
\end{equation}

With the established parameters $\gamma_1$, $\gamma_2$, $\eta$, and $\Delta_0$ we can evaluate the A$_{1\textrm{g}}$ response of the Higgs excitation following eq. S21. Details of our calculations can be found in the SI (S.2).

\newpage
\renewcommand{\thefigure}{S\arabic{figure}}
\setcounter{figure}{0}

\renewcommand{\thetable}{S\arabic{table}}
\setcounter{table}{0}

\renewcommand{\theequation}{S\arabic{equation}}
\setcounter{equation}{0}

\section{Supplementary Information}
\subsection{S.1 BCS-Dirac Correspondence} \label{S.1}
In the following, we elaborate on an analogy between the BCS- and Dirac Hamiltonian, which is repeatedly pointed out in the literature.\cite{Pekker2015, Nambu1960}\\
In natural units, the Dirac Lagrangian of relativistic Fermions in Weyl representation with the gamma matrices
\begin{equation}
    \gamma^0 = \begin{pmatrix}
        0&\mathbb{1}\\\mathbb{1}&0
    \end{pmatrix}\quad \text{and} \quad \gamma^i = \begin{pmatrix}
        0&\sigma^i\\-\sigma^i&0
    \end{pmatrix}\quad \text{is}\quad \mathcal{L}^\text{Dirac} = \bar{\Psi}\left(i\slashed{\partial} - m\right)\Psi\,.
\end{equation}
  Legendre transformation using the canonical momentum $\frac{\delta \mathcal{L}}{\delta \partial_0\Psi} = i \Psi^\dagger$ results in the Dirac-Hamiltonian\cite{Peskin} 
\begin{equation}
    \mathcal{H}^\text{Dirac} = \Psi^\dagger \left(i\bm{\sigma}\cdot \bm{\nabla} \hat{\sigma}_3 + m \hat{\sigma}_1\right)\Psi=\begin{pmatrix}
        \Psi^\dagger_L& \Psi^\dagger_R
    \end{pmatrix}
    \begin{pmatrix}
        i\bm{\sigma}\cdot \bm{\nabla}  &  m \\ m & -i\bm{\sigma}\cdot \bm{\nabla} 
    \end{pmatrix}
    \begin{pmatrix}
        \Psi_L\\\Psi_R
    \end{pmatrix} \,.
\end{equation}
Here we use $\hat{\sigma}_i = \sigma_i \otimes \mathbb{1}_{2\text{x}2}$ to distinguish the extension of the Pauli matrices to $4\times4$ block matrices.\\
The eigenvalues of this Hamiltonian are $E_\textbf{p} = \pm\sqrt{\textbf{p}^2c^2 + m^2c^4}$.\\

Using Nambu spinor notation the BCS-Hamiltonian (without local $U(1)$ symmetry i.e. uncharged BCS) may be rewritten into a Bogoliubov-deGennes Hamiltonian, 
\begin{equation}
    \mathcal{H}^{\text{BdG}}= \sum_\textbf{k}\Psi^\dagger_\textbf{k}\left(\xi_\textbf{k}\sigma_3 + \Delta_\textbf{k} \sigma_1\right)\Psi_\textbf{k} = \sum_\textbf{k} \begin{pmatrix} \Psi^\dagger_{ \textbf{k},\uparrow}&\Psi_{ -\textbf{k},\downarrow} \end{pmatrix}
    \begin{pmatrix}
        \xi_\textbf{k} & \Delta_\textbf{k}\\
        \Delta_\textbf{k} & -\xi_\textbf{k}
    \end{pmatrix}
    \begin{pmatrix} \Psi_{ \textbf{k},\uparrow}\\\Psi^\dagger_{ -\textbf{k},\downarrow} \end{pmatrix}\,.
\end{equation}
The BdG Hamiltonian's eigenvalues are $E_\textbf{k}=\pm \sqrt{\xi^2_\textbf{k}+\Delta_\textbf{k}^2}$. Unless the electronic dispersion $\epsilon_\textbf{k}$ is linear in momentum, $E_\textbf{k}$ is not a Lorentz invariant dispersion of the quasiparticles. However, in the vicinity of the Fermi surface $\abs{\xi_\textbf{k}} \ll W$, the bandwidth, we may approximate $\xi_\textbf{k}\approx v_F (\abs{\textbf{k}}-k_F)$. For s-wave symmetry ($\Delta_\textbf{k}=\Delta$) one can make the analogy explicit by introduction of $m_\Delta=\frac{\Delta}{v_F^2}$ and obtains
\begin{equation}
    E^\text{BCS}_\textbf{k}=\sqrt{(\abs{\textbf{k}}-k_F)^2v_F^2 + m_\Delta^2 v_F^4}\,,
\end{equation}
resembling the Dirac Hamiltonian's relativistic dispersion relation with the speed of light replaced by the Fermi velocity $v_F$.\\

It is worth mentioning that the (Bogoliubov quasi-)particle-hole symmetry of the BdG-Hamiltonian, in contrast to that of the Dirac Hamiltonian, is only a formal one \cite{Zirnbauer2021}. Ultimately, this is because in constructing the BdG-Hamiltonian one is doubling the Hilbert space $V\rightarrow W = V\oplus V^*$, for which the space of positive energy states $W_+ = V_+\oplus V_-^*$ and that of negative energy states $W_- = V_-\oplus V_+^*$ are always isomorphic (using the Fréchet-Riesz isomorphism). It is then easy to show, that $h^{BdG}$ satisfies particle-hole symmetry independently of physical properties like the dispersion $\xi_\textbf{k}$, gap $\Delta$ or whether $V_+ \simeq V_-$. In particular, this is also true for $\Delta=0$. Therefore, this formal particle-hole symmetry is not due to superconductivity and should not be confused with a physical symmetry. On the contrary, it may be viewed as a constraint, enforcing the dependence between the Nambu spinors $\Psi^\dagger$ and $\Psi$, which a Hamiltonian must satisfy to be a BdG-Hamiltonian \cite{Chiu2016}. The physical particle-hole symmetry mentioned in the context of superconductors is an approximate symmetry restricted to the states of energies $E\sim \Delta$ close to the Fermi surface. It is essentially just a consequence of Taylor expanding the density of states to zeroth order, which is often a sufficiently good approximation in conventional superconductors because $\Delta \ll W$.  


\def\ik{\textbf{\textit{k}}}
\def\iA{\textbf{\textit{A}}}
\def\iv{{i\nu_n}}
\def\io{{i\omega_n}}
\def\ioo{(i\omega_n)^2}

\def\TK{{\Tilde{K}}}
\def\TE{{\Tilde{E}}}
\def\TI{{{\mathcal{I}}}}

\subsection{S.2 Electronic Raman Response in a BCS Weak-Coupling Theory}

With an energy of 3 eV and an effective bandwidth of only 1 eV in Bi-2212 \cite{markiewicz_one-band_2005}, the probe pulse is resonant. To account for the full Raman cross-section, we would need to consider every band in this energy range (see \cite{devereaux_inelastic_2007} for a general review of electronic Raman scattering theory). Such a many-band calculation for resonant Raman scattering is outside the scope of this article. Instead, we work with a simple single-band toy model (equation (2) of the main text) capturing the properties of superconductivity, and parametrize the Raman vertices in terms of generalized Fermi surface harmonics, which are solely based on symmetry restrictions. It is assumed that the pairing interaction is restricted to some relatively narrow hull around the Fermi surface, just like in conventional superconductivity. This makes the susceptibilities \eqref{eq:generalSusceptibility} less sensitive to the dispersion, which we exploit by replacing it with a simple (but not constant) approximation for the density of states around the Fermi surface. We introduce the expansion coefficients $(\gamma_0,\gamma_1,\gamma_2,\gamma_b,b_0,b_1)$ as fitting parameters, capturing the dispersion and particle-hole asymmetry. Thus, we are able to make statements based entirely on the gap- and Fermi surface symmetries.\\

Both the pair-breaking and the Higgs response are constructed from common building blocks, the homogeneous polarization/susceptibility diagrams, which we define as

\begin{fmffile}{GeneralSusceptibility}
\fmfset{arrow_len}{3mm}
\begin{equation}\label{eq:generalSusceptibility}
    \chi_{\Gamma_1,\Gamma_2} (\io) =
 \eqgraph{5ex}{0ex}{\fmfframe(15,10)(15,25){
    \begin{fmfgraph*}(50,30)
    \fmfleft{i}
    \fmfright{o}
    \fmf{plain,tension=2}{i,v1}
    \fmf{plain_arrow,tension=0.1,left=1,width=1,label=$G_\ik(\iv + \io)$}{v1,v2}
    \fmf{plain_arrow,tension=0.1,left=1,width=1,label=$G_\ik(\iv )$}{v2,v1}
    \fmf{plain,tension=2}{v2,o}
    \fmfdot{v1,v2}
    \fmfv{l=$\Gamma_1$}{v1}
    \fmfv{l=$\Gamma_2$}{v2}
  \end{fmfgraph*}} }
  = -\frac{1}{2\beta}\sum_{\ik,\iv} \Tr[\Gamma_1 G_\ik(\iv+\io)\Gamma_2G_\ik(\iv) ]\,,
  \end{equation} 
\end{fmffile}

with $\io = 2n\pi/\beta$ denoting the bosonic and $\iv= (2n+1)\pi/\beta$ the fermionic Matsubara frequencies. We calculate the homogeneous ($\textbf{q}=0$) functions because we can safely use the dipole approximation given the probe wavelength of $400\, \text{nm}$ and coherence length $\xi \lesssim 2\, \text{nm} $ \cite{Hwang2021}.
The electronic Green's functions of the BCS model written in the basis of Nambu spinors are defined as
\begin{equation}
    G_\ik(\iv) = \frac{\iv + \xi_\ik \sigma_3 + \Delta_\ik \sigma_1}{(\iv)^2 - E_\ik^2} = \frac{1}{(\iv)^2 - E_\ik^2}\begin{pmatrix}
        \iv +  \xi_\ik & \Delta_\ik \\ \Delta_\ik & \iv -  \xi_\ik
    \end{pmatrix}\,,
\end{equation}
where $\Delta_\ik = \Delta_0 f_\ik$ and $f_\ik = \cos2\phi$ for d-wave pairing, $E_\ik = \sqrt{\xi^2_\ik + \Delta^2_\ik}$ is the Bogoliubov-quasiparticle dispersion and $\sigma_i$ are the Pauli matrices.  
$\Gamma_1$ and $\Gamma_2$ are 2x2-matrix-valued vertices, selecting the appropriate components of the matrix-valued Green's function upon contraction. In this work, there are three types of vertices, $\Gamma_i \in \{ f_\ik \sigma_1, \sigma_3, \gamma_\ik \sigma_3\}$ ($f_\ik$ and $\gamma_\ik$ are scalar-valued functions) which are distinguished by different vertex shapes in Feynman diagrams.\\ For the light-matter coupling, we use the Raman vertex $\gamma_\ik$ which we expand in terms of irreducible representations (irreps) of the lattice symmetry group, e.g. $\gamma_{\Ag}, \;\gamma_{\Bg} $. 

We absorb the matrix structure of the $\gamma$ vertices to improve readability since it is the same for all irreps and write e.g. $\chi_{\Ag \Gamma_2} = \chi_{\gamma_{\Ag}\sigma_3, \Gamma_2}$. Similarly, we adapt the notation $\chi_{\Delta \Gamma_2} = \chi_{f_\ik \sigma_1, \Gamma_2}$ and define an expectation value wrt. the Tsuneto function,
\begin{equation}\label{eq:kSpaceExpVal}
    \expval{\alpha(\ik)} = \sum_\ik  \frac{2\Delta_\ik^2\alpha(\ik)}{E_\ik (4E_\ik^2 - \ioo)}\tanh(\beta E_\ik/2) \,.
\end{equation}
This allows us to write compactly,

\begin{table}[h!]
\centering
\caption{Summary of the elementary susceptibilities which the observables are constructed from.}\label{tab:suscTable}
\bgroup
\def\arraystretch{2}
\begin{tabular}{|l|c|}
\hline
     bare $\Ag$ pair breaking response & $\chi_{\Ag \Ag} = \expval{(\gamma_{\Ag})^2}$ \\
     \hline
     anomalous susceptibility& $\chi_{\Delta \Delta} = \frac{1}{\Delta_0^2}\expval{\xi_\ik^2}$\\
     \hline
     bare light-Higgs coupling& $\chi_{\Delta \Ag} = \chi_{\Ag \Delta} = -\frac{1}{\Delta_0}\expval{\xi_\ik \gamma_{\Ag}}$\\
     \hline
     bare charge-fluctuation susceptibility&$\chi_{\sigma_3 \sigma_3  } =  \expval{1}$ \\
     \hline
     charge-fluctuation-Higgs coupling& $\chi_{\sigma_3\Delta} = \chi_{\Delta \sigma_3} = -\frac{1}{\Delta_0}\expval{\xi_\ik}$\\
     \hline
     light-charge coupling & $\chi_{\Ag \sigma_3} = \chi_{\sigma_3 \Ag} = \expval{\gamma_\Ag}$\\
     \hline
\end{tabular}
\egroup
\end{table}

where “bare” refers to the lack of Coulomb screening and $\Ag$ may be replaced at any point it appears by another irrep to obtain the respective susceptibility.\\

\subsubsection{Ansatz for Raman vertices and density of states}

In the simplified one-band model of the d-wave superconductors, the square lattice symmetry of the cuprates typically leads to Fermi surfaces that are either connected around $\Gamma = (0,0)$, or $M = (\pm\pi,\pm\pi)$. We choose polar coordinates whose angle $\phi$ is defined as described in Fig. \ref{fig:S14} with respect to the point that the Fermi surface encloses. Following the ansatz proposed in \cite{Devereaux1995}, we simplify the Raman vertices by parameterizing their angular dependencies with basis functions in the irreps of the tetragonal $D^{4h}$ point group: $\gamma_{\Ag} = \sum_{L=0} \gamma^{\Ag}_{L} \cos(4L\phi)$, $\gamma_{\Bg} = \sum_{L=1} \gamma^{\Bg}_{L}\cos{([4L-2]\phi)}$. We follow previous research \cite{Devereaux1995} by keeping up to $L=2$ for the $\Ag$ response, and only the $L=1$ term for the $\Bg$ response. In this work we lift the approximate particle-hole symmetry of superconductivity (\ref{S.1}) by introducing weak energy dependencies, $\gamma_{A1g}(\xi,\phi)$,  $N(\xi)$, in the $\Ag$ vertex and the density of states $N$.

To summarize, the ansatz for the Raman vertices and the density of states is:
\begin{equation}
    \begin{aligned}
        \gamma_{\Ag} &= (1+b_0\xi)(\gamma_0 + \gamma_1 \cos(4\phi) + \gamma_2 \cos(8\phi)) \\ 
        \gamma_{\Bg} &= \gamma_{b} \cos(2\phi)\\
        N(\xi) &= N_F(1+b_1\xi)
    \end{aligned}
\end{equation}
here $\gamma_0, \gamma_1, \gamma_2$ are the coefficients in the generalized Fermi surface harmonic expansion of $\gamma_{\Ag}$, likewise $\gamma_b$ is the coefficient in the expansion of $\gamma_{\Bg}$. Energy dependence of $\gamma_{\Ag}$ is parametrized by $b_0$. The deviation from a constant density of state in the thin energy hull around the Fermi surface, inside which we assume the net-attractive electron-electron interaction, is represented by $b_1$. We assume that the energy dependencies $b_0$ and $b_1$ are small $b_0\Delta_0, b_1 \Delta_1 \ll 1$. In this parametrization, the average becomes
\begin{equation}\label{eq:EnergyRepExpVal}
    \expval{\alpha(\xi(\ik),\phi(\ik))} = \int_{-\infty}^{\infty} d\xi \int_0^{2\pi}d\phi   \frac{N_F(1 + b_1 \xi)2\Delta(\phi)^2\alpha(\xi,\phi)}{\sqrt{\xi^2 + \Delta(\phi)^2} (4(\xi^2 + \Delta(\phi)^2) - \ioo)}
\end{equation}

where we dropped the $\tanh$ encoding temperature dependence from \eqref{eq:kSpaceExpVal}, whereby we approximate $T=0$ for the electronic degrees of freedom. For the highest pump fluence in the experiment, which we use to extract the Higgs signal (see Fig. 4 main text), the effective quasi-equilibrium temperature due to heating is determined to be $T_\text{eff}\approx 98 \pm 13.75\,\text{K}$. The thermal energy $k_B T \approx 8 \text{ meV}$ is still significantly smaller than the lowest energy eigenvalue $ E_{\textbf{k}=0} = \Delta_0 \approx 30 \text{ meV}$ introducing an error of $ 5 \%$ for $E_{\textbf{k}=0}$ that decays exponentially for higher energy eigenvalues. For the unpumped data, used to fit the pair-breaking peaks, with $T_\text{eff} \approx 17\,\text{K}$ ($8\,\text{K}$ base temperature plus $9\,\text{K}$ probe heating (see section \ref{sec:temperature})) the zero temperature approximation is even better.

\subsubsection{$\TI$ and $\MM$ functions} 
We can express the susceptibilities in terms of two classes of integrals, the first of which is given by
\begin{equation}
\begin{aligned}
        \expval{\cos^{2(n-1)}(2\phi)} &= N_F \int_{-\xi_D}^{\xi_D} d\xi \int_0^{2\pi}d\phi  \frac{2\Delta_\ik^2\cos^{2(n-1)}(2\phi)}{E_\ik (4E_\ik^2 - \ioo)}  
       \\
       &= N_F \int_0^{2\pi} d\phi \frac{f_\ik^{2n}/x^2}{\sqrt{f_\ik^2/x^2-1}} \tan^{-1} \left( \frac{x}{\sqrt{f_\ik^2-x^2}}\right)\\
       &=  2 N_F \int_0^1 dt \frac{t^{2n}/x^2}{\sqrt{1-t^2}\sqrt{1-t^2/x^2}} \mathcal{D}(x,t)  \coloneq \TI_n(x) \\
       \text{with}\quad \mathcal{D}(x,t) &= \left( \text{sgn}(\Re(x))i\pi + \ln \left[ \frac{1-\sqrt{1-t^2/x^2}}{1+\sqrt{1-t^2/x^2}}\right]\right)\,.
\end{aligned}
\end{equation}

In the first equality, we inserted the definition \eqref{eq:EnergyRepExpVal} of the expectation value and exploited that the $b_1$ term vanishes for energy-independent functions in the expectation value. In the second step, we exploit that $f_\textbf{k}=\cos{2\phi}$ and extended $\xi_D \rightarrow \infty$, introducing an error $\mathcal{O}\left(\frac{\Delta_0^2}{\xi_D^2}\right)$, to evaluate the energy integral. This is valid in the $\xi_D \gg \Delta_0$ limit. We use $\xi_D/\Delta_0 = 5$. Furthermore, we performed analytical continuation of the frequency and introduced $x = \omega/2\Delta_0 + i\eta$. Finally, we make a substitution $t = \cos(2\phi)$. The sgn($x$) function gives the sign of $x$. 
The second class of integrals is
\begin{equation}
    \MM_n(\xi_D,x) \coloneq \frac{1}{\Delta_0^2}\expval{\xi^2 \cos^{2(n-1)} (2\phi)} \, .
\end{equation}
This function is logarithmically divergent in the energy cut-off $\xi_D$. Such a divergence exists in the usual BCS-like weak coupling theory and is the consequence of the point-like attractive interaction $V$. The dependence on the energy cut-off is equivalent to a dependence in the magnitude of the interaction V and is necessary for numerical simulations. We have verified that in the range $\xi_D \gg \Delta_0$ results do not change significantly.

\subsubsection{B$_{1\textrm{g}}$ Pair-Breaking Response}
\label{sec:B1g}
The lowest order contribution to the B$_{1\textrm{g}}$ pair-breaking susceptibility is not Coulomb-screened \cite{Devereaux1995} and can be diagrammatically represented by:

\begin{fmffile}{B1gPB}
\fmfset{arrow_len}{3mm}
\fmfset{dot_len}{1.5mm}
\begin{equation}
\chi_{\Bg\Bg} = 
 \eqgraph{1.5ex}{0ex}{\fmfframe(15,10)(15,10){
    \begin{fmfgraph*}(80,30)
    \fmfleft{i1,i2}
    \fmfright{o1,o2}
    \fmf{wiggly}{i1,v1,i2}
    \fmf{wiggly}{o1,v2,o2}
    \fmf{plain_arrow,tension=0.4,left=1,width=1}{v1,v2}
    \fmf{plain_arrow,tension=0.4,left=1,width=1}{v2,v1}
    \fmfv{d.sh=square,d.filled=empty,d.size=7}{v1,v2}
    \fmfv{l=$\gamma_{\text{B}_{1\mathrm{g}}} \sigma_3$}{v1}
    \fmfv{l=$\gamma_{\text{B}_{1\mathrm{g}}} \sigma_3$}{v2}
  \end{fmfgraph*}} }\,,
  \end{equation} 
\end{fmffile}
where the wiggly lines symbolize light and the plain lines represent electronic propagators. The information about the light-matter interaction and the symmetry channel, in particular, is entirely contained in the Raman vertex function, $\gamma_{\mathrm{B}_{1\mathrm{g}}}\sigma_3$. Algebraically,

\begin{equation}
    \chi_{\Bg \Bg}(\iq=0,x)/N_F = \expval{\gamma_{\Bg}^2} =\gamma_b^2 \TI_2(x)
\end{equation}

with the dimensionless frequency $x = {\omega}/{2\Delta_0} + i\eta$.

\subsubsection{$\text{A}_{1\textrm{g}}$ Pair-Breaking Response}

The A$_{1\textrm{g}}$ response is modified by Coulomb screening of the charge fluctuations \cite{monien_theory_1990}. These are not present in the B$_{1\mathrm{g}}$ response because the Coulomb interaction lies in the A$_{1\textrm{g}}$ channel and therefore cannot couple to B$_{1\mathrm{g}}$. Diagrammatically, the screening shows up as an additional contribution capturing the screened Coulomb interaction between the charge fluctuations due to the Raman probe.

\begin{fmffile}{ScreenedA1gPB}
\fmfset{arrow_len}{3mm}
\fmfset{dot_len}{1.5mm}
\begin{equation}
\chi_{\Ag\Ag}^{sc} = 
 \eqgraph{1.5ex}{0ex}{\fmfframe(15,10)(15,10){
    \begin{fmfgraph*}(80,30)
    \fmfleft{i1,i2}
    \fmfright{o1,o2}
    \fmf{wiggly}{i1,v1,i2}
    \fmf{wiggly}{o1,v2,o2}
    \fmf{plain_arrow,tension=0.4,left=1,width=1}{v1,v2}
    \fmf{plain_arrow,tension=0.4,left=1,width=1}{v2,v1}
    \fmfv{d.sh=square,d.filled=empty,d.size=7}{v1,v2}
    \fmfv{l=$\gamma_\Ag \sigma_3$}{v1}
    \fmfv{l=$\gamma_\Ag \sigma_3$}{v2}
  \end{fmfgraph*}} }+
  \eqgraph{1.5ex}{0ex}{\fmfframe(0,5)(0,5){
    \begin{fmfgraph*}(150,30)
    \fmfleft{i1,i2}
    \fmfright{o1,o2}
    \fmf{wiggly}{i1,v1,i2}
    \fmf{wiggly}{o1,v4,o2}
    \fmf{plain_arrow,tension=0.4,left=1,width=1}{v1,v2}
    \fmf{plain_arrow,tension=0.4,left=1,width=1}{v2,v1}
    \fmf{dbl_dots,label=$V^{sc}_{C}(0)$,tension=0.5,l.side=left}{v2,v3}
    \fmf{plain_arrow,tension=0.4,left=1,width=1}{v3,v4}
    \fmf{plain_arrow,tension=0.4,left=1,width=1}{v4,v3}
    \fmfv{d.sh=square,d.filled=empty,d.size=7}{v1,v4}
    \fmfv{d.sh=triangle,d.filled=empty,d.size=7,d.angle=-30}{v2}
    \fmfv{d.sh=triangle,d.filled=empty,d.size=7,d.angle=30}{v3}
    \fmfv{l=$\sigma_3$,l.a=-60}{v2}
    \fmfv{l=$\sigma_3$,l.a=-120}{v3}
  \end{fmfgraph*}} }\,,
  \end{equation} 
\end{fmffile}
with the screened Coulomb interaction at zero momentum given by $V^{sc}_C(0) = - \chi_{\sigma_3 \sigma_3}^{-1}$.
The algebraic expression for the screened susceptibility,\cite{monien_theory_1990}
\begin{equation}
{\chi^{sc}_{\Ag \Ag} = \chi_{\Ag \Ag} - \frac{\chi_{\Ag \sigma_3} \chi_{\sigma_3 \Ag}}{\chi_{\sigma_3 \sigma_3}}}\,,
\end{equation}
 becomes lengthy upon Fermi-surface harmonic expansion. 
\begin{equation}
    \begin{split}
    \chi_{\Ag\Ag}
         &= \expval{(\gamma_{\Ag})^2 } 
         = \expval{(1+b_0\xi)^2(\gamma_0 + \gamma_1 \cos(4\phi) + \gamma_2 \cos(8\phi))^2}\\
        &=64\gamma_2^2\TI_5 + (32\gamma_1 \gamma_2 - 128\gamma_2^2)\TI_4  
         + (4\gamma_1^2 + 16\gamma_0\gamma_2 - 48\gamma_1\gamma_2 + 80\gamma_2^2)\TI_3 \\
         &\quad+ (4\gamma_0\gamma_1 -4\gamma_1^2 -16\gamma_0\gamma_2 + 20\gamma_1\gamma_2 - 16\gamma_2^2) \TI_2 +(\gamma_0^2 - 2\gamma_0\gamma_1 + \gamma_1^2 + 2\gamma_0\gamma_2 - 2\gamma_1\gamma_2 + \gamma_2^2)\TI_1\\
        &\quad+  (b_0^2+2b_0b_1)\Delta_0^2 \big\{ 64\gamma_2^2\MM_5 + (32\gamma_1 \gamma_2 - 128\gamma_2^2)\MM_4  
        + (4\gamma_1^2 + 16\gamma_0\gamma_2 - 48\gamma_1\gamma_2 + 80\gamma_2^2)\MM_3 \\ 
         &\quad+ (4\gamma_0\gamma_1 -4\gamma_1^2 -16\gamma_0\gamma_2 + 20\gamma_1\gamma_2 - 16\gamma_2^2) \MM_2 
        +(\gamma_0^2 - 2\gamma_0\gamma_1 + \gamma_1^2 + 2\gamma_0\gamma_2 - 2\gamma_1\gamma_2 + \gamma_2^2)\MM_1 \big\}\\
        \chi_{\Ag\sigma_3} &= 8\gamma_2 \TI_3 + (2\gamma_1-8\gamma_2)\TI_2 + (\gamma_0-\gamma_1) \TI_1 + b_0b_1\Delta_0^2 \big\{ 8\gamma_2 \MM_3 + (2\gamma_1-8\gamma_2)\MM_2 + (\gamma_0-\gamma_1) \MM_1\big\}\\
        \chi_{\sigma_3\sigma_3} &= \TI_1
    \end{split}
\end{equation}
However, all terms containing $\gamma_0$ but not $b_0$ cancel between the bare response and the screening since $\expval{\gamma_0 f(\phi)} = \frac{\expval{\gamma_0}\expval{f(\phi)}}{\expval{1}}$. Because terms linear in $\xi$ vanish in the integral due to symmetry, only terms proportional to $b_i b_j$ with $i,j\in\{0,1\}$ appear. Furthermore, the $\MM_i$ terms proportional to $b_0$ and $b_1$ are of the same order as the $\TI_i$. Since $b_i b_j \Delta_0^2\ll 1$, the terms due to energy dependence give a negligible contribution to the $\Ag$ pair breaking response. Then neglecting the terms proportional to $b_i b_j$ and simplifying the rest results in the more compact expression,
\begin{equation}
    \begin{split}
        \chi^\text{sc}_{\Ag\Ag}
        &=64\gamma_2^2\TI_5 + (32\gamma_1 \gamma_2 - 128\gamma_2^2)\TI_4  
         + (4\gamma_1^2 - 48\gamma_1\gamma_2 + 80\gamma_2^2)\TI_3 \\
         &\quad + (-4\gamma_1^2 + 20\gamma_1\gamma_2 - 16\gamma_2^2) \TI_2 
         +(\gamma_1^2  - 2\gamma_1\gamma_2 + \gamma_2^2)\TI_1\\
         &\quad - \frac{\left(8\gamma_2 \TI_3 + (2\gamma_1-8\gamma_2)\TI_2 - \gamma_1 \TI_1\right)^2}{\TI_1}\,.
    \end{split}
\end{equation}
The peak position of the pair-breaking excitation in A$_{1\textrm{g}}$ symmetry is strongly dependent on the admixture of the higher-order Fermi-surface harmonics $\gamma_1, \gamma_2$. Experimentally, it is known that the pair-breaking peak changes as a function of the incident photon energy.\cite{Budelmann2005} Changing the ratio between $\gamma_1$ and $\gamma_2$ accommodates this effect.

\subsubsection{Higgs Response}
It is well established that the lowest-order contribution to the Higgs response is given by the diagram \cite{Cea2014}\cite{puviani_current-assisted_2020}
\begin{fmffile}{BareHiggsResponse}
\fmfset{arrow_len}{3mm}
\begin{equation}
 \chi^\text{Higgs}_{\Ag \Ag} = 
  \eqgraph{1.5ex}{0ex}{\fmfframe(5,5)(5,5){
    \begin{fmfgraph*}(180,40)
    \fmfleft{i1,i2}
    \fmfright{o1,o2}
    \fmf{wiggly}{i1,v1,i2}
    \fmf{wiggly}{o1,v4,o2}
    \fmf{plain_arrow,tension=0.4,left=1,width=1}{v1,v2}
    \fmf{plain_arrow,tension=0.4,left=1,width=1}{v2,v1}
    \fmf{dbl_dashes,label=$H$,tension=0.5,l.side=left}{v2,v3}
    \fmf{plain_arrow,tension=0.4,left=1,width=1}{v3,v4}
    \fmf{plain_arrow,tension=0.4,left=1,width=1}{v4,v3}
    \fmfv{d.sh=square,d.filled=empty,d.size=7}{v1,v4}
    \fmfv{d.sh=diamond,d.filled=empty,d.size=7}{v2}
    \fmfv{d.sh=diamond,d.filled=empty,d.size=7}{v3}
    \fmfv{l=$f_\ik \sigma_1$,l.a=-60}{v2}
    \fmfv{l=$f_\ik \sigma_1$,l.a=-120}{v3}
    \fmfv{l=$\gamma_\Ag \sigma_3$}{v1,v4}
  \end{fmfgraph*}} }\,.
  \end{equation} 
\end{fmffile}
The Higgs propagator is mathematically equivalent to the attractive pairing $V$ dressed at RPA level, much like how plasmons can be understood as an RPA-dressed Coulomb interaction. Diagrammatically, it takes the form \cite{Cea2016}

\begin{fmffile}{HiggsPropagator}
\fmfset{arrow_len}{3mm}
\begin{equation}
 \eqgraph{1.5ex}{0ex}{\fmfframe(5,10)(5,10){
    \begin{fmfgraph*}(40,30)
    \fmfleft{i}
    \fmfright{o}
    \fmf{dbl_dashes,label=$H$,label.side=left}{i,o}
  \end{fmfgraph*}} }
  = \eqgraph{1.5ex}{0ex}{\fmfframe(5,10)(5,10){
    \begin{fmfgraph*}(40,30)
    \fmfleft{i}
    \fmfright{o}
    \fmf{dashes,label=$V$,label.side=left}{i,o}
  \end{fmfgraph*}} }+
  \eqgraph{1.5ex}{0ex}{\fmfframe(5,10)(5,10){
    \begin{fmfgraph*}(100,30)
    \fmfleft{i}
    \fmfright{o}
    \fmf{dashes,label=$V$,label.side=left}{i,v1}
    \fmf{plain_arrow,tension=0.4,left=1,width=1}{v1,v2}
    \fmf{plain_arrow,tension=0.4,left=1,width=1}{v2,v1}
    \fmf{dbl_dashes,label=$H$,label.side=left}{v2,o}
    \fmfv{d.sh=diamond,d.filled=empty,d.size=7}{v1,v2}
    \fmfv{l=$f_\ik \sigma_1$,l.a=-120}{v1}
    \fmfv{l=$f_\ik \sigma_1$,l.a=-60}{v2}
  \end{fmfgraph*}} }\,.
  \end{equation} 
\end{fmffile}

The homogeneous Higgs propagator, assuming an isotropic Fermi-surface and at low temperature can be expressed in terms of the integrals $\TI_1$ and $\TI_2$:
\begin{equation}
\begin{aligned}
    H^{-1}(\io,\iq=0) &= \frac{1}{V} - \chi_{\Delta \Delta} = \sum_\ik \frac{f_\ik^2(4\Delta_\ik^2-\io^2)}{2E_\ik(4E^2_\ik-\io^2)}\tanh(\beta E_\ik/2)\\
                    &\overset{T\approx 0}{\approx}\frac{2N_F}{4} \int \frac{t^4/x^2 - t^2}{\sqrt{1-t^2}\sqrt{1-t^2/x^2}} \mathcal{D}(x,t) dt \\ 
        &= \frac{N_F}{4} \left( \TI_2 - x^2\TI_1 \right)
\end{aligned}
\end{equation}
The density of states introduces an extra $(1+b_1\xi)$ factor inside all $\expval{\cdot}$. Terms odd in $\xi$ vanish under the integral, thus the Higgs propagator is not affected by the introduction of $b_1$.\\
The A$_{1\textrm{g}}$ channel of the Higgs response is also affected by Coulomb screening \cite{Cea2016}. The \emph{screened Higgs response} $\chi^{\text{Higgs, sc} }_{\Ag\Ag}$ is obtained by screening both the Higgs propagator and the light-Higgs coupling according to
\begin{equation}\label{eqn:sc_higgs}
    \begin{aligned}
        \chi^{\text{Higgs, sc}}_{\Ag\Ag} = \frac{(\chi_{{\Ag}\Delta} - {\chi_{{\Ag} \sigma_3} \chi_{\sigma_3\Delta}}/{\chi_{\sigma_3 \sigma_3}})^2}{H^{-1} - \chi_{\Delta\sigma_3}\chi_{\sigma_3 \Delta}/{\chi_{\sigma_3\sigma_3}}} = (\chi^\text{sc}_{{\Ag}\Delta})^2 H^\text{sc}\,,
    \end{aligned}
\end{equation}
with
\begin{equation}\label{eqn:chi13}
\begin{split}
        \chi_{{\Ag}\Delta} &= -\frac{1}{\Delta_0}\expval{\gamma_{\Ag} \xi} \\ 
        &= -\frac{1}{\Delta_0}\expval{(1+b_0\xi)(\gamma_0 + \gamma_1 \cos(4\phi) + \gamma_2 \cos(8\phi))\xi}\\
        &=-\Delta_0 (b_0+b_1) \bigg(  [\gamma_0-\gamma_1+\gamma_2]\MM_1 + [2\gamma_1-8\gamma_2]\MM_2 + 8\gamma_2 \MM_3 \bigg) \\ 
        \chi_{\sigma_3\Delta} &= -\frac{1}{\Delta_0}\expval{\xi} = -b_1\Delta_0\MM_1\,.
\end{split}
\end{equation}
As can be seen above the light-Higgs coupling, $\chi_{{\Ag}\Delta}$ scales with the sum $b_0+b_1$. However, unlike $b_0$ which is a property only of the Raman vertex, $b_1$ measures the deviation from a constant density of states and thus the breaking of particle-hole symmetry. Charge fluctuations will couple to the pairing channel, allowing screening of the Higgs mode by the background charges. Notice that in \eqref{eqn:chi13} the coupling of the Higgs mode to charge fluctuations $\chi_{\sigma_3\Delta}$, which leads to Coulomb screening, scales only with $b_1$. We introduce a parameter $r = b_1/(b_0+b_1)$, controlling the ratio between Coulomb screening and light coupling of the Higgs mode. Increasing the parameter $r$ shifts the peak of the full Higgs response towards smaller energies.

\subsubsection{Determining $\gamma$ and $b$ parameters}

In the above approximation, information of the band structure $\xi_\ik$ is encoded in the Raman vertices $\gamma_{\Ag}, \gamma_{\Bg}$ and the parameters $b_0,b_1$. We assume that $b_0$ and $b_1$ are small: $\Delta_0 b_0, \Delta_0 b_1 \ll 1$; therefore $\Ag$ has neglegible dependence on $b_0, b_1$. Since the $\TI_n(x)$ are only functions of frequency $x = \omega/2\Delta_0 + i\eta$, the susceptibilities $\chi_{\Ag\Ag},\chi_{\Bg\Bg}$ are functions of: $\chi_{\Ag\Ag}(\gamma_0,\gamma_1,\gamma_2, \eta,\Delta_0,\omega)$, $\chi_{\Bg\Bg}(\gamma_b,\eta,\Delta_0,\omega)$. 

As a first step of fitting, we extract the parameters $(\gamma_b, \eta, \Delta_0)$ from the $\Bg$ signal (see Fig. 4a). We use a  Trust Region Reflective (trf) algorithm (scipy.optimize.
least\_squares) to fit the data. Electronic lifetime and the superconducting order parameter are characterized by $\eta$ and $\Delta_0$, which are expected to be shared amongst all response functions. 

With the knowledge of $\eta$ and $\Delta_0$ we proceed to fit the $\Ag$ pair-breaking response to acquire $(\gamma_1,\gamma_2)$ (see Fig. 4b). No information of $\gamma_0$ can be obtained due to the complete screening of the isotropic $\Ag$ component.\cite{Devereaux1995} 

We have known parameters $(\eta,\Delta_0,\gamma_1,\gamma_2)$ and free parameters $(b_0,b_1,\gamma_0)$ in fitting the Higgs response (see Fig. 4c). These unknown values control the magnitude and lineshape of the Higgs Raman response function. It is well known that the diamagnetic response of the Higgs mode is small due to the approximate particle-hole symmetry near the Fermi surface. In a non-equilibrium experiment, however, redistribution of particles in response to an external optical pump will influence the state occupation, and thus the Raman signal magnitude. After 3 ps, the electronic degrees of freedom have relaxed to a quasi-equilibrium state, as can be seen by the establishment of thermal distribution on both Stokes and anti-Stokes sides except for the Higgs response. We therefore assume that the susceptibilities in the pumped states are well approximated by their equilibrium counterparts. In our equilibrium-calculation carried out in the clean limit we find that the Higgs susceptibility is about 3 orders of magnitudes weaker compared to the PB peak. This is due to the weak breaking of particle-hole symmetry in our model, which assumes a weak linear deviation in the Raman vertex and the energy dependence of the density of states.


\subsection{S.3 Population Inversion}\label{sec:PopulationInv}
The Higgs modes are the lowest-energy collective excitations of the superconductor in the Cooper channel. In addition, the Higgs modes are metastable. This leads to a three-level picture of population inversion as schematically shown in Fig. \ref{fig:S9}, motivating the NEARS experiment and outlining a basic understanding of the NEARS mechanism.

The rate equation for a three-level system \cite{shimoda2013} is given as

\begin{equation}
    \frac{(N_2 - N_1)}{N}=\frac{(1-\frac{\tau_{32}}{\tau_{21}})\cdot W_p \tau_{21} - 1}{(1-\frac{\tau_{32}}{\tau_{21}})\cdot W_p \tau_{21} + 1}
\end{equation}

with N representing the total number of particles, $N_1$ being the number of particles in the ground state, and $N_2$ the measure for population of the metastable energy state (here the Higgs state). Via pumping, state 3 will be populated (Quench of the Mexican hat), with a short lifetime $\tau_{32}$, subsequently populating the state 2 with a longer lifetime $\tau_{21}>\tau_{32}$. Assuming $\tau_{32}$ is much smaller than $\tau_{21}$ leads to $\frac{\tau_{32}}{\tau_{21}} \rightarrow 0$. $W_p$ represents a measure of the strength of the pump, or $W_p \tau_{21} \propto F$ with $F$ as pump fluence. 

Population inversion occurs if $N_2 > N_1$, and $\frac{(N_2 - N_1)}{N} > 0$. Therefore, the anti-Stokes intensity of the metastable Higgs excitation, which corresponds to the anti-Stokes NEARS difference intensity shown in Fig. 2 of the main text, scales with the population ratio following

\begin{equation}
\label{popinv}
    I_{AS}    
    \begin{cases}
    \propto \frac{(N_1 - N_2)}{N}=\frac{F/F_{\textrm{crit}} - 1}{F/F_{\textrm{crit}} + 1} & \text{for } F > F_{\textrm{crit}} \rightarrow N_2 > N_1\\
    = 0 & \text{for } F < F_{\textrm{crit}} \rightarrow  N_2 < N_1 \ .
   \end{cases}
\end{equation}

In fact, the integrated NEARS difference intensities on the anti-Stokes side (see Fig. 2c, d, and e) as a function of fluence are in agreement with eq. \ref{popinv} as shown in Fig. \ref{fig:S9}b and Fig. 2e. $F_{\textrm{crit}}$ is the critical fluence at which population inversion is initiated. Fitting eq. \ref{popinv} to the experimental NEARS intensity integrals (see Fig. 2e main text and Fig. S9) results in $F_{\textrm{crit}}$ of $21.7 \pm 5.3$~µJ~cm$^{-2}$ for A$_{1\textrm{g}}$ and $31.6 \pm 2.3$~µJ~cm$^{-2}$ for B$_{1\textrm{g}}$ symmetry.


\subsection{S.4 Higgs Response in Ginzburg-Landau Theory}

We describe the charged bosonic condensate using a Klein-Gordon like Lagrangian.\cite{Greiter2005, Pekker2015} The potential can be calculated with a Ginzburg-Landau Mexican-hat potential $F(\Psi) = \alpha |\Psi|^2 + \frac{\beta}{2} |\Psi|^4$ ($\alpha < 0$). For small fluctuations of the Higgs amplitude ($H$) around the ground state $|\Psi_0| = \sqrt{\frac{-\alpha}{\beta}}$ the Lagrangian can be written as \cite{puviani_current-assisted_2020}

\begin{eqnarray}\label{Klein_Gordon_extended}
\mathcal{L}^{KG} = (\partial_{\mu}H) (\partial^{\mu}H)  + 2 \alpha H^2 - \frac{1}{4} F_{\mu\nu}F^{\mu\nu} + q^2 \Psi_0^2 A_{\mu} A^{\mu} + 2 q^2 \Psi_0 A_{\mu} A^{\mu}  H \ ,
\end{eqnarray}

with the Cooper-pair charge $q = -2e$ and the electromagnetic field tensor $F_{\mu\nu}=\partial_{\mu}A_{\nu}-\partial_{\nu}A_{\mu}$. Due to the Anderson-Higgs mechanism, phase fluctuations ($\Theta$) are no longer present in this expression.\cite{Anderson1958, Anderson1963} Instead, the low energy excitation spectrum in the condensate is only characterized by the Higgs mode. The photon field acquires a mass term leading to the Meißner effect when evaluating the Euler-Lagrange equations for the vector potential.\cite{Greiter2005} Furthermore, there is no linear coupling of the Higgs amplitude to a vector potential. The Higgs mode is Raman active and couples quadratically to the vector potential. We can calculate the equation of motion for the Higgs mode by using the Euler-Lagrange equations and we find in the \textit{optical} $q \rightarrow 0$ limit

 \begin{equation}\label{Higgs_Equation}
	 \left(\frac{d^2}{dt^2} + 2 \abs{\alpha} \right) H(t) =  e^2 |{\Psi}_0| A^2,
\end{equation}

where we have set the electrostatic potential to 0.

In order to account for the non-equilibrium experimental conditions, we can quench the order parameter within the Ginzburg-Landau theory by quenching $\alpha$, $\beta$, or both of them, since $\Psi_0$ itself depends on the ratio of $\sqrt{\frac{|\alpha|}{\beta}}$. If we quench $\alpha$ we will change the frequency of the Higgs mode to lower energy and reduce the coherence length $\xi= \sqrt{\hbar^2/({|\alpha|4m^*})}$ of the Cooper pairs, which is not observed in the NEARS experiment as shown in Fig. 2f. If we quench $\beta$, we will reduce the superfluid density and not change the frequency of the Higgs mode. This case fits our experimental observations and has a particularly simple solution. We can calculate the Green's function in the ${\bf q} \rightarrow 0$ case by assuming a $\delta$ function-like quench due to a change in $\beta$. We then obtain the following Green's function:

\begin{equation}
\label{eq:Greens}
	G_H(\omega) = \frac{1}{\omega^2-2\abs{\alpha} +i \gamma \omega} =  \frac{- i\gamma \omega}{(\omega^2-2\abs{\alpha})^2+(\gamma \omega)^2} + \frac{(2\abs{\alpha}-\omega^2)}{(\omega^2-2\abs{\alpha})^2+(\gamma \omega)^2} \ ,
\end{equation}

where we have added a phenomenological damping $\gamma$. This is identical to the Green's function of a harmonic oscillator with an eigenfrequency $\omega_0^2 = 2 \abs{\alpha}$.

It has been shown that in non-equilibrium the response function determining the Raman intensity $I(\omega)$ is proportional to the spectral function, i.e. the imaginary part of the Green's function (eq. \ref{eq:Greens}).\cite{Freericks2021, Deveraux2018}
Accordingly, in the proximity of the Higgs-mode energy we approximate the Raman intensity as

\begin{equation}
\label{eq:GL_Higgs_res}
	I(\omega) = I_0 \frac{\gamma \omega}{(\omega^2-2\abs{\alpha})^2+(\gamma \omega)^2},
\end{equation}

where $I_0$ summarizes proportionalities to statistical factors, the strength of the quench, the excited state fraction of the experiment.
 
This is effectively a Lorentzian line shape with a strength of the response being proportional to the strength of the quench. In Ginzburg-Landau theory there is a direct connection
between the energy of the Higgs mode encoded in $\alpha$ and the coherence length of a superconductor $\xi= \sqrt{\hbar^2/({|\alpha|4m^*})}$ as well as the penetration depth $\lambda= \sqrt{{m^* c^2}/(4\pi e^2 \Psi_0^2)}$.\cite{Cyrot1973} A quench of $\Psi$ by $\beta$ changes the penetration depth but leaves the coherence length constant, whereas the other two scenarios change both coherence length and penetration depth. In a strong coupling superconductor with a short coherence length we can expect our results to be close to a $\beta$ quench, i.e. changing the suprafluid density alone. In more traditional weak coupling superconductors we would expect a change in position of the Higgs mode as $\alpha$ gets quenched. In mean field weak coupling approximation $\alpha$ is tied to $2\Delta$.

We fit the experimentally derived Higgs mode (see Fig. 2 main text) using eq. \ref{eq:GL_Higgs_res}. The result is presented in Fig. 2c and d. In order to achieve consistent units in our experimental fit, we apply $\omega$, $\gamma$, and $\alpha$ in units of $2 \Delta_0$, as determined by fitting the B$_{1\textrm{g}}$ pair-breaking peak (see \ref{sec:B1g}, and Fig. 4 a). With this, the fit of the A$_{1\textrm{g}}$ Higgs data at a fluence of 113 µJ cm$^{-2}$ results in $2\abs{\alpha} = (0.165 \pm 0.004) \ 2\Delta_0 = 10.07 \pm 0.24$ meV. This corresponds to an energy of the Higgs mode of $\omega_H = \sqrt{2\abs{\alpha}} = 0.41\cdot 2\Delta_0 = 25$ meV. Using established values m*/me for optimally-doped cuprates of the order of $m^* = 10 \ m_e$,\cite{Legros2019} we obtain in-plane coherence lengths of smaller than 5 nm.


\subsection{S.5 Equilibrium Temperature Calculation}\label{sec:temperature}
Raman spectroscopy enables the determination of an effective equilibrium sample temperature by the linking of Stokes and anti-Stokes spectra via the temperature-dependent Bose-function and thus makes a subtraction of laser-heating contributions from the original data possible. The effective temperature was identified using an algorithm based on the following equation:\cite{Bock1995}

\begin{equation}
\label{AS_S}
\frac{I_\textrm{AS}(\Delta E)}{I_\textrm{S} (\Delta E)} = \left(\frac{E_l+\Delta E}{E_l-\Delta E}\right)^4 \textrm{exp}\left(-\frac{\Delta E}{k_\textrm{B} T}\right)
\end{equation}

$I_\textrm{AS}$ and $I_\textrm{S}$ describe the measured anti-Stokes and Stokes Raman intensities as functions of Raman shift, $E_l$ refers to the photon energy of the incident probe laser (3.09 eV in this study), $\Delta E$ is the Raman shift, $k_\textrm{B}$ describes the Boltzmann constant, and $T$ is the effective equilibrium temperature. Fig. \ref{fig:S8} (a) shows the measured Stokes data (blue) and anti-Stokes data (red) together with the Stokes data mirrored to the anti-Stokes side by equation \ref{AS_S} (black). The base temperature for this measurement was 293 K (room temperature), and an effective temperature of 302 K had to be considered to achieve agreement between Stokes and anti-Stokes spectra. This corresponds to a heating of 9 $\pm$ 0.25 K for a probe laser power of 4.8 mW. Fig. \ref{fig:S8} (b) shows the difference of Stokes and anti-Stokes integral as a function of heating for the exemplary measurement shown in (a) and defines the best estimate of the effective sample temperature as the minimum of the difference. To do this, we evaluate the integral difference in the energy region around two dominant phonons, as marked in Fig. \ref{fig:S8} (a). To value the higher-energy integral with respect to the drastic reduction in anti-Stokes intensity as a function of energy due to the Bose function $n(\omega, T)$, we weight the higher-energy integral with the ratio $n(\omega = 14.4\,\textrm{meV}, T)/n(\omega = 57.5\,\textrm{meV}, T)$ (see Fig. \ref{fig:S8}). For the heating caused by the pump, the fluence dependent data set was used to define a heating rate (K/mW). As shown in Fig. \ref{fig:S8} (c), we find a heating rate of 3 $\pm$ 0.5 K/mW for the pump. This leads to errors between 2.5 K and 13.75 K for the applied pump fluences (5 to 27 mW). The highest pump fluence at 113 µJ cm$^{-2}$ (27 mW), therefore, corresponds to an effective equilibrium temperature of 98 $\pm$ 13.75 K, which equals T$_\textrm{C}$ within its error bar. We validate this result in the superconducting state by analyzing the strength of the superconductivity-induced features of the Stokes spectra as a function of fluence. Fig. \ref{fig:S12} shows the ratios of Stokes Raman spectra at 8 K base temperature divided by the 100 K probe-only data. As a function of fluence, the pair-breaking feature gets suppressed and gap-filling occurs. To obtain a measure for the system's state, we evaluate the integral of the absolute value of these ratios compared to the normal state, in other words $\int{\abs{\textrm{8K data / 100 K probe data}-1}}$. This is shown in the inset of Fig. \ref{fig:S12}. The 100 K baseline intersects with a linear regression fit at a fluence of $139.3 \pm 28.5$ µJ cm$^{-2}$. 

In summary, from both methods we conclude that (I) the sample pumped with a fluence of 113~µJ~cm$^{-2}$ has an effective equilibrium temperature of 98 $\pm$ 13.75~K (see Table \ref{tab:tabS1}), and (II) $T_C$ is reached at a fluence of $139.3 \pm 28.5$~µJ~cm$^{-2}$. Both error bars overlap. Since we observe a clear pair-breaking feature at our highest pump fluence (see Fig. \ref{fig:S12} and Fig. \ref{fig:S6}), we conclude that the sample remains in its superconducting state during our measurements.

\subsection{S.6 Data Parameterization}\label{sec:parameterization}
Stokes Raman intensities are described by the following fit function:

\begin{equation}
\label{stokesfitting}
I_\textrm{S}=(n(\omega,T)+1)\cdot \left[ y_0\cdot \textrm{tanh}\left(\frac{\omega}{\omega_\textrm{c}}\right)+\frac{A_{2\Delta} \, \omega \, \tilde{\Gamma}_{2\Delta}}{(\omega^2-\omega_{0,2\Delta}^2)^2+\tilde{\Gamma}_{2\Delta}^2 \omega^2} + \sum_m \frac{A_{m} \, \omega \, \tilde{\Gamma}_{m}}{(\omega^2-\omega_{0,m} ^2)^2+\tilde{\Gamma}_{m}^2 \omega^2}  \right]
\end{equation}

where $(n(\omega,T)+1)$ represents the thermal population factor (Bose-function, see Fig. \ref{fig:S4}):

\begin{equation}
n(\omega,T) + 1 = \frac{1}{\textrm{exp}\left(\frac{\hbar \omega}{k_\textrm{B} T} \right) - 1}+1 = \frac{1}{1-\textrm{exp}\left(-\frac{\hbar \omega}{k_{\textrm{B}}T} \right)}
\end{equation}

In the superconducting state, the gap feature is parameterized by a tanh-function $y_0\cdot \textrm{tanh}\left(\frac{\omega}{\omega_\textrm{c}}\right)$ with the critical frequency $\omega_\textrm{c}$.\cite{Budelmann2005} According to \cite{Budelmann2005}, the pair-breaking peak is described by a Lorentzian, with $A_{2\Delta}$ being the amplitude, $\tilde{\Gamma}_{2\Delta}$ the damping and $\omega_{0,2\Delta}$ corresponding to the gap energy. This parameterization of the pair-breaking feature (tanh + Lorentzian) represents a model-independent analysis of the electronic susceptibility. The quality of the electronic features derived from our fit model and parameterized by a tanh-function together with Lorentzian can be evaluated by comparison with a microscopic theory as shown in Fig. 4 (main text). The electronic responses are extracted from the data by subtracting the phonons as determined by the fit (eq. \ref{stokesfitting}, see also Fig. \ref{fig:S10}). In fact, the parameterization results show perfect agreement with the respective theory calculations. The phonons ($m=9$) are fitted as Lorentzians, where $A_{m}$ denotes the amplitude, $\tilde{\Gamma}_{m}$ is the phonon damping, and $\omega_{0,m}$ is the phonon frequency of the $m^{\textrm{th}}$ phonon.

To parameterize the anti-Stokes spectra, phonon frequencies and widths, background, and pair-breaking feature are kept constant between Stokes and anti-Stokes fits. The Bose-function changes from $(n(\omega,T)+1)$ to $n(\omega,T)$ (see Fig. \ref{fig:S4}). Phonon amplitudes are allowed to adapt to the measured anti-Stokes phonon intensities.


\subsection{S.7 Higgs Particle and Other SC Excitations} 

We now discuss our results in the context of the Higgs mode and how it can be discriminated from alternative excitations of the superconductor. The important candidates are the pair-breaking (quasiparticle) excitations, the Josephson plasmon modes, and the Bardasis-Schrieffer modes. \emph{Pair-breaking excitation:} This is the dominant feature on the Stokes side of the spectra. It occurs due to the breaking of Cooper-pairs and generates quasiparticles in the single-particle channel, i.e. holes in Bi-2212 (see Fig. 1c). It is symmetry-dependent due to the inherent Coulomb screening, which is fully symmetric and affects only the A$_{1\textrm{g}}$ channel, reducing the A$_{1\textrm{g}}$ susceptibility and shifting the excitation to lower energies compared to B$_{1\textrm{g}}$ symmetry (see Fig. 4).
In S.2 and the main text (Fig. 4), we discuss in detail the pair-breaking response which is measured by NEARS and calculated within the framework of a microscopic theory. \emph{Excitations of in-plane Josephson plasmon modes:} They exhibit A$_{1\textrm{g}}$ and B$_{1\textrm{g}}$ symmetry.\cite{Gabriele2021} However, we would expect the Josephson plasmon to exhibit an excitation energy that shifts to zero frequency as we increase the pump fluence.\cite{Gabriele2021} Even our strongly-pumped A$_{1\textrm{g}}$ data do not show this behavior. Therefore, they can be ruled out as well. \emph{Bardasis-Schrieffer mode:} This mode represents a subdominant pairing channel that could be activated by pumping the SC state. Its excitation energy would be below the binding energy $2\Delta$ of the dominant pairing channel and this mode would also not be subject to screening. Hence, these two arguments do not rule out this mode. However, the Bardasis-Schrieffer mode would leave a fingerprint on the Stokes side, as observed in pnictides.\cite{Scalapino2009, Bardasis1961, Chubukov2009} We see no indication of this mode. Furthermore, it would be considered in our analysis, which uses the Stokes channel to calculate the anti-Stokes intensity. In order to explain the observation of a B$_{1\textrm{g}}$ excitation energy below the A$_{1\textrm{g}}$ excitation energy, a highly unconventional order parameter of the subdominant pairing mechanism would be required. This leaves the Higgs particle as the best explanation for the NEARS feature.

\begin{table}
	\caption{\label{tab:tabS1}Laser power measured by PowerMax USB PS19 (Coherent, see power meter in Fig. \ref{fig:S1}), spot size at sample position measured by a DFK 23GM021 industrial camera (The Imaging Source) with a pixel size of 3.75 µm, repetition rate of the used Tsunami system with a pulse width of FWHM = 1.2 $\pm$ 0.1 ps, corresponding fluence for probe and pump, and effective equilibrium temperature following Fig. \ref{fig:S8}}.
\begin{tabular}{ |p{1.1cm}|p{2.4cm}|p{2.8cm}|p{1.8cm}|p{3.2cm}|p{2.4cm}|} 
	\hline
	 & \bf{Laser Power} & \bf{Spot Area} & \bf{Rep. Rate} & \bf{Fluence} & \bf{Effective Temp.}\\
	\hline
	\bf{Probe} & 4.75 $\pm$ 0.15 mW  & 169.23 $\pm$ 10.88 µm$^2$ & 80 MHz & 35.08 $\pm$ 1.15 µJ cm$^{-2}$ & 17 $\pm$ 0.25 K \\ 
	\hline
	 & 5.0 $\pm$ 0.1 mW & 298.65 $\pm$ 19.19 µm$^2$ & 80 MHz & 20.97 $\pm$ 0.93 µJ cm$^{-2}$ & 32 $\pm$ 2.75 K\\
	 & 10.0 $\pm$ 0.1 mW & 298.65 $\pm$ 19.19 µm$^2$ & 80 MHz & 41.95 $\pm$ 2.27 µJ cm$^{-2}$ & 47 $\pm$ 5.25 K\\ 
	\bf{Pump} & 14.0 $\pm$ 0.1 mW & 298.65 $\pm$ 19.19 µm$^2$ & 80 MHz & 58.72 $\pm$ 3.35 µJ cm$^{-2}$ & 59 $\pm$ 7.25 K \\
	 & 18.0 $\pm$ 0.1 mW & 298.65 $\pm$ 19.19 µm$^2$ & 80 MHz & 75.50 $\pm$ 4.43 µJ cm$^{-2}$ & 71 $\pm$ 9.25 K\\
	 & 27.0 $\pm$ 0.1 mW & 298.65 $\pm$ 19.19 µm$^2$ & 80 MHz & 113.25 $\pm$ 6.86 µJ cm$^{-2}$ & 98 $\pm$ 13.75 K\\
	\hline
\end{tabular}
\end{table}

\begin{figure}[h]
        \centering
	\includegraphics[width=\linewidth
	]{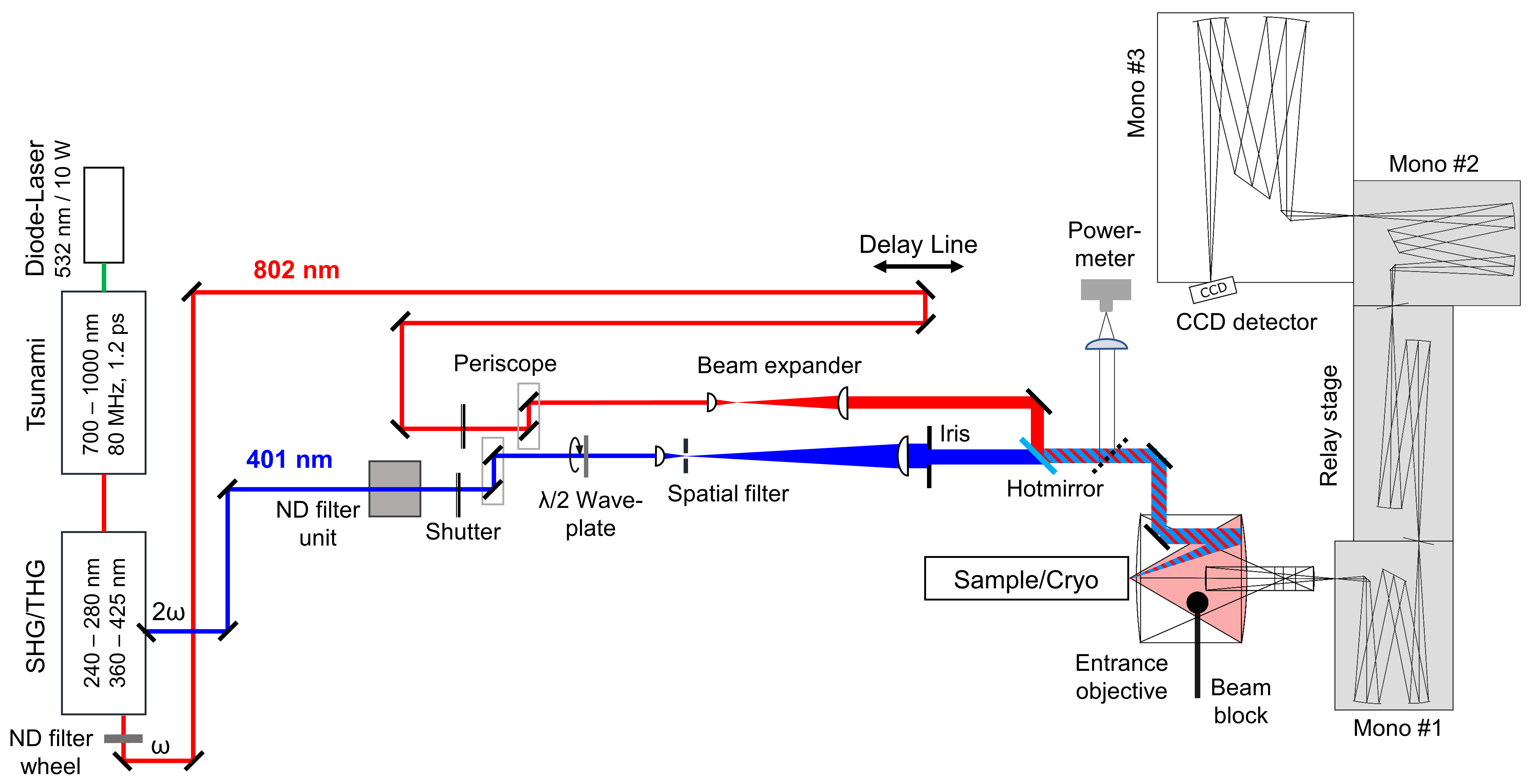}
	\caption{\label{fig:S1}Schematic view on the Raman setup. A Tsunami Ti:Sapphir system with a pulse duration of 1.2~ps and a repetition rate of 80~MHz is used as laser source at a fundamental wavelength of 802~nm (pump). From a second harmonic generation (SHG) unit, a 402~nm beam is used as the probe. The beam path includes ND filter units, shutters, a $\lambda /2$ waveplate to change the linear polarization of incidence light, a beam expander and spatial filter, a delay line for temporal overlap and scanning, and a hot mirror to overlay both beams. The entrance objective with its large numerical aperture of 0.5 is used to focus the light on the sample, and couples the Raman light into the first monochromator of the UT-3 spectrometer.\cite{Schulz2005} A beam block is used to block the specular reflex.}
\end{figure}

\begin{figure}[h]
        \centering
	\includegraphics[width=0.65\linewidth
	]{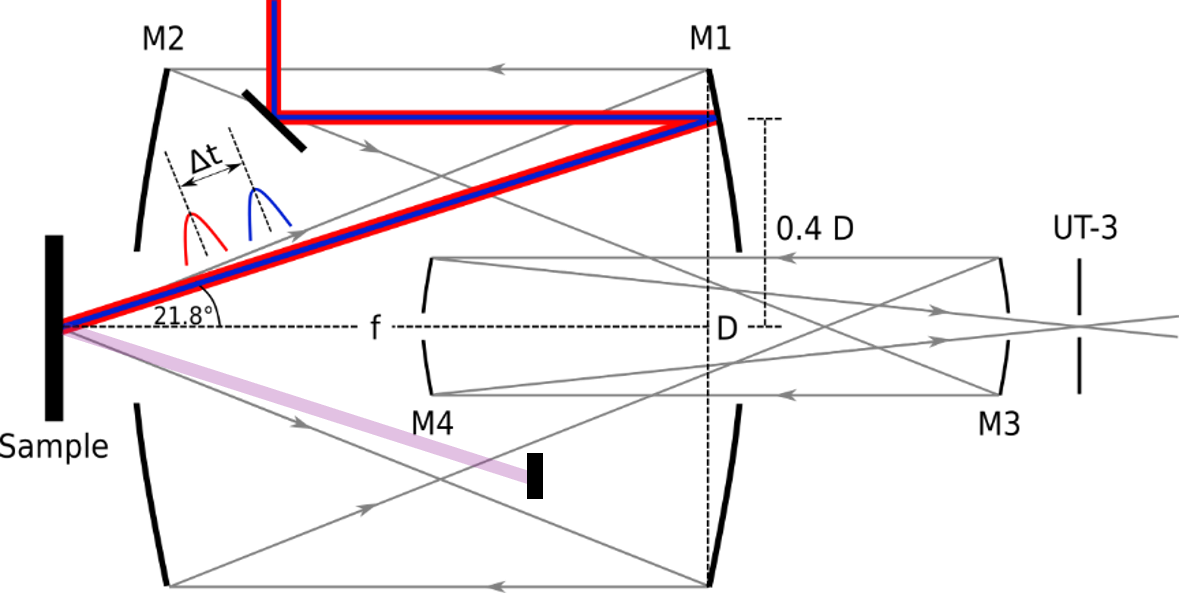}
	\caption{\label{fig:S2}Detailed view on the entrance objective of the UT-3. The Raman signal is collected with M1, collimated onto M2, which focuses the light in front of M3. This mirror collimates the light to provide a parallel beam section for the insertion of an analyzing beam cube. Pump and probe beam hit the sample with an angle of 21.8$^\circ$ to the vertical. By this, we apply a finite in-plane momentum to the sample, which causes symmetry breaking and activation of the Higgs mode in B$_{1\textrm{g}}$ symmetry. A beam dump blocks the reflected light in 21.8$^\circ$.}
\end{figure}

\begin{figure}[h]
        \centering
	\includegraphics[width=0.55\linewidth
	]{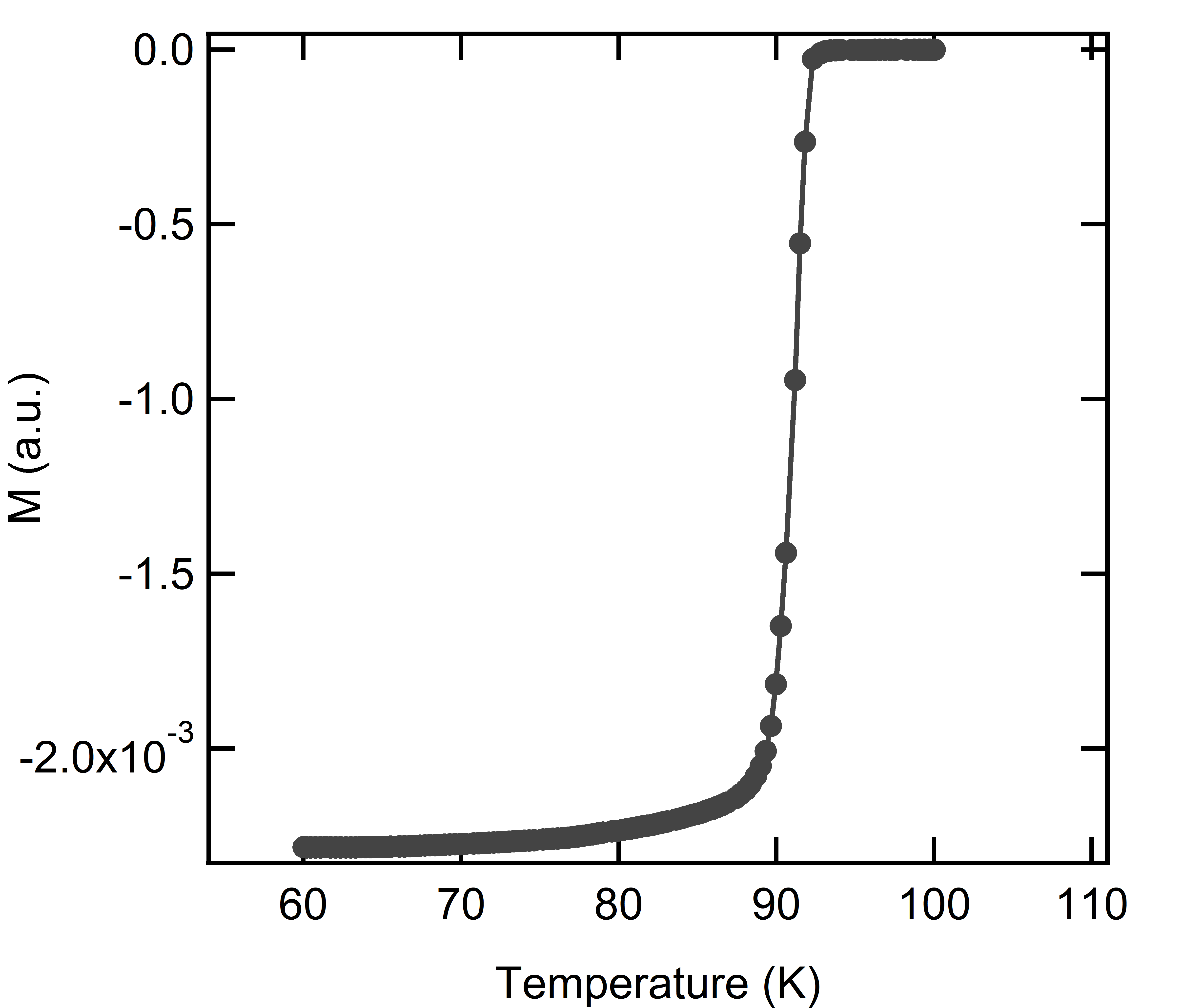}
	\caption{\label{fig:S3}Susceptibility curves of the measured Bi-2212 crystal with a T$_\textrm{c}$ of 92 K.}
\end{figure}

\begin{figure}[h]
        \centering
	\includegraphics[width=0.8\linewidth
	]{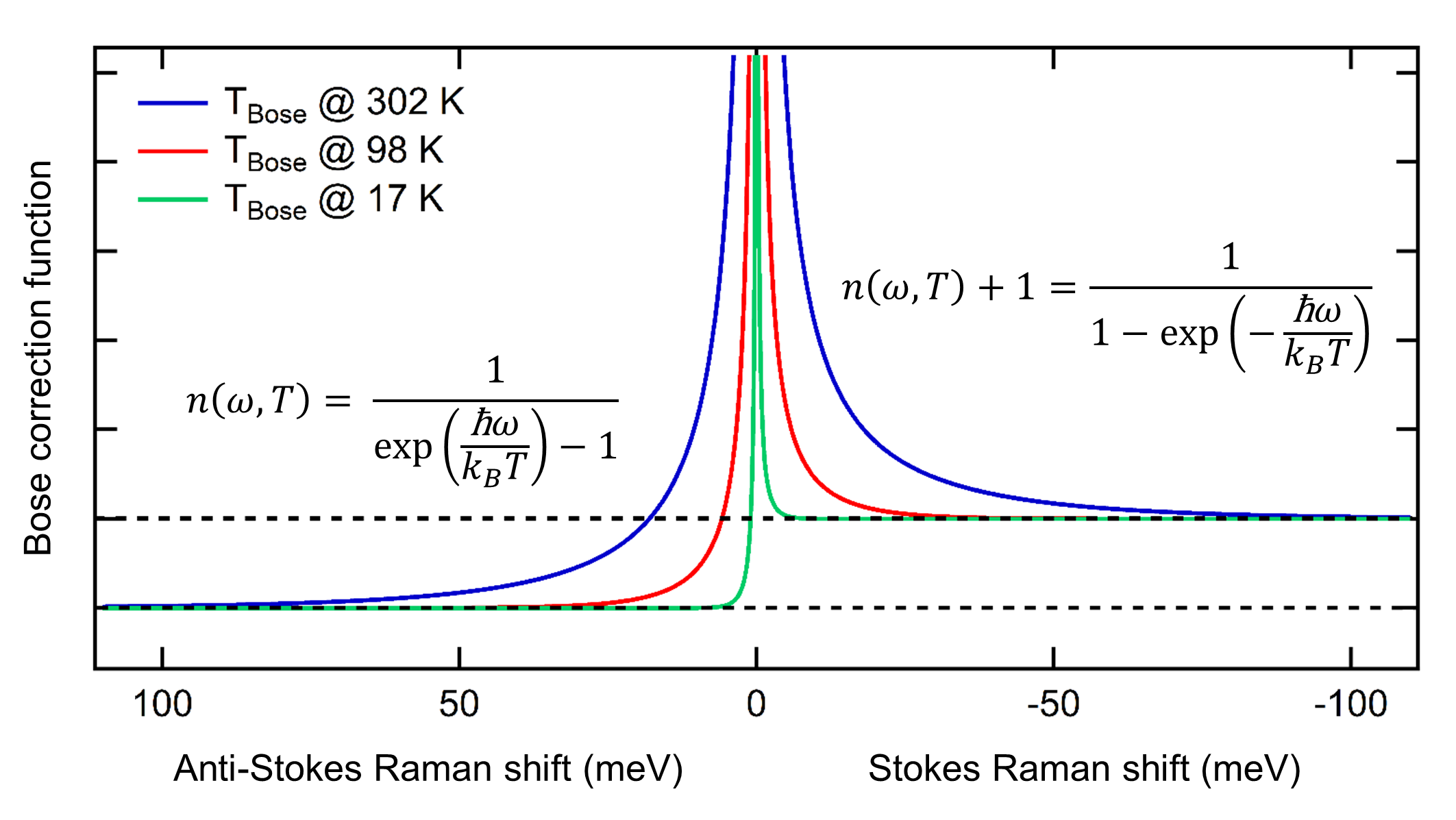}
	\caption{\label{fig:S4}Bose function for anti-Stokes Raman scattering $(n(\omega,T))$ und Stokes Raman scattering $(n(\omega,T)+1)$ for three exemplary temperatures. On the Stokes side, the Bose function converges to 1 for high energies, resulting in a non-zero Raman intensity. In comparison, on the anti-Stokes side, the Raman intensity aspires to become zero. Depending on the thermal population, an experimental observation window at low Raman shifts exists.}
\end{figure}

\begin{figure}[h]
        \centering
	\includegraphics[width=0.65\linewidth
	]{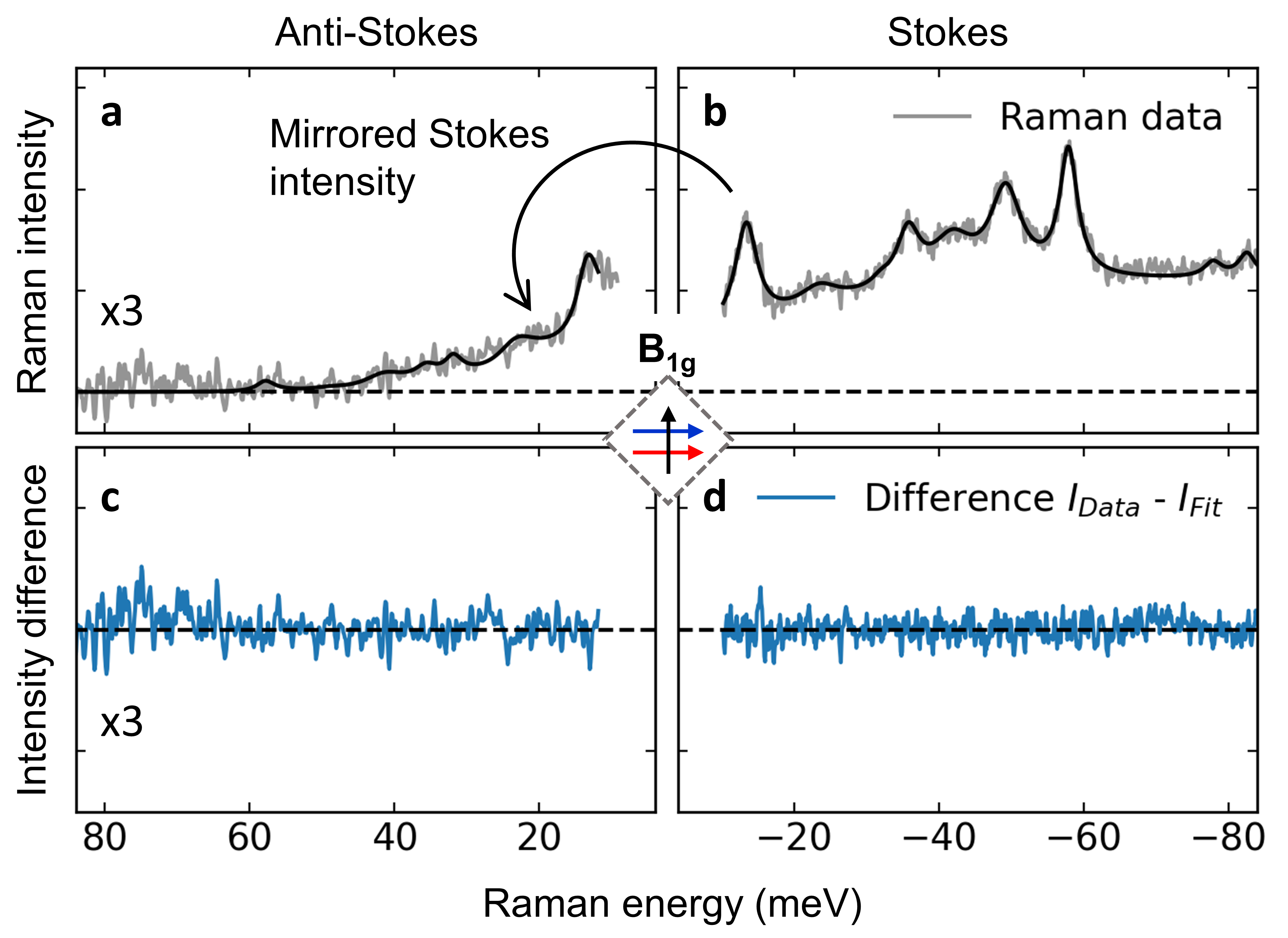}
	\caption{\label{fig:S5}Bi-2212 anti-Stokes (a) and Stokes (b) probe-only Raman intensity for B$_{1\textrm{g}}$ geometry as a function of Raman shift at 100 K base temperature. The data is shown in gray. The black solid line represents a fit to the data, for which the Stokes Raman intensity (b) has been parameterized as described in section S.4. The Stokes intensity fit was then applied to the anti-Stokes side by using the anti-Stokes Bose function $n(\omega,T)$ instead of $n(\omega,T)+1$. A temperature $T = 109$ K was used (100 K base temperature and 9 K probe heating, see Fig. \ref{fig:S8}). Phonon widths and frequencies were kept constant after fitting to the Stokes side. (c) and (d) show the difference between the data (gray) and the fit (black) shown in (a) and (b). The dashed line marks zero.}
\end{figure}

\begin{figure}[h]
        \centering
	\includegraphics[width=0.85\linewidth
	]{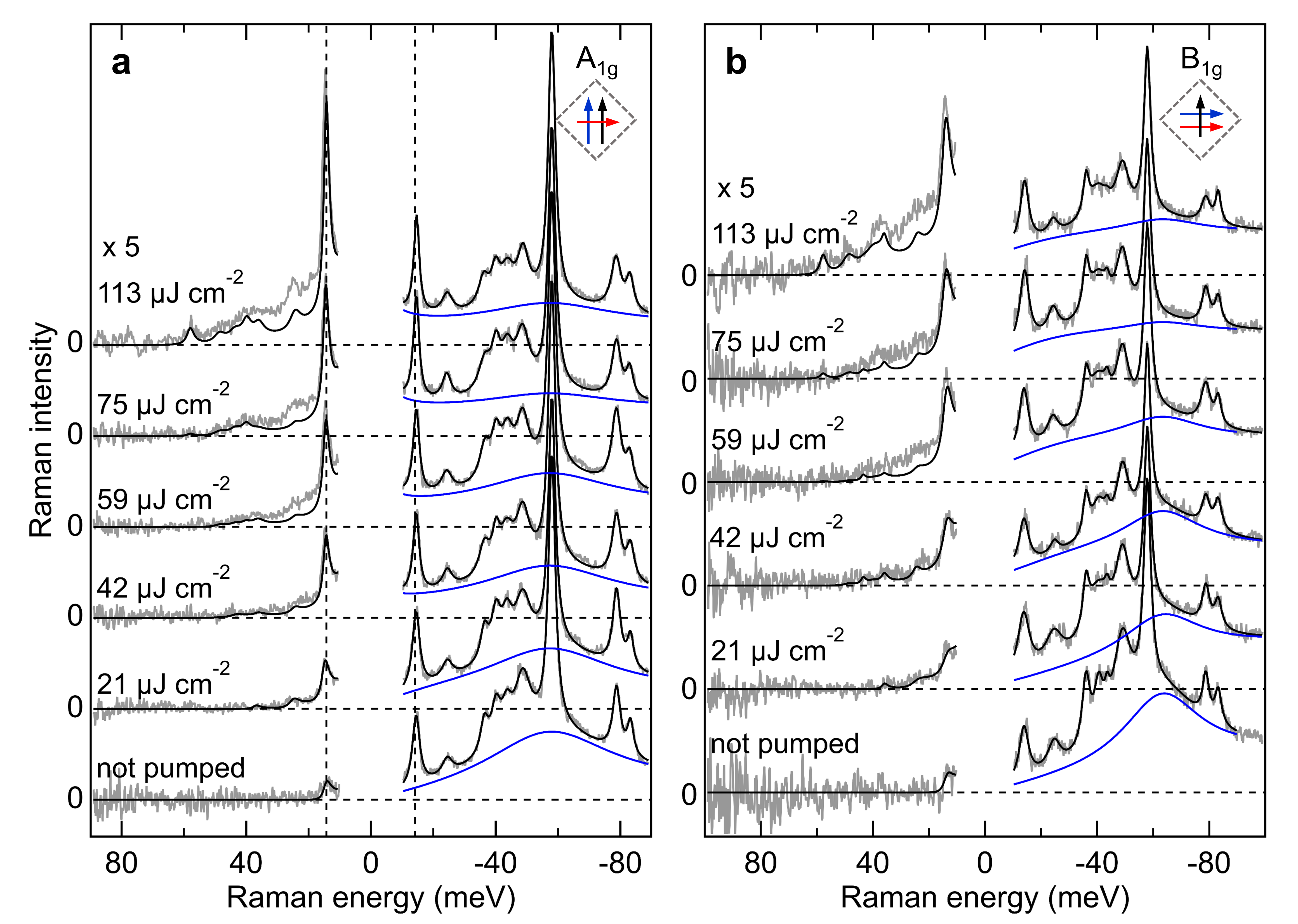}
	\caption{\label{fig:S6}Complete data sets corresponding to Fig. 2 of the main text. (a) A$_{1\textrm{g}}$ Pump-probe Stokes and anti-Stokes Raman spectra at 8~K base temperature and a time delay of 3~ps for fluences between 0~µJ~cm$^{-2}$ (not pumped) and 113~µJ~cm$^{-2}$. Anti-Stokes Raman intensities are multiplied with a factor of 5 for better visibility. Data is shown in gray and black solid lines represent fits to the Stokes Raman intensities. The dashed horizontal lines mark the zero intensity for the data sets at different fluences. The PB peak on the Stokes side is highlighted in blue. At fluences larger than 50~µJ~cm$^{-2}$ the soft quench overpopulates the Higgs excitation and additional spectral weight is detected on the anti-Stokes side. (b) Analogous data set for B$_{1\textrm{g}}$ configuration.}
\end{figure}

\begin{figure}[h]
        \centering
	\includegraphics[width=0.6\linewidth
	]{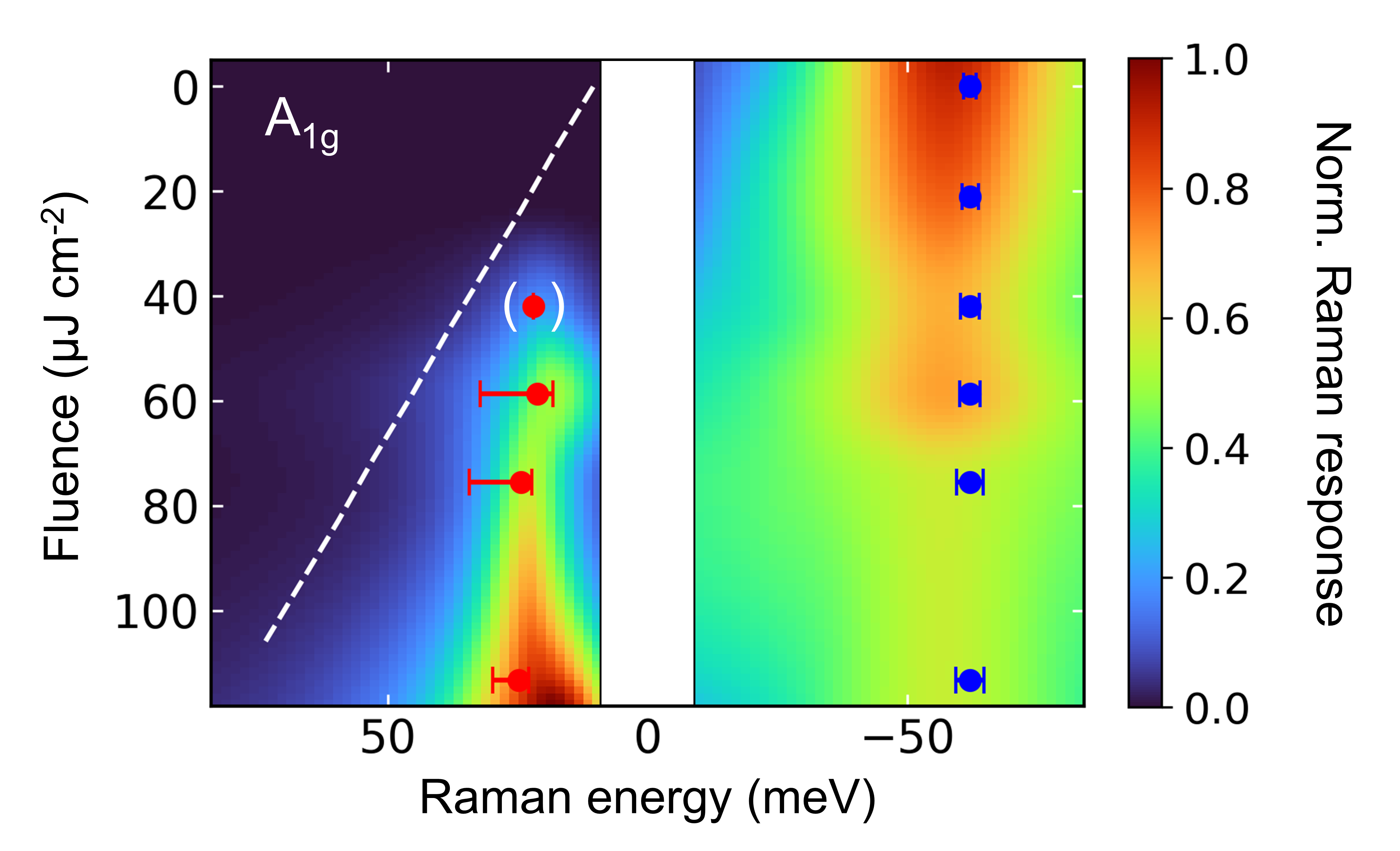}
	\caption{\label{fig:S7}Interpolated 2D color plot of the superconductivity-induced Raman susceptibilities obtained from the fit to the NEARS data in A$_{1\textrm{g}}$ geometry. In the right (energy-loss) side, the pair-breaking feature from the Stokes data is shown (tanh-function plus Lorentzian as extracted from our fit to the Raman intensities). It decreases with increasing fluence. The blue data points represent the fit results (see Fig. 2 main text). On the left (energy-gain) side, one can observe the NEARS feature, representing the difference signal visible on the anti-Stokes side compared to the Stokes data. This difference signal can be identified, since Stokes and anti-Stokes data are linked via the Bose-function and, therefore, the Stokes data can be mirrored to the anti-Stokes side. The white dashed line represents the cutoff of the thermal observation window on the anti-Stokes side at the derived equilibrium temperatures. In correspondence with our instrumental resolution, we determine this cutoff to be at a critical energy as a function of effective sample temperature, where the Bose-function $n(\omega,T)$ brings the Raman intensity down to 10 \%. With this, we derive the NEARS map (see Fig. 3 main text) as a superposition of the two features as a function of absolute value of the Raman energy, corresponding to the excitation energy.}
\end{figure}

\begin{figure}[h]
        \centering
	\includegraphics[width=0.99\linewidth
	]{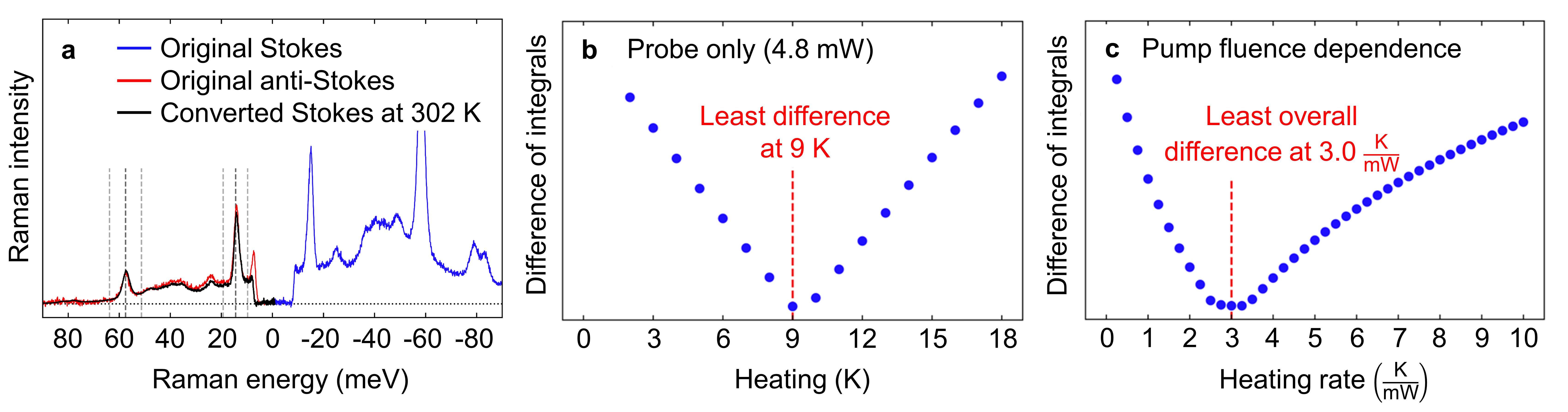}
	\caption{\label{fig:S8}(a) Converted probe-only Stokes spectrum (black) calculated from the original Stokes data (blue) according to eq. \ref{AS_S} at an effective temperature of 302~K (base temperature = 293~K). To determine the effective sample temperature of 302~K, the integral difference between the original anti-Stokes data (red) and the converted Stokes data around the dominant phonon modes for different temperatures was evaluated. Integral bounds are shown as vertical light gray dashed lines and the locations used for determining the thermal weighting factor for the high-energy integral are shown as gray dashed lines in the middle of the two integral bounds, respectively. Zero intensity is marked as horizontal dotted line. (b) Difference of the integrals of original anti-Stokes data and converted Stokes data as introduced in (a) as a function of assumed heating for the probe-only data. The effective heating is equal to the difference between the temperature used to convert the Stokes data to the anti-Stokes side and the base temperature of 293~K. One can clearly identify a minimum at a heating of 9~$\pm$ 0.25~K (302~K sample temperature). (c) Result of the algorithm performed on the entire series of fluence-dependent pump-probe measurements at a base temperature of 293~K, showing a heating rate of 3~$\pm$~0.5~K/mW for the pump.}
\end{figure}

\begin{figure}[h]
        \centering
	\includegraphics[width=0.9\linewidth
	]{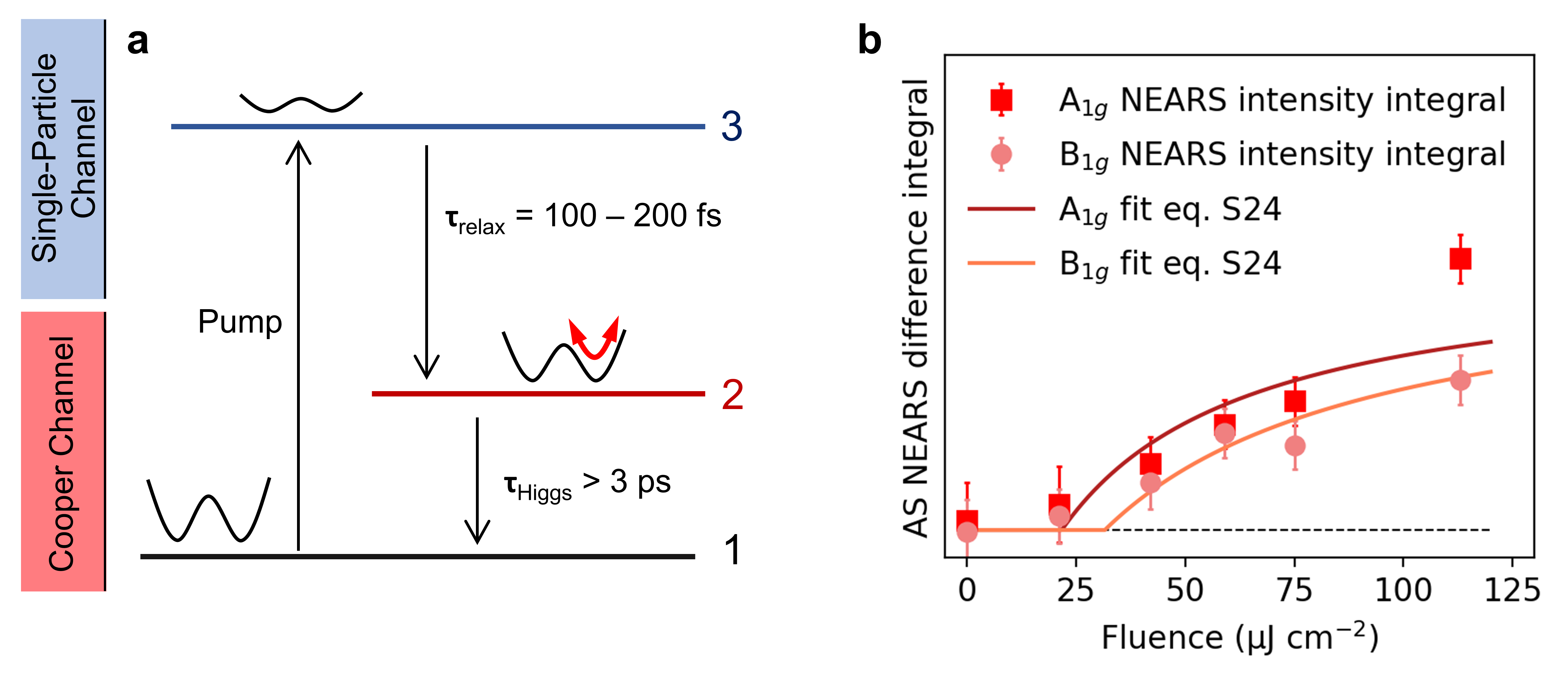}
	\caption{\label{fig:S9}a) Cooper channel vs. single-particle channel in the NEARS experiment on Bi-2212 highlighting the specific excitation process activating the Higgs mode in the NEARS channel. The Mexican-Hat potential representing the superconducting ground state (1) is quenched upon pumping (3). Concomitantly, the reduction of the superfluid density and gap-filling occurs. On a time-scale of 100-200 fs\cite{Perfetti2007} the Mexican-Hat potential relaxes. The holes from the single-particle channel form again bosonic Cooper pairs, which oscillate in the relaxing Mexican-Hat potential representing an excited and metastable Higgs state (2). This population inversion is probed by NEARS leading to a strongly enhanced Raman susceptibility on the anti-Stokes side. It is important to note that this represents a three-level system in which the excited transient Higgs state ($N_2$) has a higher population than the ground state ($N_1$) leading to population inversion and a stronger anti-Stokes than Stokes response. This is only possible in superconductors due to the interplay between single-particle and Cooper channels and the metastable character of the Higgs excitations as the lowest-energy collective excitation of the Cooper channel. Conventional excitations such as phonons cannot show this behavior. (b) Integrated anti-Stokes NEARS difference intensity (Higgs mode intensity) (see Fig. 2 main text) for A$_{1\textrm{g}}$ (red squares) and B$_{1\textrm{g}}$ (light red circles) symmetry, together with the strength of population inversion following equation \ref{popinv} with a critical fluence of $21.7 \pm 5.3$~µJ~cm$^{-2}$ for A$_{1\textrm{g}}$ (dark red solid line) and $31.6 \pm 2.3$~µJ~cm$^{-2}$ for B$_{1\textrm{g}}$ symmetry (light red solid line). The dashed horizontal line marks zero.}
\end{figure}

\begin{figure}[h]
       \centering
	\includegraphics[width=0.5\linewidth
	]{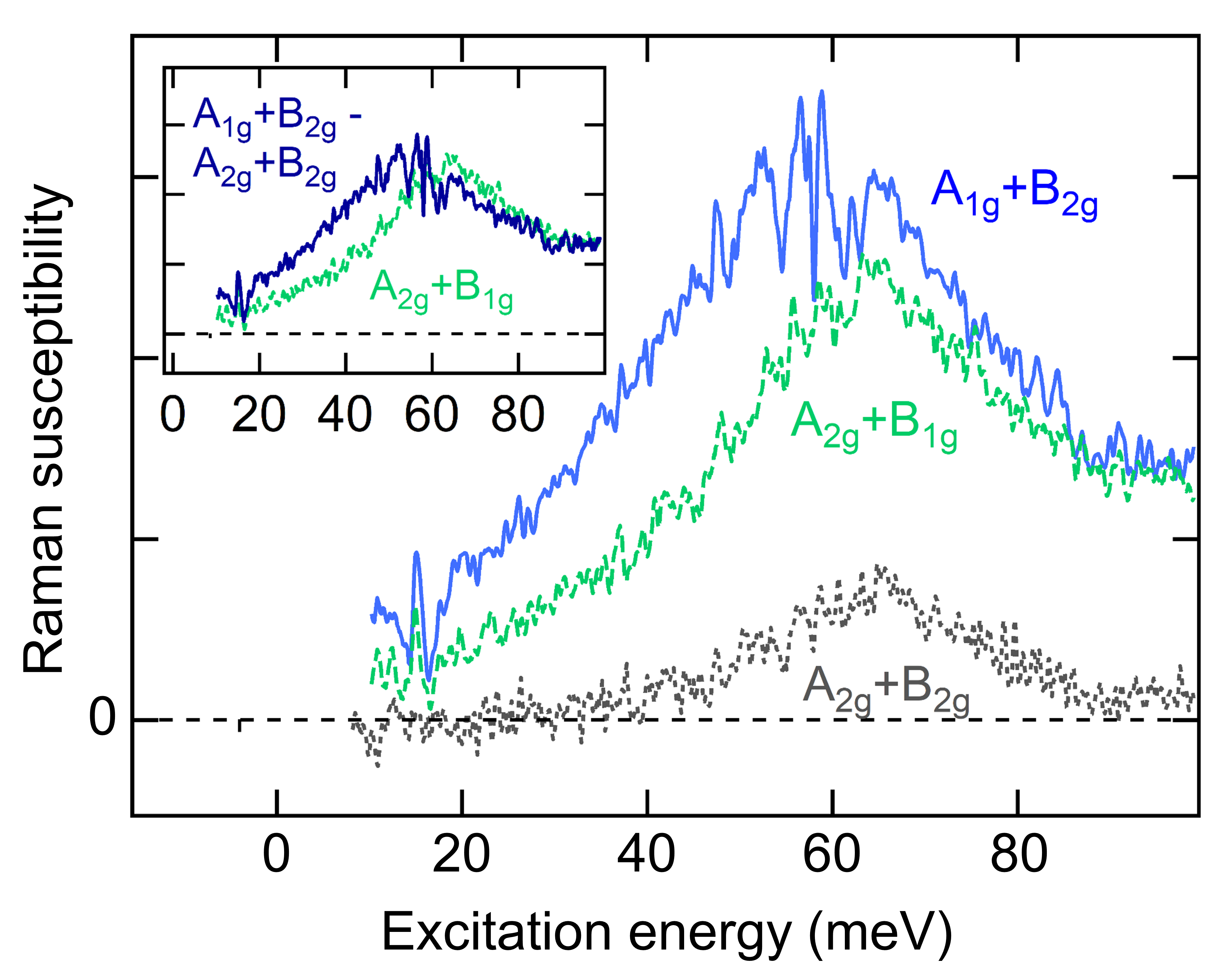}
	\caption{\label{fig:S10}Raman susceptibilities of the pair-breaking feature in the experimental scattering configurations A$_{1\textrm{g}}$+B$_{2\textrm{g}}$ (blue), A$_{2\textrm{g}}$+B$_{1\textrm{g}}$ (green), and A$_{2\textrm{g}}$+B$_{2\textrm{g}}$ (gray). To extract these curves from our Raman intensities, we corrected for the Bose function and subtracted all phonons as determined by our parameterization approach (see eq. \ref{stokesfitting}). The inset shows the comparison between the electronic B$_{1\textrm{g}}$ feature (green, A$_{2\textrm{g}}$+B$_{1\textrm{g}}$) together with the subtracted feature A$_{1\textrm{g}}$+B$_{2\textrm{g}}$ - A$_{2\textrm{g}}$+B$_{2\textrm{g}}$ (dark blue).}
\end{figure}

\begin{figure}[h]
       \centering
	\includegraphics[width=0.5\linewidth
	]{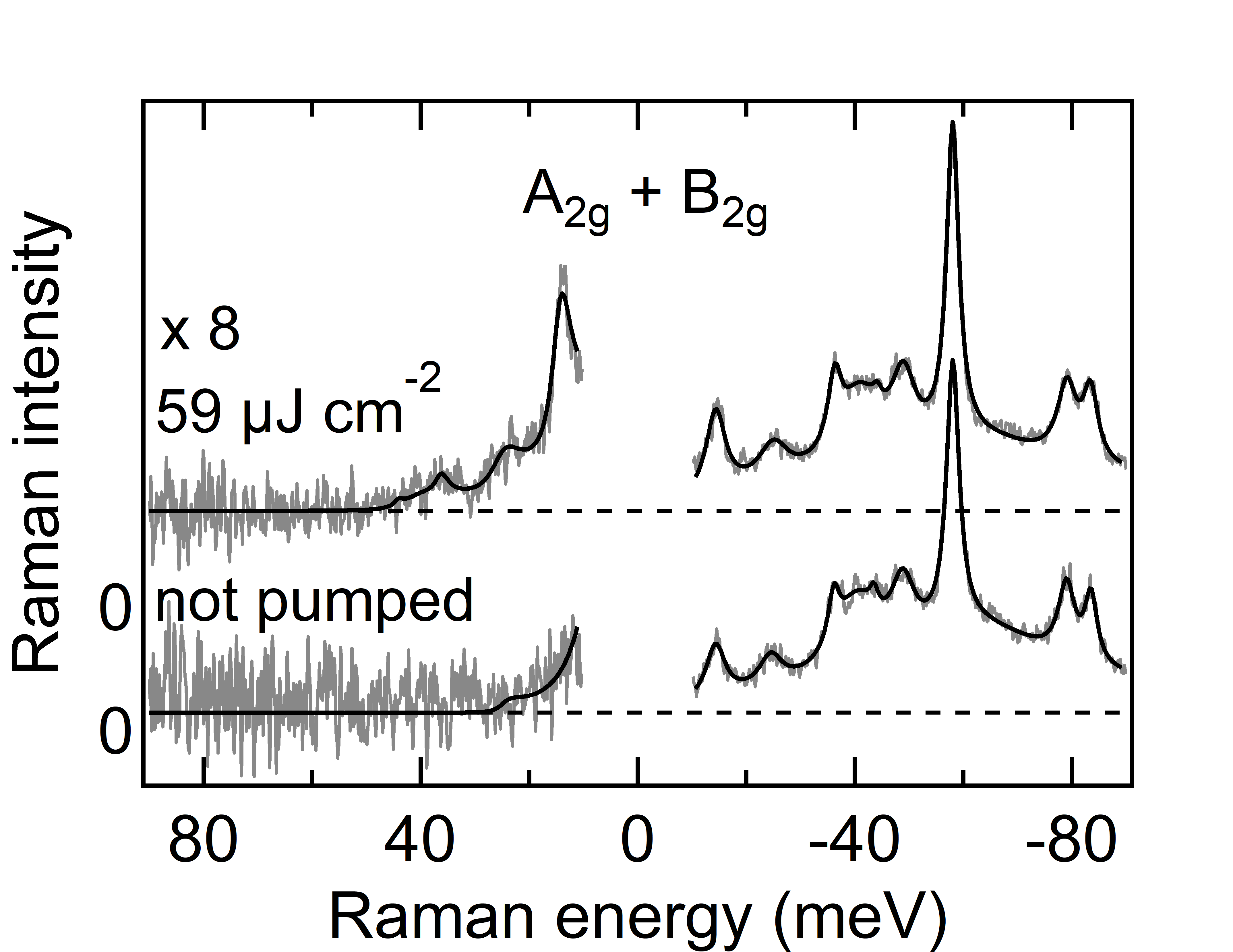}
        \caption{\label{fig:S11}A$_{2\textrm{g}}$+B$_{2\textrm{g}}$ Stokes and anti-Stokes Raman data (gray) and fits as discussed in S.4 and Fig. 1 and 2 (main text). The bottom spectra show probe-only (equilibrium) data. For the upper spectra a pump fluence of 59~µJ~cm$^{-2}$ was applied. In contrast to the A$_{1\textrm{g}}$ and B$_{1\textrm{g}}$ data presented in the main text, we cannot identify a Higgs mode here. The anti-Stokes data in the puped state can be fully described by the phonons and electronic background fitted to the Stokes side. No difference occurs.}
\end{figure}

\begin{figure}[h]
       \centering
	\includegraphics[width=0.6\linewidth
	]{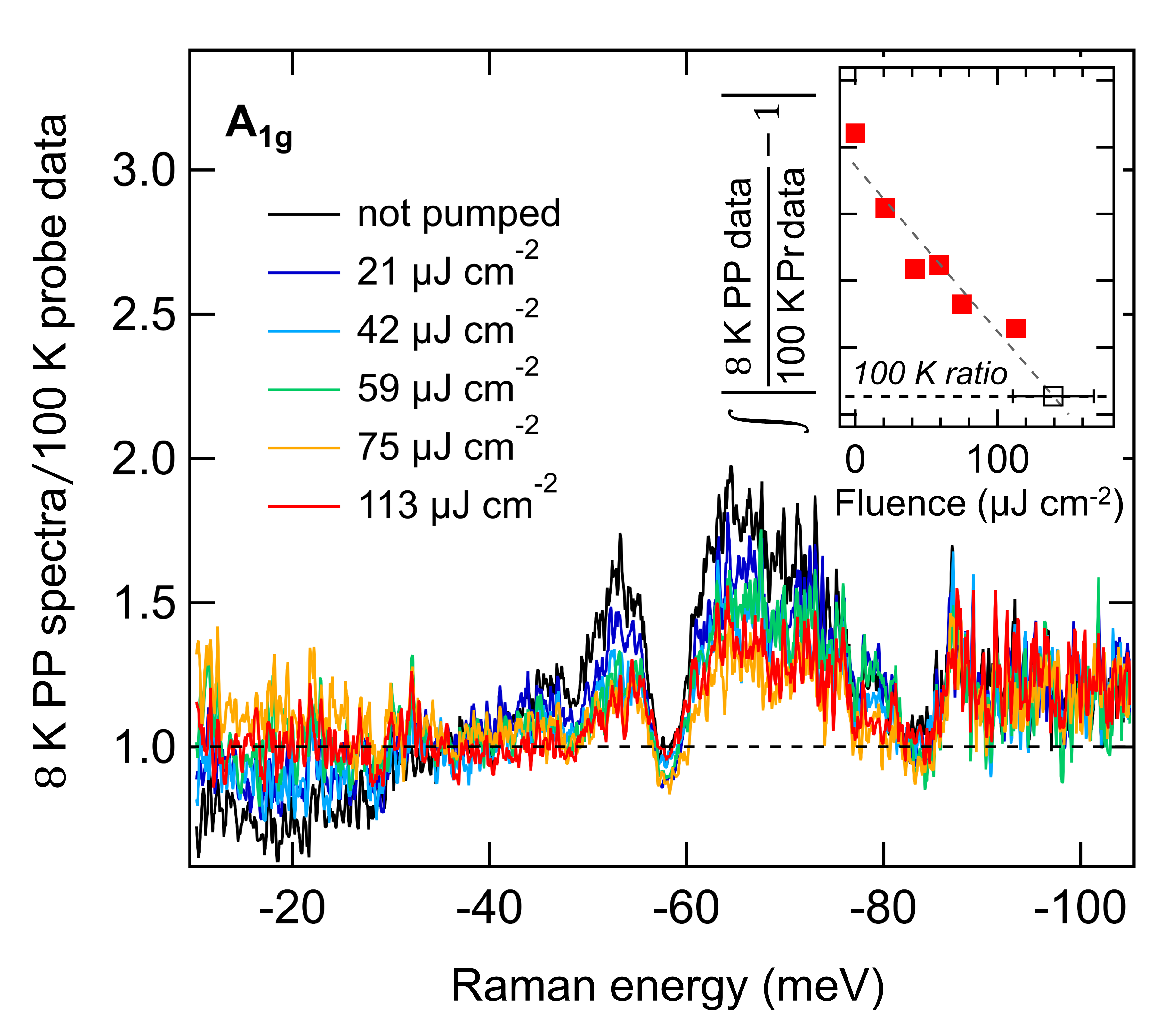}
	\caption{\label{fig:S12} Ratios between 8 K Stokes and 100 K Stokes Raman spectra. Bi-2212 A$_{1\textrm{g}}$ Stokes Raman susceptibilities at 8 K base temperature and different pump fluences are divided by the 100 K non-pumped (probe-only) susceptibility. This comparison shows the pair-breaking feature around 60~meV and the superconducting gap below 30~meV (see also Fig. \ref{fig:S13}). As a function of pump fluence, the pair-breaking peak gets suppressed and gap-filling occurs. The dashed horizontal line marks the ratio of 1. The inset shows the integrated values for the absolute values of the displayed ratios ($\int{\abs{\textrm{8 K PP spectra / 100 K Probe}-1}}$) representing a measure for the strength of the superconductivity-induced feature on the Stokes side. A linear regression fit intersects with the base line for the ratio determined in the normal state at a fluence of $139.3 \pm 28.5$~µJ~cm$^{-2}$ (black square with error bar in the inset). This is an alternative method to derive the effective sample temperature compared to $T_C$ and it's result is in agreement with our method shown in Fig. \ref{fig:S8}. For the latter, we obtain an equilibrium temperature for the sample at 8 K base temperature and with a pump fluence of 113~µJ~cm$^{-2}$ of $98 \pm 13.75$ K, which corresponds to $T_C$ within the error bar. Here, $T_C$ is reached at a fluence of  $139.3 \pm 28.5$~µJ~cm$^{-2}$, which includes 113~µJ~cm$^{-2}$ within the error bar.}
\end{figure}

\begin{figure}[h]
       \centering
	\includegraphics[width=1\linewidth
	]{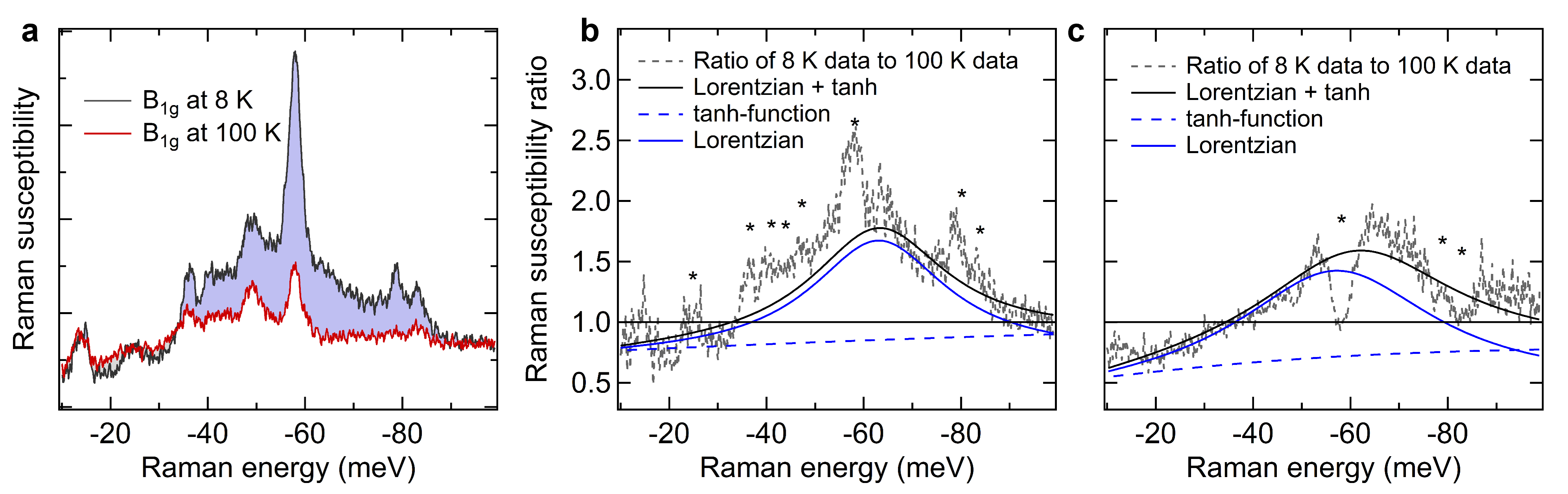}
	\caption{\label{fig:S13}(a) Bi-2212 B$_{1\textrm{g}}$ Stokes Raman spectra (not pumped) at 8 K (gray) and 100 K (red) base temperatures. The comparison between 8 K and 100 K shows the pair-breaking feature (highlighted in blue) and a slight gap feature at this excitation wavelength of 400 nm (3.1 eV). This result is in agreement with previous studies.\cite{Budelmann2005} (b) Ratio between 8 K and 100 K B$_{1\textrm{g}}$ Raman susceptibilities as shown in (a) (gray dashed). The horizontal line marks the ratio of 1. Stars mark phonon positions which lead to artifacts in the ratio due to slight differences in phonon intensity and width at the different temperatures. A model-independent parameterization of the electronic susceptibility can be achieved via a tanh-function (dashed blue) together with a Lorentzian (solid blue). The sum of both curves is shown as black solid line. (c) Ratio of 8 K to 100 K data for A$_{1\textrm{g}}$ symmetry. The parameterization of the electronic susceptibility is presented in the same way as shown in (b).}
\end{figure}

\begin{figure}[h]
        \centering
	\includegraphics[width=.6\linewidth
	]{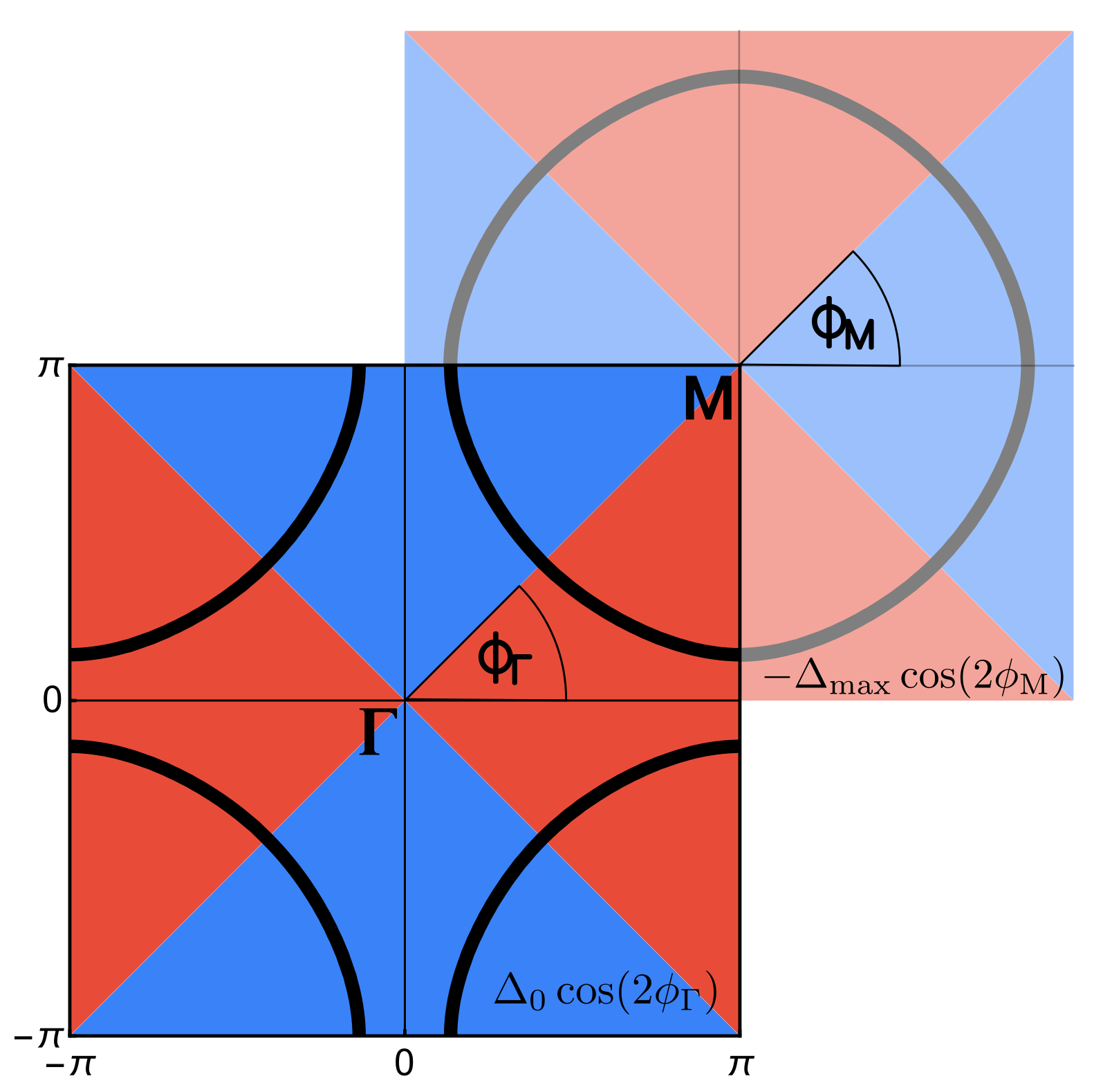}
	\caption{\label{fig:S14}Schematic view of the typical Fermi surface of a cuprate SC like Bi-2212. The d-wave symmetry of the gap gives an angular dependence in reciprocal space of the form $\Delta_\textbf{k} = \Delta_0 \cos{2\phi}$, the sign of which is indicated in red and blue. Due to the first Brillouin zone's periodicity, the Fermi surface can be viewed as connected around the point $M = (\pi,\pi)$. In terms of the angle defined around the $M$-point the gap is described by $\Delta_\textbf{k} \approx -\Delta_\text{max} \cos{2\phi}$. The inversion of the absolute phase is irrelevant since it is gauge-dependent. The appearance of the new parameter $\Delta_\text{max}\leq\Delta_0$ is due to the distance of the Fermi surface from the coordinate axes (when viewed around $\Gamma$), where $\cos 2\phi_\Gamma = 1$. However, for the M enclosing Fermi surface $\Delta_\text{max}$ is the only physical gap amplitude and is therefore called $\Delta_0$ as is conventional.}
\end{figure}

\end{document}